\documentclass[twoside, 11pt]{article}

\usepackage{geometry}
\usepackage{amsmath, amsthm, amssymb, mathrsfs, BOONDOX-cal, dsfont, upgreek}
\usepackage{graphicx, caption, float, multirow, arydshln}

\usepackage{chngcntr, etoolbox}

\usepackage[hidelinks]{hyperref}
\usepackage[dvipsnames]{xcolor}
\hypersetup{colorlinks=true, linkcolor=NavyBlue, citecolor=NavyBlue}

\usepackage{txfonts}
\usepackage[T1]{fontenc}

\usepackage{fancyhdr}
\counterwithin*{equation}{section}
\counterwithin*{equation}{subsection}
\renewcommand\theequation{\ifnumgreater{\value{subsection}}{0}{\thesubsection.}{\thesection.}\arabic{equation}}

\allowdisplaybreaks

\newtheoremstyle{theorem}
  {10pt}
  {10pt}
  {\sl}
  {\parindent}
  {\bf}
  {. }
  { }
  {}
\theoremstyle{theorem}
\newtheorem{theorem}{Theorem}
\newtheorem{proposition}{Proposition}
\newtheorem{corollary}{Corollary}

\renewcommand{\Lambda}{\varLambda}

\begin{document}

\title{\bf Crypto Inverse-Power Options and Fractional \\ Stochastic Volatility}
\author{Boyi Li\footnote{Department of Finance, Boston University, Boston, MA 02215, USA. Email: \underline{liboyi@bu.edu}}\and Weixuan Xia\footnote{Department of Mathematics, University of Southern California, Los Angeles, CA 90089, USA. \newline\indent Corresponding author. Email: \underline{weixuanx@usc.edu}}}
\date{2024}
\maketitle

\thispagestyle{plain}

\begin{abstract}
  Recent empirical evidence has highlighted the crucial role of jumps in both price and volatility within the cryptocurrency market. In this paper, we integrate price--volatility co-jumps and volatility short-term dependency into a coherent model framework, featuring fractional stochastic volatility. We particularly focus on inverse options, including the emerging Quanto inverse options and their power-type generalizations, aiming at mitigating cryptocurrency exchange rate risk and adjusting inherent risk exposure. Characteristic function-based pricing--hedging formulas are derived for these inverse options. The model framework is applied to asymmetric Laplace jump-diffusions and Gaussian-mixed tempered stable-type processes, employing three types of fractional kernels, for an extensive empirical analysis involving model calibration on two independent Bitcoin options data sets, during and after the COVID-19 pandemic. Key insights from our theoretical analysis and empirical findings include: (1) the superior performance of fractional stochastic-volatility models compared to various benchmark models, including those incorporating jumps and stochastic volatility, along with high computational efficiency when utilizing a piecewise kernel, (2) the practical necessity of considering jumps in both price and volatility, along with rough volatility, in pricing and hedging cryptocurrency options, (3) stability of calibrated parameter values in line with stylized facts. \medskip
\end{abstract}

\newcommand{\dd}{{\rm d}}
\newcommand{\pd}{\partial}
\newcommand{\ii}{{\rm i}}
\newcommand{\as}{{\rm a.s.}}
\newcommand{\PP}{\mathsf{P}}
\newcommand{\Q}{\mathsf{Q}}
\newcommand{\E}{\mathsf{E}}
\newcommand{\Var}{\mathsf{Var}}
\newcommand{\Cov}{\mathsf{Cov}}
\newcommand{\Skew}{\mathsf{Skew}}
\newcommand{\Kurt}{\mathsf{Kurt}}
\newcommand{\1}{\mathds{1}}
\newcommand{\T}{\mathcal{T}}
\newcommand{\Gf}{\mathrm{\Gamma}}
\newcommand{\sgn}{\mathrm{sgn}}
\newcommand{\F}{\mathrm{F}}
\newcommand{\Li}{\mathrm{Li}}

\renewcommand{\Re}{\mathrm{Re}}
\renewcommand{\Im}{\mathrm{Im}}
\renewcommand{\Delta}{\varDelta}
\renewcommand{\Pi}{\varPi}

\def\crypto{%
  \leavevmode
  \vtop{\offinterlineskip 
    \setbox0=\hbox{C}%
    \setbox2=\hbox to\wd0{\hfil\hskip.05em
    \vrule height .3ex width .15ex\hskip.06em
    \vrule height .3ex width .15ex\hfil}
    \vbox{\copy2\box0}\box2}}

\vspace{0.1in}

\section{Introduction}

In the present paper we introduce a class of fractional stochastic-volatility models that are capable of reconciling various prevailing stylized facts in financial markets. Broadly speaking, our main objective is to design a general model framework that accommodates both flexible jump dynamics and certain path-dependence properties, and simultaneously achieve fast, semi-analytical valuation. The proposed models address the shortcomings of various existing ones that struggle to provide comprehensive modeling and calibration within a reasonable time frame, while they also maintain parametric parsimony by adding only one or two optional parameters, depending on the desired jump behavior. Following this model framework, we revisit the pricing--hedging problem of derivatives in the cryptocurrency market, where significant price and volatility jumps have been well-documented. Particular attention is drawn to inverse contracts, namely those that are settled in cryptos (\crypto) despite the underlying being valued in a fiat currency (e.g., USD (\$)), which account for the majority of actively traded options across crypto exchanges (including the Deribit exchange).

The recent paper [Alexander et al., 2023] \cite{ACI} gave an interesting discussion on the valuation of crypto derivatives under the Black--Scholes model. The authors there argued the prevalence of inverse contracts (futures and options) in the crypto market and discussed their pricing--hedging implications, especially their importance in risk management for international investors denominated in currencies different from the USD. The study also introduced a class of generalizations called ``Quanto inverse options,'' which, instead of converting the settlement price to fiat at the spot rate, use a predetermined conversion rate and provide flexibility in terms of lifting market frictions in crypto exchanges, besides accounting convenience. A closely related work is [Lucic and Sepp, 2024] \cite{LS}, which explained the generic connections between crypto inverse contracts and their direct (vanilla) counterparts using the num\'{e}raire invariance principle, apart from introducing crypto-based accounting rules for measuring performances of delta-hedged strategies.\footnote{We also refer to [Deng et al., 2021] \cite{DPZZ} for a study of the optimal trading problem of Bitcoin spot and inverse futures in a similar diffusion-driven market, revealing the benefits from trading inverse futures when Bitcoin spot volatility is high.}

From an empirical viewpoint, we will consider both traditional crypto inverse options and these Quanto inverse options under the proposed fractional stochastic-volatility model, while further putting forward a new class of exotic options, termed ``inverse-power options.'' Despite posing a completely new class of exotic derivatives, these options are directly comparable to traditional power options (see [Blenmann and Clark, 2005] \cite{BC}, [Blenman et al., 2022] \cite{BB-GC}, [Xia, 2017] \cite{X1}, [Xia, 2019] \cite{X2}, and [Hussain, 2023] \cite{H2}, among others), albeit with non-positive powers on the underlying price -- in the same way as how an inverse option reformats its direct vanilla counterpart written on the same underlying. These new contract types provide insights into a concrete pricing mechanism for the Quanto inverse options discussed in [Alexander et al., 2023] \cite{ACI}, while they also share similar functionalities with the Quanto types in adjusting risk exposure for fiat currency-based investors, but in a more flexible way -- incorporating nonlinearities.

\subsection{Jumps in crypto prices and volatility}

It is well-known that crypto prices are highly prone to large fluctuations over short time periods, and models with jumps and stochastic volatility are naturally preferred over continuous, diffusion-driven ones for describing crypto price dynamics. This phenomenon has been ascribed to the market's highly speculative and unregulated nature ([Blau, 2017] \cite{B} and [Scaillet et al., 2020] \cite{STT}), market illiquidity ([Kang and Kim, 2019] \cite{KK} and [Zhang and Li, 2023] \cite{ZL}), as well as investor attention ([Troster et al., 2019] \cite{TTSM} and [Philippas et al., 2019] \cite{PRGG}).

From a derivative pricing perspective, crypto (especially Bitcoin) price dynamics have been explored extensively in recent studies. To mention a few, [Madan et al., 2019] \cite{MRS} considered calibrating a series of Markov models on Bitcoin options, concluding that models with infinitely active jumps and volatility clustering are generally favorable in terms of pricing errors. This aspect was also strengthened in [Hou et al., 2020] \cite{HWCH}, which evidenced the importance of jumps in crypto prices and volatility, as well as their co-movements, by considering both statistical estimation and option pricing under two stochastic-volatility models with correlated jumps from [Duffie et al., 2000] \cite{DPS} and [Bandi and Ren\`{o}, 2016] \cite{BR}, respectively. Significantly, [Hou et al., 2020] \cite{HWCH} has documented an inverse leverage effect, indicating positive serial correlations between crypto price and volatility jumps. Notably, while the diffusive leverage effect may remain negative or insignificant, the inverse leverage effect is particularly pronounced during periods of large fluctuations, which phenomenon is uniquely captured by jump models. We refer to [Huang et al., 2022] \cite{HNX} for a detailed discussion based on analyzing crypto time series data. Besides, the papers [Jalan et al., 2021] \cite{JMA} and [Hilliard and Ngo, 2022] \cite{HN} also stressed, respectively, the importance of stochastic volatility and price jumps by comparing the pricing performances of a regime-switching stochastic-volatility model and a jump-diffusion model with stochastic convenience yield to that of the Black--Scholes model. Furthermore, [Cao and Celik, 2021] \cite{CC} studied the valuation of Bitcoin options under an equilibrium setting with various sources of jump risks, again highlighting the undervaluation of deeply-in-the-money options by the Black--Scholes model.

\subsection{Short-range volatility dependence}

The importance of price and volatility jumps aside, cryptos have also been documented to exhibit anti-persistence, or short-range dependence, in their (instantaneous) volatility; see [Takaishi, 2020] \cite{T} for empirical evidence regarding Bitcoin. Equivalently speaking, the conditional volatility of Bitcoin tends to reverse its moving directions more frequently than what a usual autoregressive process implies. In the equity market, such a property, also known as ``rough volatility'' since the pioneering work [Gatheral et al., 2018] \cite{GJR}, is regularly modeled through a class of affine Volterra processes (including the rough Heston model as a special case) proposed by [Abi-Jaber et al., 2019] \cite{A-JLP}, under which semi-closed option pricing formulas based on characteristic functions are available ([El Euch and Rosenbaum, 2019] \cite{ER2}); we also refer to [Bondi et al., 2024] \cite{BLP} for generalizations to include volatility jumps. With this additional aspect in mind, volatility models with a short-range dependence feature should be further preferred over usual stochastic-volatility models for describing crypto price dynamics.

On a general level, the paper [Wang and Xia, 2022] \cite{WX} made the first attempt to reconcile these two important properties of instantaneous volatility, namely jumps and short-range dependence (not to mention suitable jump activity), by writing the instantaneous (squared) volatility as a fractional L\'{e}vy-driven Ornstein--Uhlenbeck process with a square-integrable kernel. While the paper revolved around the equity market, the proposed methodology easily extends to any asset classes. Indeed, the basic idea was to infuse jumps of a broad spectrum of activity into the volatility behavior without jeopardizing short-range dependence. Superior calibration performance has been achieved \text{ibid.} on Volatility Index (VIX) options even under abnormal market conditions during the COVID-19 pandemic.

A major theoretical contribution of the present paper will be to formally combine this volatility structure with the evolution of the underlying and demonstrate its compatibility with a wide range of L\'{e}vy models, subject to flexible correlations, or (direct or inverse) leverage effect.\footnote{Such a combination was briefly discussed in [Wang and Xia, 2022, \text{Sect.} 7] \cite{WX}, where the underlying is assumed to have a diffusion component.} The resultant model framework, possessing fractionally-driven instantaneous volatility, can be seen as a generalized Barndorff-Nielsen--Shephard model with heavy-tailed innovations governing both price and volatility movements and a square-integrable Volterra-type kernel for temporal dependence in volatility and automatically has the capacity of incorporating price--volatility co-jumps, as discussed above. It is also the first model framework, to our knowledge, that is capable of simultaneously accounting for both of these empirical facts. In view of computational efficiency, rather than Monte Carlo methods, we will focus on a pricing--hedging procedure based on characteristic functions, which can be dynamically achieved with the aid of crypto variance swaps.\footnote{Crypto variance swaps have been traded on-chain for several years; see [Alexander and Imeraj, 2020] \cite{AI} and [Woebekking, 2021] \cite{W} for a detailed construction of (VIX-like) crypto volatility measures.}

As an important takeaway from this study, we suggest replacing the widely adopted exponential-Riemann--Liouville kernel for constructing Volterra processes with the piecewise kernel proposed in [Wang and Xia, 2022] \cite{WX}. Through a comparative empirical analysis based on Bitcoin options, we shall explain why the piecewise kernel should be the default kernel choice in constructing fractional models, allowing for computational efficiency, goodness-of-fit, and amenability to explicit valuation at the same time, apart from showing the general applicability and efficiency of the proposed fractional stochastic-volatility model. Towards that end, we also hint at its potential use in constructing a modified version of the rough Heston model ([El Euch and Rosenbaum, 2019] \cite{ER2}), which is particularly advantageous in dealing with volatility-of-volatility dependence on volatility; however, a formal treatment of such a model will be retained for future research.

\subsection{Organization of paper}

The remainder of this paper is organized as follows. In Section \ref{sec:2} we provide a brief review of L\'{e}vy-driven Volterra processes and formally construct our fractional stochastic-volatility model (FSV model hereinafter), discussing the (conditional) distributional properties of resultant log-crypto prices. We also briefly discuss model transformations under measure changes, which provide insights into corresponding real-world price dynamics. In Section \ref{sec:3}, we derive generic characteristic function-based pricing--hedging formulas for the crypto inverse-power derivatives and discuss their benefits on managing risk exposure. These formulas are directly applicable to the analytical pricing--hedging of Quanto inverse options as introduced in [Alexander et al., 2023] \cite{ACI}, subject to general model distributions. Section \ref{sec:4} serves to specify the FSV model for subsequent implementation and to show that it accommodates many common choices of base models and fractional kernels, and in some cases a closed-form characteristic function can be hoped for. In Section \ref{sec:5}, empirical modeling is undertaken using six combinations of specified models, with the aim of validating various stylized facts within the crypto market and assessing the overall efficiency of the FSV model framework and its associated pricing--hedging formulas. Conclusions and further research directions are given in Section \ref{sec:6}, with all mathematical proofs to follow in Appendix \ref{A}.

\medskip

\section{Model framework}\label{sec:2}

We assume that the crypto market evolves in continuous time $t\geq0$ and is governed by a stochastic basis $(\Omega,\mathcal{F},\PP,\mathbb{F}\equiv\{\mathcal{F}_{t}:t\geq0\})$. The market filtration $\mathbb{F}$ satisfies the usual conditions and contains all market information, and $\PP$ is the physical measure. All stochastic processes to appear are understood to be supported on this basis.

For a generic crypto (\crypto), our modeling approach adopts the canonical exponential framework for representing its price dynamics, which is substantially equivalent to specifying a model for the associated log-return process. To this end, we write $S\equiv(S_{t})$ for the crypto price, henceforth the crypto/USD conversion rate, and postulate that it has the following general dynamics under $\PP$:
\begin{equation}\label{2.1}
  S_{t}=S_{0}e^{\xi_{t}}\;(\$),\quad t\geq0,
\end{equation}
where $S_{0}$ is the spot rate and $\xi\equiv(\xi_{t})$ stands for the log-cumulative return, i.e., $\xi=\log(S/S_{0})$. A wide range of well-known existing models can largely be seen as outcomes of specifying a particular stochastic process for $\xi$ -- such as pure diffusion processes in the Black--Scholes and Heston models, or processes incorporating jumps, as in jump-diffusion and Gaussian mixture models; compare, e.g., [Madan, 2019, \text{Sect.} 3] \cite{MRS}.

Within the general structure (\ref{2.1}), we introduce the FSV model through a hierarchical framework that nests a base process governing return dynamics and a latent component representing stochastic volatility. These elements will be detailed in the following two subsections in proper order. With our primary focus on valuation, we conduct the analysis directly under a preselected (risk-neutral) pricing measure $\Q$, equivalent to the physical measure $\PP$.

\subsection{Fractional stochastic volatility with jumps}\label{sec:2.1}

When jumps occur in both price and volatility dynamics, a common yet powerful technique that avoids explicitly specifying an instantaneous variance (or squared volatility) process is to incorporate a time-change argument, which works by replacing the calendar time $\imath\equiv(t)$ with a random, constantly increasing business time, denoted as $\T\equiv(\T_{t})$. Intuitively speaking, the business time serves to capture variations in trading activity, incorporate return autocorrelations, and create volatility clustering. Its theoretical validity is deep-rooted in the so-called ``theorem of time change'' by [Monroe, 1978] \cite{M}, which states that any semimartingale can be represented as a time-changed Brownian motion. By extension, the time change of a generally semimartingale is valid provided that the sample paths of $\T$ are continuous $\Q$-a.s. With the monotonicity of $\T$, this implies the typical construction
\begin{equation}\label{2.1.1}
  \T_{t}=\int^{t}_{0}A_{s}\dd s,\quad t\geq0,
\end{equation}
where the process $A\equiv(A_{t})$, being nonnegative, is referred to as the instantaneous activity rate of time change and serves as a proxy for instantaneous variance, measuring the intensity of volatility clustering.\footnote{If the underlying price is purely driven by diffusion, such a proxy coincides with the instantaneous variance itself (up to indistinguishability); otherwise it is generally not exactly the same when jumps are present.} We refer to [An\'{e} and Geman, 2000] \cite{AG}, [Geman et al., 2001] \cite{GMY}, [Carr et al., 2003] \cite{CGMY}, and [Carr and Wu, 2004] \cite{CW} for more details regarding time-changed L\'{e}vy processes.

In general, since volatility is mean-reverting in the long run, a natural model choice for $A$ in the presence of jumps is a L\'{e}vy-driven Ornstein--Uhlenbeck process. In the present context, we modify the linear structure with kernel convolution for short-range dependence, by adapting [Wang and Xia, 2022, \text{Sect.} 2.2] \cite{WX}. Such modification readily leads to fractional Ornstein--Uhlenbeck processes discussed in [Wolpert and Taqqu, 2004] \cite{WT}, and these processes, having many attractive analytical properties, belong to the broader class of L\'{e}vy-driven Volterra processes detailed in [Barndorff-Nielsen et al., 2014] \cite{B-NBPV}.

More precisely, let $Y\equiv(Y_{t})$ be a square-integrable L\'{e}vy subordinator (nondeterministic by default), with (L\'{e}vy) characteristics $(\mu_{Y},0,\nu_{Y})$ under $\Q$, where $\mu_{Y}\geq0$ and the L\'{e}vy measure $\nu_{Y}$ is non-atomic and corresponds to a Poisson random measure $N_{Y}$ defined on $(\mathbb{R}_{++},\mathbb{R})$, such that $Y_{1}$ admits the characteristic exponent\footnote{By default, expectation is taken under the pricing measure $\Q$, i.e., $\E\equiv\E^{\Q}$.}
\begin{equation*}
  \log\phi_{Y_{1}}(u):=\log\E[e^{\ii uY_{1}}]=\ii u\mu_{Y}+\int^{\infty}_{0+}(e^{\ii uz}-1)\nu_{Y}(\dd z),\quad u\in\mathbb{R}
\end{equation*}
according to the L\'{e}vy--Khintchine formula. By convention, we assume that $Y_{1}$ has finite moments of all orders.

Then, consider a continuously differentiable kernel function $h:\{(t,s):t>0,s\in[0,t)\}\mapsto\mathbb{R}$ with the square-integrability property
\begin{equation*}
  \sup_{t>0}\int^{t}_{0}h^{2}(t,s)\dd s<\infty
\end{equation*}
and the tail behaviors
\begin{equation}\label{2.1.2}
  h(t+v,t)=
  \begin{cases}
    O(v^{d-1})&\quad\text{as }v\searrow0 \\
    O(e^{-\kappa v})&\quad\text{as }v\rightarrow\infty,
  \end{cases}
  \quad t>0,
\end{equation}
where $\kappa>0$ and $d\in(1/2,1)$ are a mean-reversion parameter and a fraction parameter, respectively. The fractional counterpart of $Y$ is constructed as the following $h$-convoluted stochastic integral:
\begin{equation}\label{2.1.3}
  Y^{(h)}_{t}:=\int^{t}_{0}h(t,s)\dd Y_{s}=\mu_{Y}\int^{t}_{0}h(t,s)\dd s+\int^{t}_{0}\int^{\infty}_{0+}h(t,s)zN_{Y}(\dd z,\dd s),\quad t\geq0.
\end{equation}

As mentioned in [Wang and Xia, 2022, \text{Sect.} 2.1] \cite{WX}, since $h$ satisfies the singularity property that $\lim_{s\nearrow t}h(t,s)=\infty$, $\forall t>0$, due to $d<1$, then $Y^{(h)}$ exhibits short-range dependence in the following covariance function sense:
\begin{equation}\label{2.1.4}
  \Cov\big[Y^{(h)}_{t},Y^{(h)}_{t+v}\big]=\Var\big[Y^{(h)}_{t}\big]+\Var[Y_{1}]C(t)v^{\varpi}+O(v),\quad\text{as }v\searrow0,
\end{equation}
for some deterministic function $C(t)$ and some constant $\varpi\in(0,1)$. From (\ref{2.1.3}) the L\'{e}vy--It\^{o} isometry implies that $\Cov\big[Y^{(h)}_{t},Y^{(h)}_{t+v}\big]=\Var[Y_{1}]\int^{t}_{0}h(t,s)h(t+v,s)\dd s$, $v\geq0$.

A notably simple choice of $h$ is the exponential-Riemann--Liouville product kernel (\text{a.k.a.} the gamma kernel) (see (\ref{4.1.1})), $h(t,s)\equiv h(t-s)=e^{-\kappa(t-s)}(t-s)^{d-1}/\Gf(d)$, where $\Gf(\cdot)$ is the usual Gamma function and which is widely adopted in the rough volatility literature due to its structural simplicity, including in constructing the rough Heston model (see, e.g., [El Euch and Rosenbaum, 2019] \cite{ER2}) and fractional Ornstein--Uhlenbeck processes in [Wolpert and Taqqu, 2004] \cite{WT}. We will consider other kernel forms as appearing in [Wang and Xia, 2022, \text{Sect.} 2.2] \cite{WX} later in Section \ref{sec:4.1}, and some other special forms of $h$ can be found in [Barndorff-Nielsen et al., 2014, \text{Sect.} 3.1] \cite{B-NBPV}.

The instantaneous activity rate process can then be constructed as the following generalized Ornstein--Uhlenbeck process:
\begin{equation}\label{2.1.5}
  A_{t}=A_{0}e^{-\kappa t}+m(1-e^{-\kappa t})+Y^{(h)}_{t},\quad t\geq0,
\end{equation}
where the parameters are $A_{0}>0$ (initial state of activity), $m\geq0$ (auxiliary mean reversion level), $\kappa$ (speed of mean reversion), and $d\in(1/2,1)$ (fraction) embedded in $h$. The fact that $A$ is mean-reverting can be justified by the right-tail behavior in (\ref{2.1.2}), while the short-range dependence is ensured through (\ref{2.1.4}). In particular, we have that $\lim_{t\rightarrow\infty}A_{t}=m+\E[Y_{1}]\lim_{t\rightarrow\infty}\int^{t}_{0}h(t,s)\dd s>0$ exists.\footnote{Strictly speaking, (\ref{2.1.5}) is called ``generalized'' and not ``fractional'' because $A$ need not satisfy the Ornstein--Uhlenbeck stochastic differential equation, unless $h$ is of a special form, which we will explain in Section \ref{sec:4.1}.} This construction is directly comparable to the instantaneous variance process in [Wang and Xia, 2022, \text{Sect.} 2.2] \cite{WX}.

\subsection{Time-changed processes and crypto prices}\label{sec:2.2}

Based on (\ref{2.1.1}), the business time incorporating fractional stochastic volatility is constructed as the time integral of $A$ in (\ref{2.1.5}), and a major portion of shocks to the crypto price are governed by some $\T$-changed L\'{e}vy process, denoted $X\equiv(X_{t})$. Here, $X$ is a real-valued L\'{e}vy process with $\Q$-characteristics $(\mu_{X},\sigma_{X},\nu_{X})$, admitting the characteristic exponent
\begin{equation*}
  \log\phi_{X_{1}}(u):=\log\E[e^{\ii uX_{1}}]=\ii u\mu_{X}-\frac{1}{2}u^{2}\sigma^{2}_{X}+\int_{\mathbb{R}\setminus\{0\}}(e^{\ii uz}-1-\ii uz)\nu_{X}(\dd z),\quad u\in\mathbb{R}.
\end{equation*}
Likewise, we have assumed that $X_{1}$ has finite moments. Also, to capture continual price fluctuations, it is reasonable to assume that either $\sigma_{X}>0$ or $\nu_{X}$ is infinite. The process $X$ is a base process describing return innovations without volatility clustering, which can already have jumps of flexible activity and scales. With stochastic volatility, the crypto price (crypto/USD conversion rate) under the pricing measure $\Q$ is formally given by
\begin{equation}\label{2.2.1}
  S_{t}=\frac{S_{0}e^{X_{\mathcal{T}_{t}}+\rho Y_{t}}}{\phi^{\mathcal{T}_{t}}_{X_{1}}(-\ii)\phi^{t}_{Y_{1}}(-\ii\rho)}\;(\$),\quad t\geq0,
\end{equation}
where $\rho\in\mathbb{R}$ is a leverage parameter channeling the (direct or inverse) leverage effect between prices and volatility. The existence of the (positive) quantities $\phi_{X_{1}}(-\ii)$ and $\phi_{Y_{1}}(-\ii\rho)$ is the same as the requirement that $\E[e^{X_{1}}]<\infty$ and $\E[e^{\rho^{+}Y_{1}}]<\infty$; together, these quantities act as exponential compensators to guarantee the martingale property of $S$ under $\Q$.

The price model (\ref{2.2.1}) can be seen as a generalized version of the well-known Barndorff-Nielsen--Shephard model ([Barndorff-Nielsen and Shephard, 2001] \cite{B-NS}) in which $X$ is particularly a Brownian motion and $A$ is an ordinary Ornstein--Uhlenbeck process, i.e., one with $h(t,s)=e^{-\kappa(t-s)}$, $s\in[0,t)$. An extension where $X$ is a general L\'{e}vy process has been proposed and studied by [Yamazaki, 2016] \cite{Y}, from which the key difference in the present framework mainly lies in the (singular) kernel function $h$.

To enable dynamic valuation of crypto derivatives, we need the conditional distribution of the log-price $\log S_{t}|\mathcal{F}_{t_{0}}$, for generic time points $0\leq t_{0}<t\leq T$, where $T>0$ is a predetermined expiry date. In the presence of stochastic volatility, dynamic valuation is easily achievable provided access to corresponding volatility derivatives, such as a variance swap which entitles the long party to receive a floating leg equal to the realized variance taken over the swap's life at the expiry $T>0$ in exchange for a preset fixed payment such that the swap has zero value at the time of issuance. We refer again to [Alexander and Imeraj, 2020] \cite{AI} and [Woebekking, 2021] \cite{W} for detailed discussions on crypto volatility. Let us denote by $V_{S}(t_{0},t)$ the variance swap on $S$ initiated at time $t_{0}$ and expiring at time $t$. Then, under continuous monitoring, the swap payoff to the long party is exactly the quadratic variation difference of log-returns cumulated from $t_{0}$ to $t$, with spot price given by
\begin{equation}\label{2.2.2}
  V_{S}(t_{0},t):=\E_{t_{0}}\bigg[\bigg[\log\frac{S}{S_{0}},\log\frac{S}{S_{0}}\bigg]_{t}\bigg] -\bigg[\log\frac{S}{S_{0}},\log\frac{S}{S_{0}}\bigg]_{t_{0}}\;(\$),
\end{equation}
where the notation $\E_{t_{0}}[\cdot]$ means $\E[\cdot|\mathcal{F}_{t_{0}}]$. Clearly, (\ref{2.2.2}) is well-defined because $S$ in (\ref{2.2.1}) is a semimartingale.

The following proposition expresses the conditional characteristic function of the log-price in terms of the base characteristic functions ($\phi_{X_{1}}$ and $\phi_{Y_{1}}$) and the above variance swap price.

\begin{proposition}\label{pro:1}
Let $0\leq t_{0}<t$ be fixed in the setting of (\ref{2.2.1}). Then it holds that
\begin{align}\label{2.2.3}
  \phi_{\log S_{t}|t_{0}}(u)&:=\E_{t_{0}}[e^{\ii u\log S_{t}}] \nonumber\\
  &=\exp\bigg(\ii u(\log S_{t_{0}}-(t-t_{0})\log\phi_{Y_{1}}(-\ii\rho)) \nonumber\\
  &\quad+\int^{t}_{t_{0}}\log\phi_{Y_{1}}(\rho u-H(t,s)(\ii\log\phi_{X_{1}}(u)+u\log\phi_{X_{1}}(-\ii)))\dd s \nonumber\\
  &\quad+(\log\phi_{X_{1}}(u)-\ii u\log\phi_{X_{1}}(-\ii))\bigg(\frac{V_{S}(t_{0},t)-\rho^{2}(t-t_{0})\Var[Y_{1}]}{\Var[X_{1}]}-\int^{t}_{t_{0}}H(t,s)\dd s\E[Y_{1}]\bigg)\bigg)
\end{align}
for $u\in\mathbb{R}$, with the tail-integrated kernel
\begin{equation}\label{2.2.4}
  H(t,s):=\int^{t}_{s}h(v,s)\dd v,\quad s\in[0,t).
\end{equation}
\end{proposition}

Technically, in the above expression, apart from the existence of $\phi_{X_{1}}(-\ii)$ and $\phi_{Y_{1}}(-\ii\rho)$ by assumption, note that the integrand $\phi_{Y_{1}}(\rho u-H(t,s)(\ii\log\phi_{X_{1}}(u)+u\log\phi_{X_{1}}(-\ii)))$ is automatically well-defined provided that the integrated kernel $H$ is nonnegative, which condition is satisfied by all fractional kernels $h$ to be considered. Indeed, in this case one has $(H(t,s)\Re\log\phi_{X_{1}}(u))Y_{1}\leq0$, $\Q$-a.s., for any $s\leq t$ and $u\in\mathbb{R}$. Also, $H$ is by construction bounded, hence nonsingular. In general, $H$ and its further integral (with respect to $s\in[t_{0},t)$) can both be easily written in closed form, whereas it is difficult to obtain an explicit expression for the $s$-integral involving $\log\phi_{Y_{1}}$,\footnote{The reason is that products of power and exponential functions do not produce elementary change of variables under the integral sign; see [Wang and Xia, 2022, \text{Appx.} B] \cite{WX} for details.} except when $h$ takes a piecewise form, to be shown in Section \ref{sec:4}.

While the characteristic function (\ref{2.2.3}) already fully determines the required conditional distribution, we give the next result for the existence of the associated density function as the inverse Fourier transform. This existence property will lay the foundation for justifying the general pricing--hedging formulas in Section \ref{sec:3}.

\begin{theorem}\label{thm:1}
For any fixed $0\leq t_{0}<t$, the conditional distribution of the log-price, $\mathcal{L}(\log S_{t}|\mathcal{F}_{t_{0}})$, is absolutely continuous with respect to the Lebesgue measure.
\end{theorem}

We finish this section with a brief discussion of how the crypto price dynamics (\ref{2.1}) under the original physical measure $\PP$ may be reverse-engineered from (\ref{2.2.1}). One possibility under which the L\'{e}vy properties (namely increment independency and stationarity) of $X$ and $Y$ stay is with some constant $\zeta>0$ and the Radon--Nikodym derivative
\begin{equation}\label{2.2.5}
  \frac{\dd\Q}{\dd\PP}\bigg|_{\mathcal{F}_{t}}=\mathcal{E}\bigg(\zeta\log\frac{S}{S_{0}}\bigg)_{t},\quad t\geq0,
\end{equation}
where $\mathcal{E}$ denotes the Dol\'{e}ans--Dade exponential. This type of measure change, known as the Esscher transform (see, e.g., [Madan and Milne, 1991] \cite{MM}), is workable in the present context since the log-cumulative return process $\xi=\log(S/S_{0})$ has a finite second moment.\footnote{In the language of [Jacod and Shiryaev, 2010] \cite{JS}, this means that the process $\log(S/S_{0})$ is ``exponentially special.''} We refer to [Hubalek and Sgarra, 2009] \cite{HS} for a discussion of this type of measure changes specifically for the Barndorff-Nielsen--Shephard model, which has $X$ being a Brownian motion and $h$ the exponential kernel.

In particular, under the pricing measure $\Q$, $\log(S/S_{0})=X_{\T}-\log\phi_{X_{1}}(-\ii)\T+\rho Y-\imath\log\phi_{Y_{1}}(-\ii\rho)$ is a semimartingale with characteristics\footnote{Note that the characteristic exponents $\log\phi_{Y_{1}}$ and $\log\phi_{X_{1}}$ have the integrals with respect to the L\'{e}vy measures written in different forms -- only the latter has employed an integrable truncation function (id). Also, in case $\rho=0$, $\nu_{Y}(\dd(\rho^{-1}z))$ is the zero measure.}
\begin{align*}
  &\big((\mu_{X}-\log\phi_{X_{1}}(-\ii))\T_{t}+t(\rho\mu_{Y}-\log\phi_{Y_{1}}(-\ii\rho)),\;\sigma_{X}\T_{t},\;(\nu_{X}(\dd z)A_{t}+\nu_{Y}(\dd(\rho^{-1}z)))\dd t\big)
\end{align*}
for $(t,z)\in\mathbb{R}_{+}\times(\mathbb{R}\setminus\{0\})$. Then, according to [Jacod and Shiryaev, 2010, \text{Chap.} III \text{Thm.} 7.18 and \text{Thm.} 7.23] \cite{JS}, if the measure change is achieved through (\ref{2.2.5}), it suffices that $\xi$ have the $\PP$-semimartingale characteristics
\begin{equation*}
  \big(\mu_{\tilde{X}}\T_{t}+\mu_{\tilde{Y}}t,\sigma_{\tilde{X}}\T_{t},(\nu_{\tilde{X}}(\dd z)A_{t}+\nu_{\tilde{Y}}(\dd(\rho^{-1}z)))\dd t\big)
\end{equation*}
with
\begin{align}\label{2.2.6}
  \mu_{\tilde{X}}&=\mu_{X}-\log\phi_{X_{1}}(-\ii)-\sigma^{2}_{X}\zeta+\int_{\mathbb{R}\setminus\{0\}}(e^{-\zeta z}-1)z\nu_{X}(\dd z), \nonumber\\
  \mu_{\tilde{Y}}&=\rho\mu_{Y}-\log\phi_{Y_{1}}(-\ii\rho), \nonumber\\
  \sigma_{\tilde{X}}&=\sigma_{X}, \nonumber\\
  \nu_{\tilde{X}}(\dd z)&=e^{-\zeta z}\nu_{X}(\dd z), \nonumber\\
  \nu_{\tilde{Y}}(\dd z)&=e^{-\zeta z}\nu_{Y}(\dd(\rho^{-1}z)),
\end{align}
with the understanding that the quantities $\phi_{X_{1}}(-\ii)$ and $\phi_{Y_{1}}(-\ii\rho)$ take the same values as in (\ref{2.2.1}). This implies that the price dynamics (\ref{2.1}) can be formulated as
\begin{equation}\label{2.2.7}
  S_{t}=S_{0}e^{\tilde{X}_{\T_{t}}+\tilde{Y}_{t}}\;(\$),
\end{equation}
with $\xi=\tilde{X}_{\T}+\tilde{Y}$. Here, $\tilde{X}$ and $\tilde{Y}$ are L\'{e}vy processes with respective L\'{e}vy characteristics $(\mu_{\tilde{X}},\sigma_{\tilde{X}},\nu_{\tilde{X}})$ and $(\mu_{\tilde{Y}},0,\nu_{\tilde{Y}})$, while the business time $\T$ still uses (\ref{2.1.1}) except with $Y$ replaced by $\tilde{Y}/\rho$ (provided $\rho\neq0$). The processes $\tilde{X}$ and $\tilde{Y}$ have the same path regularities -- measured \text{e.g.} by the Blumenthal--Getoor index for the jump components -- as $X$ and $\rho Y$, respectively, and the volatility leverage effect is embedded in $\tilde{Y}$, which is nonetheless no longer a $\rho$-scaled L\'{e}vy process. Therefore, (\ref{2.2.7}) also makes it possible to develop estimation methods for the proposed FSV model.

\medskip

\section{Crypto inverse-power options}\label{sec:3}

As noted in [Alexander et al., 2023] \cite{ACI}, approximately 90\% of active crypto options trading currently takes place on the Deribit exchange. A distinctive feature of Deribit's options is their inverse nature: Contracts are denominated and settled in the underlying crypto although the underlying asset is quoted in USD. Deribit reports USD-equivalent prices by multiplying the crypto-settled payoff by the crypto/USD conversion rate. For example, a Bitcoin option's profit or loss is calculated and settled in Bitcoin, even though the option prices and strike prices are displayed in USD.\footnote{For a comprehensive explanation of Deribit's inverse contract structure and profit calculation, refer to their official guide \href{https://insights.deribit.com/options-course/}{here}.}

For USD-denominated investors, such an inverse contract is substantially equivalent to a direct contract upon immediate crypto/USD conversion at expiry, whose valuation will follow the same procedures for standard derivatives. In this regard, the terminal payoff of an inverse call option written on $S$ with strike price $K$ (\$) and expiry date $T$ satisfies that\footnote{In what follows the subscripts ``(i),'' ``(ip),'' ``(p),'' ``(qi),'' and ``(qip)'' stand for ``inverse,'' ``inverse-power,'' ``power,'' ``Quanto inverse,'' and ``Quanto inverse-power,'' respectively.}
\begin{equation}\label{3.1}
  C^{\rm (i)}_{T}=\frac{(S_{T}-K)^{+}}{S_{T}}=\bigg(1-\frac{K}{S_{T}}\bigg)^{+}\;(\crypto),\quad C_{T}=S_{T}C^{\rm (i)}_{T}=(S_{T}-K)^{+}\;(\$),
\end{equation}
where, upon currency conversion, $C_{T}$ is the payoff of a similar (direct) standard call option. A Quanto inverse contract, on the other hand, serves to mitigate currency risk borne by the inverse contract via specifying a fixed conversion rate $R$ (crypto/USD) at inception; its terminal payoff is given by
\begin{equation}\label{3.2}
  C^{\rm (qi)}_{T}=RC^{\rm (i)}_{T}=R\bigg(1-\frac{K}{S_{T}}\bigg)^{+}\;(\$).
\end{equation}
Thus, the contract remains ``inverse'' in nature and can exhibit a locally concave payoff for (long) calls as shown in [Alexander et al., 2023, \text{Fig.} 4] \cite{ACI}.

In the rest of this section, we present characteristic function-based pricing--hedging formulas for these Quanto inverse contracts as well as their power-type extensions,\footnote{The idea of treating inverse options as special cases of negative-powered options is motivated from an independent study concerning Poisson functionals ([Xia, 2022] \cite{X4}). As seen from (\ref{3.1}), a standard inverse option (pre-conversion) can be regarded as a power option on the same underlying with (negative) power $-1$.} which provide significant flexibility for adjusting risk exposure.

\subsection{Inverse-power options}\label{sec:3.1}

The terminal payoff of an inverse-power call option on $S$, in the setting of (\ref{3.1}), is given by
\begin{equation}\label{3.1.1}
  C^{\rm (ip)}_{T}=\frac{(S^{p_{1}}_{T}-K^{p_{2}})^{+}}{S^{p_{1}}_{T}}=\bigg(1-\frac{K^{p_{2}}}{S^{p_{1}}_{T}}\bigg)^{+}\;(\crypto),
\end{equation}
with two predetermined power coefficients $p_{1},p_{2}\geq0$; for a similar put option, the payoff is
\begin{equation}\label{3.1.2}
  P^{\rm (ip)}_{T}=\bigg(\frac{K^{p_{2}}}{S^{p_{1}}_{T}}-1\bigg)^{+}\;(\crypto).
\end{equation}

This power mechanism provides the option investor with nonlinear leverage on the crypto/USD conversion rate in the same way as usual power options. From the above transformations, the inverse-power call option is equivalent to a put option on a powered USD/crypto conversion rate, and similarly for the put option. The power coefficient $p_{2}$ is explicitly used to ensure that the powered prices are at the same scale and hence ease comparison, though each $K^{p_{2}}$ can be treated as a new strike price. The actual impact of $p_{1}$ on risk exposure depends on the magnitude of the crypto/USD conversion rate relative to 1 as a threshold. For instance, as per status quo, if $S$ is for Bitcoin or Ethereum, then $p_{1}>1$ is most likely associated with increased levels of exposure while $p_{1}<1$ imposes a decreasing effect; however, if $S$ is for Dogecoin or Polygon (MATIC) (also with options available), this effect should be completely reversed. Therefore, the choice of the powers in designing these contracts should be made on a case-by-case basis; for the same reason, the power mechanism will be ineffective for cryptos pegged to USD at 1-to-1, such as Tether.

\begin{figure}[H]
  \centering
  \includegraphics[scale=0.475]{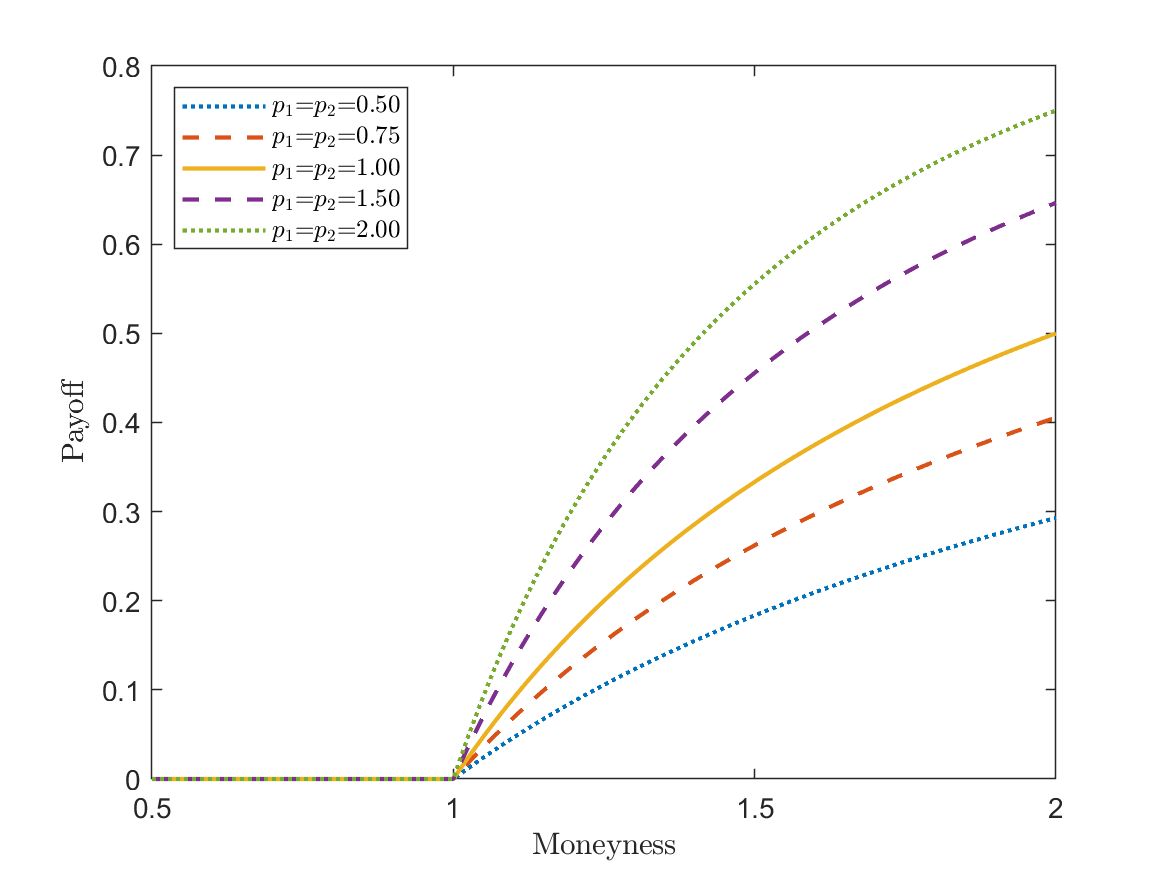}
  \includegraphics[scale=0.475]{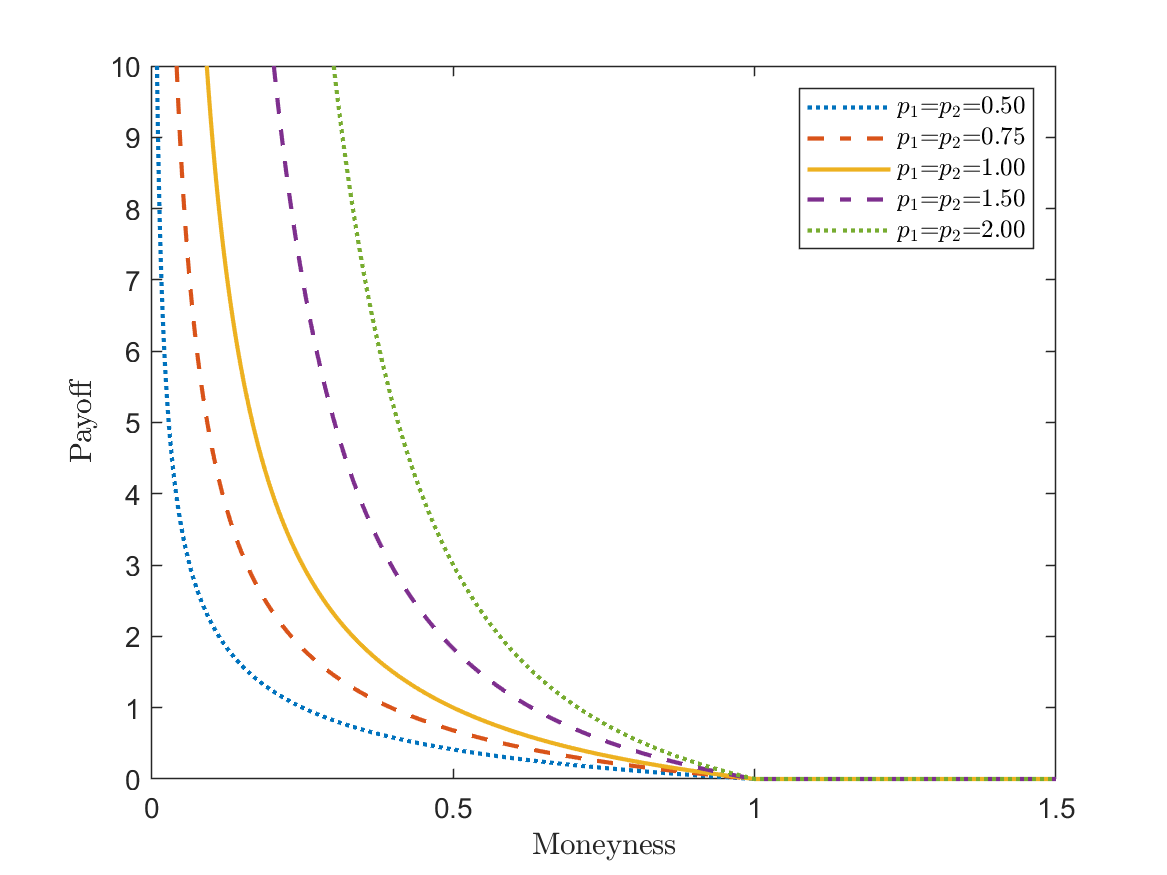}\\
  \caption{Power leverage for inverse options}
  \label{fig:1}
\end{figure}

Figure \ref{fig:1} displays the payoff functions of the inverse-power options for different values of $p_{1}=p_{2}$. We see that powers mainly alter the curvature of the nonlinear part but do not change the locally concave / convex feature, and thus they should be understood as providing an additional layer for risk management.

As discussed earlier, for USD-denominated investors, the above payoffs can be translated into USD at expiry, making the true payoff function (approximately) equal to that of the direct counterparts. In a similar manner to (\ref{3.1}), this means that
\begin{equation}\label{3.1.3}
  C^{\rm (p)}_{T}=S^{p_{1}}_{T}C^{\rm (ip)}_{T}=(S^{p_{1}}_{T}-K^{p_{2}})^{+}\;(\$).
\end{equation}
In contrast, as done in (\ref{3.2}), leveraging a predetermined conversion rate $R$ gives rise to the following Quanto inverse-type payoff:
\begin{equation}\label{3.1.4}
  C^{\rm (qip)}_{T}=R^{p_{1}}C^{\rm (ip)}_{T}=R^{p_{1}}\bigg(1-\frac{K^{p_{2}}}{S^{p_{1}}_{T}}\bigg)^{+}\;(\$),
\end{equation}
which is technically a scalar multiple of the inverse-power option payoff without conversion. This in turn shows that the pricing problem of Quanto inverse-power options from a USD-denominated investor's perspective is analogous to that of inverse-power options for a non-USD-based investor.

For the empirical analysis in Section \ref{sec:5} exclusively involving call options, we will use the (approximate) payoff to USD-denominated investors in (\ref{3.1.3}), in line with Deribit's option quotes. The value of this direct call option at time $t_{0}\in[0,T)$ is given by the formula due to [Bakshi and Madan, 2000] \cite{BM}
\begin{equation}\label{3.1.5}
  C_{t_{0}}=S_{t_{0}}\Pi_{1}-K\Pi_{2}\;(\$),
\end{equation}
where the in-the-money probabilities are
\begin{equation}\label{3.1.6}
  \Pi_{1}=\frac{1}{2}+\frac{1}{\pi}\int^{\infty}_{0}\Re\bigg[\frac{K^{-\ii u}\phi_{\log S_{T}|t_{0}}(u-\ii)}{\ii u\phi_{\log S_{T}|t_{0}}(-\ii)}\bigg]\dd u,\quad \Pi_{2}=\frac{1}{2}+\frac{1}{\pi}\int^{\infty}_{0}\Re\bigg[\frac{K^{-\ii u}\phi_{\log S_{T}|t_{0}}(u)}{\ii u}\bigg]\dd u.
\end{equation}
Equivalently, it can be expressed using the adjusted formula due to [Carr and Madan, 1999] \cite{CM2},
\begin{equation}\label{3.1.7}
  C_{t_{0}}=\frac{1}{\pi K^{\alpha}}\int^{\infty}_{0}\Re\bigg[\frac{K^{-\ii u}\phi_{\log S_{T}|t_{0}}(u-\ii(\alpha+1))}{\alpha^{2}+\alpha-u^{2}+\ii(2\alpha+1)u}\bigg]\dd u\;(\$),
\end{equation}
for a damping factor $\alpha>0$, provided that $\E\big[S^{\alpha+1}_{T}\big]<\infty$, which formula was also used in [Madan et al., 2019] \cite{MRS} for the calibration exercise.

Despite these well-known formulas, for our calibration purposes we choose to work with the parity-implied formula
\begin{equation}\label{3.1.8}
  C_{t_{0}}=S_{t_{0}}-K\bigg(\frac{1}{2}+\frac{1}{\pi}\int^{\infty}_{0}\Re\bigg[\frac{K^{-\ii u}\phi_{\log S_{T}|t_{0}}(u)}{u^{2}+\ii u}\bigg]\dd u\bigg)\;(\$),
\end{equation}
which has the advantage of avoiding strictly complex arguments in the model characteristic function. As a result, this formula proves to be much faster and stabler than (\ref{3.1.5}) with (\ref{3.1.6}), especially when the characteristic function involves non-elementary functions; compared with (\ref{3.1.7}), it is also stabler for general parameter values while only requiring $\E[S_{T}]<\infty$.

For the Quanto inverse-power options (again, with USD-denomination), we give general characteristic function-based pricing formulas in the next proposition, accommodating a broad range of underlying models. Similar to their direct option counterparts (see, e.g., [Xia, 2017, \text{Sect.} 5] \cite{X1}), these formulas are very amenable to efficient implementations and elude the need for Monte Carlo simulations.

\begin{proposition}\label{pro:2}
The value of the Quanto inverse-power call option with terminal payoff (\ref{3.1.4}) at time $t_{0}\in[0,T)$ is given by
\begin{equation}\label{3.1.9}
  C^{\rm (qip)}_{t_{0}}=R^{p_{1}}\bigg(\frac{1}{2}+\frac{p_{1}}{\pi}\int^{\infty}_{0}\Re\bigg[\frac{K^{-\ii up_{2}/p_{1}}\phi_{\log S_{T}|t_{0}}(u)}{\ii p_{1}u-u^{2}}\bigg]\dd u\bigg)\;(\$),
\end{equation}
and the time-$t_{0}$ value of a similar inverse-power put option is
\begin{equation}\label{3.1.10}
  P^{\rm (qip)}_{t_{0}}=C^{\rm (qip)}_{t_{0}}+R^{p_{1}}\big(K^{p_{2}}\phi_{\log S_{T}|t_{0}}(\ii p_{1})-1\big)\;(\$).
\end{equation}
\end{proposition}

In the above formula, the quantity $\phi_{\log S_{T}|t_{0}}(\ii p_{1})=\E_{t_{0}}\big[S^{-p_{1}}_{T}\big]$, assumed finite, represents the contemporaneous value of an ``inverse-power forward'' on $S$, or a futures contract on the powered USD/crypto rate. As the term suggests, such a forward is not to be confused with an inverse futures, the latter being the inverse of crypto/USD futures rates.

The pricing formulas for usual Quanto inverse options (recalling (\ref{3.2}) e.g.) are immediately obtained by setting $p_{1}=p_{2}=1$ in (\ref{3.1.9}) and (\ref{3.1.10}), and we have (in USD)
\begin{equation*}
  C^{\rm (qi)}_{t_{0}}=R\bigg(\frac{1}{2}+\frac{1}{\pi}\int^{\infty}_{0}\Re\bigg[\frac{K^{-\ii u}\phi_{\log S_{T}|t_{0}}(u)}{\ii u-u^{2}}\bigg]\dd u\bigg),\quad
  P^{\rm (qi)}_{t_{0}}=C^{\rm (qi)}_{t_{0}}+R(K\phi_{\log S_{T}|t_{0}}(\ii)-1),
\end{equation*}
the computational complexity of which is at the same level as (\ref{3.1.8}) for the direct options.

\subsection{Dynamic hedging}\label{sec:3.2}

It is widely known that perfect replication of options is possible when the underlying price is driven by diffusions or Poisson-type jumps. Direct implications on market incompleteness notwithstanding, given stochastic volatility and price--volatility co-jumps, we aim to construct partial hedges for the Quanto inverse-power options following Proposition \ref{pro:2}, by utilizing the underlying crypto/USD conversion rate ($S$) and the associated variance swap ($V_{S}$), both being investible at present.

From Proposition \ref{pro:1}, $S$ and $V_{S}$ can be taken as the only state variables for the Quanto inverse-power option value, noted that $R$ is predetermined. This makes it legitimate to define a deterministic function $C^{\rm (qip)}:[0,T]\times\mathbb{R}_{++}\times\mathbb{R}_{++}\mapsto\mathbb{R}_{+}$ such that
\begin{equation*}
  C^{\rm (qip)}_{t}=C^{\rm (qip)}(t,S_{t},V_{S}(t,T)),\quad t\in[0,T].
\end{equation*}
Here, recall that $V_{S}(t,T)$, for every $t$, represents the initial value of a (different) variance swap expiring at time $T$, rather than the time-$t$ value of the same variance swap issued at time 0. Hence, the necessary hedging instruments for the Quanto inverse-power option consist of the spot conversion rate and a flow of variance swaps available prior to $T$, the latter being equivalent to the collection of forward variance curves, $\{(\pd_{T}V_{S}(v,T))_{v\in[t,T)}:t\in[0,T)\}$, which is well-defined (see the proof of Corollary \ref{cor:1} in \ref{A}), parallel to the case of the rough Heston model ([El Euch and Rosenbaum, 2018] \cite{ER1}).

By specifying the (martingale) dynamics of the Quanto inverse-power option value $(C^{\rm (qip)}_{t})$, Corollary \ref{cor:1} forms the basis for constructing the required hedges.

\begin{corollary}\label{cor:1}
For any $t\in[0,T]$, the value of the Quanto inverse-power call option satisfies that
\begin{align}\label{3.2.1}
  C^{\rm (qip)}_{t}&=C^{\rm (qip)}_{0}+\int^{t}_{0}\dot{C}^{\rm (qip)}_{s}\dd s+\int^{t}_{0}(\pd_{S}C^{\rm (qip)})_{s-}\dd S_{s}+\int^{t}_{0}(\pd_{V_{S}}C^{\rm (qip)})_{s-}\dd V_{S}(s,T) \nonumber\\
  &\quad+\frac{1}{2}\int^{t}_{0}(\pd_{SS}C^{\rm (qip)})_{s}\dd\langle S^{\rm c},S^{\rm c}\rangle_{s}+\sum_{s\leq t}\big(\Delta C^{\rm (qip)}_{s}-\pd_{S}C^{\rm (qip)}_{s-}\Delta S_{s}-\pd_{V_{S}}C^{\rm (qip)}_{s-}\Delta V_{S}(s,T)\big),
\end{align}
where
\begin{align}\label{3.2.2}
  \dot{C}^{\rm (qip)}_{s}&=\frac{p_{1}R^{p_{1}}}{\pi}\int^{\infty}_{0}\Re\bigg[\frac{K^{-\ii up_{2}/p_{1}}\phi_{\log S_{T}|s}(u)}{\ii p_{1}u-u^{2}}\bigg(\ii u\log\phi_{Y_{1}}(-\ii\rho) \nonumber\\
  &\quad-\log\phi_{Y_{1}}(\rho u-H(T,s)(\ii\log\phi_{X_{1}}(u)+u\log\phi_{X_{1}}(-\ii))) \nonumber\\
  &\quad+(\log\phi_{X_{1}}(u)-\ii u\log\phi_{X_{1}}(-\ii))\bigg(\frac{\rho^{2}\Var[Y_{1}]}{\Var[X_{1}]}+H(T,s)\E[Y_{1}]\bigg)\bigg)\bigg]\dd u, \nonumber\\
  (\pd_{S}C^{\rm (qip)})_{s}&=\frac{p_{1}R^{p_{1}}}{\pi S_{s}}\int^{\infty}_{0}\Re\bigg[\frac{K^{-\ii up_{2}/p_{1}}\phi_{\log S_{T}|s}(u)}{\ii u+p_{1}}\bigg]\dd u, \nonumber\\
  (\pd_{SS}C^{\rm (qip)})_{s}&=\frac{p_{1}R^{p_{1}}}{\pi S^{2}_{s}}\int^{\infty}_{0}\Re\bigg[\frac{(\ii u-1)K^{-\ii up_{2}/p_{1}}\phi_{\log S_{T}|s}(u)}{\ii u+p_{1}}\bigg]\dd u, \nonumber\\
  (\pd_{V_{S}}C^{\rm (qip)})_{s}&=\frac{p_{1}R^{p_{1}}}{\pi\Var[X_{1}]}\int^{\infty}_{0}\Re\bigg[\frac{K^{-\ii up_{2}/p_{1}}(\log\phi_{X_{1}}(u)-\ii u\log\phi_{X_{1}}(-\ii))\phi_{\log S_{T}|s}(u)}{\ii p_{1}u-u^{2}}\bigg]\dd u, \nonumber\\
  \Delta C^{\rm (qip)}_{s}&=\frac{p_{1}R^{p_{1}}}{\pi}\int^{\infty}_{0}\Re\bigg[\frac{K^{-\ii up_{2}/p_{1}}\phi_{\log S_{T}|s-}(u)}{\ii p_{1}u-u^{2}}\big(\exp\big(\ii u(\Delta X_{\T_{s}}+\rho\Delta Y_{s}) \nonumber\\
  &\quad+(\log\phi_{X_{1}}(u)-\ii u\phi_{X_{1}}(-\ii))H(T,s)\Delta Y_{s}\big)-1\big)\bigg]\dd u,
\end{align}
and
\begin{align*}
  \langle S^{\rm c},S^{\rm c}\rangle_{s}=\sigma^{2}_{X}\int^{s}_{0}S^{2}_{v}A_{v}\dd v,\quad\Delta S_{s}=S_{s-}(e^{\Delta X_{\T_{s}}+\rho\Delta Y_{s}}-1),\quad\Delta V_{S}(s,T)=H(T,s)\Delta Y_{s}\Var[X_{1}].
\end{align*}
\end{corollary}

On the right side of (\ref{3.2.1}), the jump components $\Delta S_{s}$, $\Delta V_{S}(s,T)$, and $\Delta C^{\rm (qip)}_{s}$ all contain non-hedge-able jump risks present in both crypto prices and volatility. Also, since the precise activity rate $A$ is not observed,\footnote{For the same reason, integrated variance is not directly observable, despite being estimable from corresponding time series.} unless $X$ is purely discontinuous ($\sigma_{X}=0$), the quadratic variation term $\langle S^{\rm c},S^{\rm c}\rangle_{s}$ will induce additional volatility risk that cannot be perfectly hedged. Still, this term may be estimated by resorting to compound Poisson approximation techniques which truncate the jump space and coalesce small jumps into the Brownian component; we refer to [Schoutens, 2003, \text{Sect.} 8.2.1] \cite{S1} for details.\footnote{However, note that such estimation should be done under some (induced) physical measure $\PP$ and should go through the characteristic changes specified in (\ref{2.2.6}).} All the other terms in (\ref{3.2.1}) constitute perfect hedges given the accessibility to $S$ and $V_{S}$. In particular, while the integral with respect to $S$ is simply a delta hedge requiring periodic investments in the underlying crypto (with USD), the integral with respect to $V_{S}$, being a vega hedge, consists in buying and selling (or shorting and buying back) newly issued variance swaps over consecutive periods, whose practical validity should depend on the frequency of issuance.

An alternative hedge, which is less complete but easier to implement, is to trade a single variance swap $U$ issued at time 0 and expiring at $T$ instead of the flow of variance swaps mentioned above. Put precisely, since $C^{\rm (qip)}$ is by construction a $\Q$-martingale and so is the value process of this single swap, namely
\begin{equation*}
  U_{t}:=\E_{t}\bigg[\bigg[\log\frac{S}{S_{0}},\log\frac{S}{S_{0}}\bigg]_{T}\bigg]=V_{S}(t,T)+\bigg[\log\frac{S}{S_{0}},\log\frac{S}{S_{0}}\bigg]_{t}, \quad t\in[0,T],
\end{equation*}
the dynamics (\ref{3.2.1}) can be rewritten as
\begin{equation*}
  C^{\rm (qip)}_{t}=C^{\rm (qip)}_{0}+\int^{t}_{0}(\pd_{S}C^{\rm (qip)})_{s-}\dd S_{s}+\int^{t}_{0}(\pd_{U}C^{\rm (qip)})_{s-}\dd U_{s}+M_{t},\quad t\in[0,T],
\end{equation*}
where $(\pd_{U}C^{\rm (qip)})_{s}=(\pd_{V_{S}}C^{\rm (qip)})_{s}$ specifies the periodic amounts to be invested in $U$ and $M\equiv(M_{t})$ is a $\Q$-martingale containing all non-hedged (or ignored) jump and volatility risks.

\medskip

\section{Specification analysis}\label{sec:4}

We now attend to the issue of specifying the fractional stochastic-volatility model proposed in Section \ref{sec:2} for efficient implementation. We will consider special forms of the fractional kernel $h$ in (\ref{2.1.3}) according to the discussions in [Wang and Xia, 2022, \text{Sect.} 2.2] \cite{WX}, all of which have the same parametrization $\{\kappa>0,d\in(1/2,1]\}$. We will also adopt some of the new L\'{e}vy models developed in [Fei and Xia, 2024] \cite{FX} for the background-driving processes $X$ and $Y$. Again, the main goal is to select a suitable class of models that capture the key traits of crypto price dynamics, including inherent price jumps, volume impact, stochastic volatility with short-range dependence, as well as volatility jumps of suitable activity and their co-movements with the prices. The analysis will focus on obtaining various explicit expressions for the ingredients in the pricing--hedging formulas.

\subsection{Fractional kernels and activity rate}\label{sec:4.1}

For the activity rate process established in (\ref{2.1.5}), we only need to specify the fractional kernel $h$. Subject to the tail behaviors in (\ref{2.1.2}), we start with the aforementioned exponential-Riemann--Liouville kernel which has been mentioned in the same section. This kernel, referred to as the type-I kernel in [Wang and Xia, 2022] \cite{WX}, is stationary and unconditionally positive, taking the form
\begin{equation}\label{4.1.1}
  h_{1}(t,s)\equiv h_{1}(t-s)=\frac{e^{-\kappa(t-s)}(t-s)^{d-1}}{\Gf(d)}.
\end{equation}
In this case, the long-run average of the activity rate is given by $\lim_{t\rightarrow\infty}\E[A_{t}]=m+\E[Y_{1}]/\kappa^{d}$. The resulting tail-integrated kernel $H$ is automatically stationary and positive, and is given by
\begin{equation*}
  H_{1}(t,s)\equiv H_{1}(t-s)=\frac{\Gf(d)-\Gf(d,\kappa(t-s))}{\kappa^{d}\Gf(d)},
\end{equation*}
where $\Gf(\cdot,\cdot)$ is the (upper) incomplete gamma function. Then, using the primitive $\int\Gf(d,z)\dd z=z\Gf(d,z)-\Gf(d+1,z)$, $z\in\mathbb{R}$, along with the fact that $\Gf(d+1)=d\Gf(d)$, the integral of $H$ is also easily computed to be
\begin{align*}
  J_{1}(t-t_{0})&:=\int^{t}_{t_{0}}H_{1}(t-s)\dd s \\
  &=\frac{\kappa\Gf(d)(t-t_{0})-(\Gf(d+1)+\kappa(t-t_{0})\Gf(d,\kappa(t-t_{0}))-\Gf(d+1,\kappa(t-t_{0})))}{\kappa^{d+1}\Gf(d)},
\end{align*}
for fixed $0\leq t_{0}<t$.

The type-II kernel appearing in [Wang and Xia, 2022] \cite{WX} was designed to match the usual Ornstein--Uhlenbeck process driven by a fractional process with the Riemann--Liouville kernel. In particular, by choosing
\begin{equation*}
  h_{2}(t,s)\equiv h_{2}(t-s)=\frac{(t-s)^{d-1}+(-\kappa)^{1-d}e^{-\kappa(t-s)}(\Gf(d)-\Gf(d,-\kappa(t-s)))}{\Gf(d)},
\end{equation*}
one has the Volterra-type stochastic differential equation ($\Q$-a.s.):
\begin{equation*}
  A_{t}=A_{0}+\kappa\int^{t}_{0}\big(m-A_{s}\big)\dd s+\int^{t}_{0}\frac{(t-s)^{d-1}}{\Gf(d)}\dd Y_{s},\quad t\geq0.
\end{equation*}
The long-run average in this case is $\lim_{t\rightarrow\infty}\E[A_{t}]=m+\E[Y_{1}]\1_{\{d=1\}}/\kappa$, which always coincides with the mean-reverting level in the short-range dependence case ($d<1$). Notably, unlike $h_{1}$, $h_{2}$ involves complex values for the powers and the incomplete gamma functions and extra care must be taken for its computation; it may also take negative values, despite the integrability of its right tail. As before, using the primitive of the incomplete gamma function we obtain the tail-integrated kernel
\begin{equation*}
  H_{2}(t,s)\equiv H_{2}(t-s)=\frac{(-\kappa(t-s))^{d}+e^{-\kappa(t-s)}(\Gf(d+1)-\Gf(d+1,-\kappa(t-s)))}{(-\kappa)^{d}\Gf(d+1)},
\end{equation*}
in which the incomplete gamma function is also generally complex-valued. The integral of $H_{2}$ is similarly found to be ($0\leq t_{0}<t$)
\begin{equation*}
  J_{2}(t-t_{0}):=\int^{t}_{t_{0}}H_{2}(t-s)\dd s=\frac{e^{-\kappa(t-t_{0})}(\Gf(d+2)-(d+1)\Gf(d+1,-\kappa(t-t_{0})))}{(-\kappa)^{d+1}\Gf(d+2)}.
\end{equation*}

The third type of kernel, which was proposed in [Wang and Xia, 2022, \text{Sect.} 2.2] \cite{WX} for restricting gamma-type functions to the parametric level and thus enhancing computational efficiency, is a stationary piecewise function in the following form:
\begin{equation}\label{4.1.2}
  h_{3}(t,s)\equiv h_{3}(t-s)=
  \begin{cases}
    \displaystyle \frac{(t-s)^{d-1}-((1-d)/\kappa)^{d-1}}{\Gf(d)}-\frac{\kappa^{1-d}}{(1-d)^{2-d}\Gf(d-1)}&\displaystyle \text{if }t-s<\frac{1-d}{\kappa}\\
    \displaystyle -\frac{e^{1-d-\kappa(t-s)}}{(1-d)\Gf(d-1)}\bigg(\frac{1-d}{\kappa}\bigg)^{d-1}&\displaystyle \text{if }t-s\geq\frac{1-d}{\kappa}.
  \end{cases}
\end{equation}
The two pieces subtly separate the rough behavior of the activity rate at small time intervals from its mean-reverting behavior in the long run, with the long-run average being $\lim_{t\rightarrow\infty}\E[A_{t}]=m+\E[Y_{1}]((1-d)/\kappa)^{d}/((1-d)\Gf(d+1))$. The tail-integrated kernel is given by
\begin{equation}\label{4.1.3}
  H_{3}(t,s)\equiv H_{3}(t-s)=
  \begin{cases}
    \displaystyle \frac{(t-s)^{d}}{\Gf(d+1)}&\quad\displaystyle \text{if }t-s<\frac{1-d}{\kappa}\\
    \displaystyle \frac{1-de^{1-d-\kappa(t-s)}}{(1-d)\Gf(d+1)}\bigg(\frac{1-d}{\kappa}\bigg)^{d}&\quad\displaystyle \text{if }t-s\geq\frac{1-d}{\kappa},
  \end{cases}
\end{equation}
which also does not involve non-elementary functions on the time variable $s$. This feature points to the major computational advantage of the type-III kernel when evaluating the $s$-integral involving the log-characteristic function in (\ref{2.2.3}). As we shall see further, with the type-III kernel, this integral can actually be explicitly evaluated when $X$ and $Y$ are chosen to be many common L\'{e}vy processes whose generic-time characteristic functions admit closed-form expressions. The associated integral of $H_{3}$ over $s\in[t_{0},t)$ is given by ($0\leq t_{0}<t$)
\begin{align*}
  J_{3}(t-t_{0})&:=\int^{t}_{t_{0}}H_{3}(t-s)\dd s \\
  &=
  \begin{cases}
    \displaystyle \frac{(t-t_{0})^{d+1}}{\Gf(d+2)}&\quad\displaystyle \text{if }t-t_{0}<\frac{1-d}{\kappa}\\
    \displaystyle \frac{d(d+1)e^{1-d-\kappa(t-t_{0})}+\kappa(d+1)(t-t_{0})-d(3-d)}{(1-d)^{2}\Gf(d+2)}\bigg(\frac{1-d}{\kappa}\bigg)^{d+1}&\quad\displaystyle \text{if }t-t_{0}\geq\frac{1-d}{\kappa}.
  \end{cases}
\end{align*}
It is worth noting that unlike the other two, the type-III kernel is not infinitely smooth (despite positivity); in fact, it is only ensured that $h_{3}\in\mathcal{C}^{1}(\mathbb{R}_{++})$ and so $H_{3}\in\mathcal{C}^{2}(\mathbb{R}_{++})$.

It is easily seen that in the limit as $\kappa\searrow0$, all three types of kernels reduce to the Riemann--Liouville kernel $(t-s)^{d-1}/\Gf(d)$, while as $d\nearrow1$ they become the exponential kernel $h_{0}(t-s)=e^{-\kappa(t-s)}$, with tail integral $H_{0}(t-s)=(1-e^{-\kappa(t-s)})/\kappa$, exactly as in a usual Ornstein--Uhlenbeck process. These properties ensure their comparability and the latter limiting case will also be considered in a benchmark model for evaluation.

Figure \ref{fig:2} compares the three types of kernels and their tail integrals as functions of time $s\in[0,1)$ for $t=1$, under the parameter choices $\kappa=5$ and $d=0.6$. We observe that the kernels coincide, irrespective of their types, at the right end (corresponding to the small-time behavior in (\ref{2.1.2})), and that all tail-integrated kernels are bounded.

\begin{figure}[H]
  \centering
  \includegraphics[scale=0.475]{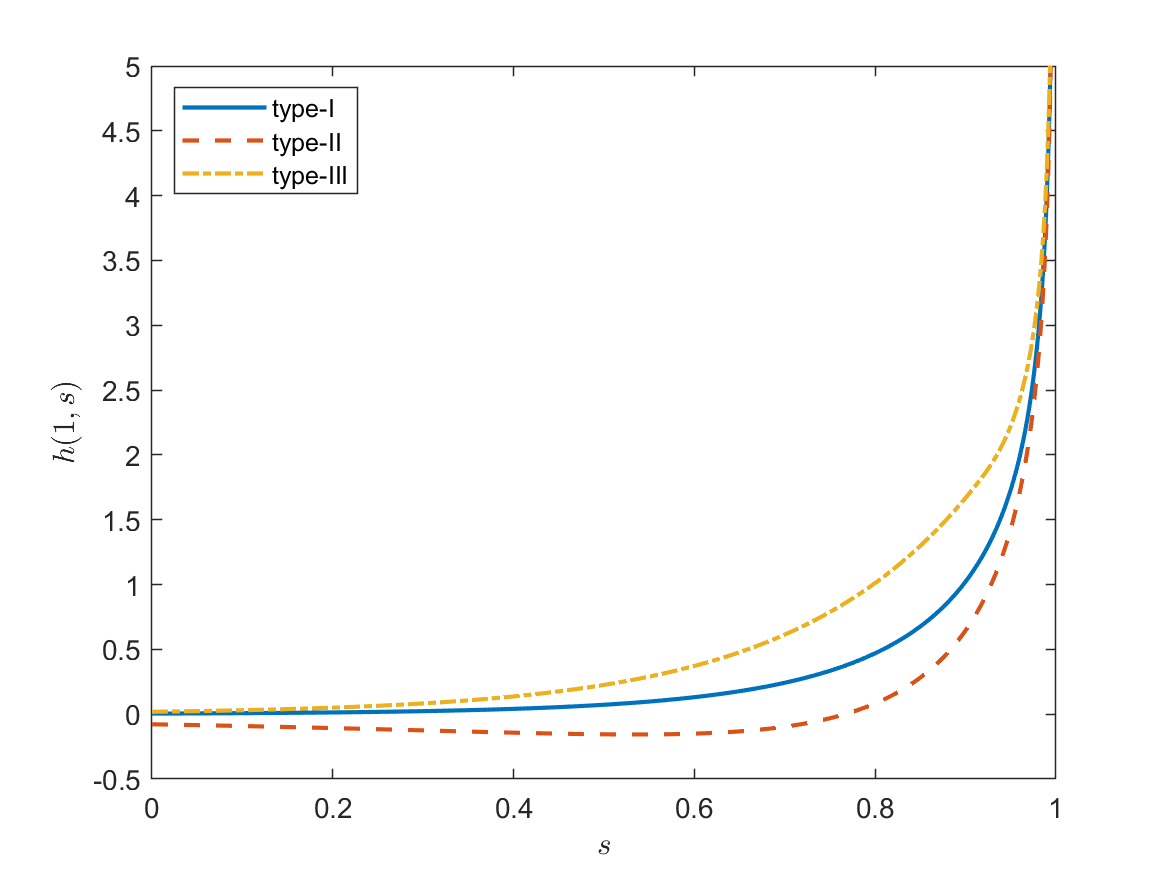}
  \includegraphics[scale=0.475]{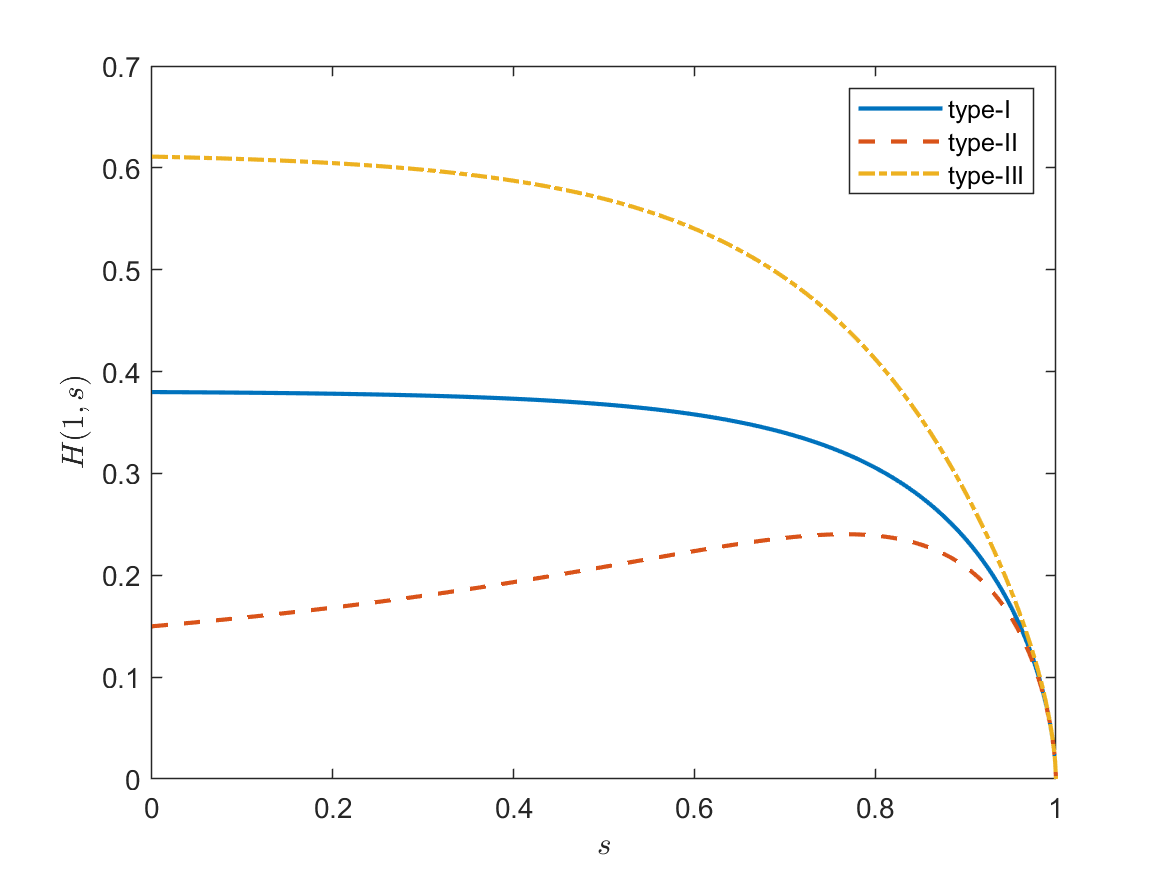}\\
  \caption{Fractional kernels and corresponding tail-integrated kernels}
  \label{fig:2}
\end{figure}

\subsection{Regulating kernels and base processes}\label{sec:4.2}

We shall consider two classes of models for the base processes $X$ and $Y$: asymmetric Laplace jump-diffusions and Gaussian-mixed tempered stable processes. The former class is based on a variation of the Kou model ([Kou, 2002] \cite{K}), with the implication that infrequent large fluctuations in crypto prices are captured by the jump part (with $\nu_{X}$ and $\nu_{Y}$ being finite), while continual small price movements are governed by a Brownian component (with $\sigma_{X}>0$). In contrast, the latter class is a rich family of L\'{e}vy models with no Brownian component (namely $\sigma_{X}=0$) but infinitely active jumps (with $\nu_{X}$ and $\nu_{Y}$ both being infinite), in which case frequent small price movements are also described by the jump part; a detailed analysis can be found in [K\"{u}chler and Tappe, 2013] \cite{KT}. Aside from the overall flexibility, an important reason to consider tempered stable processes is to build useful connections to the so-called ``regulating kernels'' proposed in [Fei and Xia, 2024] \cite{FX} aiming to reconcile high trading volumes with extreme tail risks in the presence of large speculations.

Unlike stationary fractional kernels, regulating kernels are stationary in log-times and thus focused on temporal ratios rather than differences. To briefly explain, let $Z$ be an arbitrary L\'{e}vy process and $\varkappa:\mathbb{R}_{+}\times(0,1]\mapsto[0,1]$ be a function strictly decreasing in both variables. Define for a fixed degree $n\in\mathbb{R}_{+}$ the following kernel-modulated process:
\begin{equation*}
  \tilde{Z}^{(n)}_{t}=\int^{t}_{0}\varkappa\bigg(n,\frac{s}{t}\bigg)\dd Z_{s},
\end{equation*}
whose finite-dimensional distribution is infinitely divisible, hence giving rise to a new L\'{e}vy process, $Z^{(n)}$, with the distributional equivalence $\mathcal{L}(Z^{(n)}_{t})=\mathcal{L}(\tilde{Z}^{(n)}_{t})$ for any $t\geq0$ and $n\geq0$.

Allowing for specificity, in the present paper we only consider the above regulation operation at the (crypto) price level, which has strong empirical support in [Fei and Xia, 2024] \cite{FX}. In theory, the same idea also applies to randomness in stochastic volatility, or $Y$ as well. Besides, we only employ the third type of regulating kernel discussed ibid., which is computationally the simplest of the three. This specific regulating kernel reads\footnote{For complete information, the type-I and type-II kernels, reading $\varkappa_{1}(n,s/t)=1-\Gf(n-\log(s/t))/\Gf(n)$ and $\varkappa_{2}(n,s/t)=e^{-\mathrm{Q}(n,1-s/t)}$, respectively, where $\mathrm{Q}\equiv\mathrm{Q}(\cdot,\cdot)$ denotes the inverse regularized gamma function, are only guaranteed to preserve explicitness when $n\in\{0,1\}$ and $n\in\mathbb{Z}_{+}$, respectively.}
\begin{equation*}
  \varkappa_{3}\bigg(n,\frac{s}{t}\bigg)=\frac{(1-s/t)^{n}}{\Gf(n+1)}.
\end{equation*}

In the jump-diffusion case, $X$ has asymmetric Laplace-distributed jumps and $Y$ just has exponentially distributed jumps.\footnote{In this case, $Y$ is exactly the background-driving process for a gamma-Ornstein--Uhlenbeck process; see, e.g., [Schoutens, 2003, \text{Sect.} 8.4.6] \cite{S1}.} An immediate implication is that the activity rate processes are subject to only finitely many shocks over any time interval. The corresponding time-1 characteristic functions are given by
\begin{equation}\label{4.2.1}
  \phi_{X_{1}}(u)=\exp\bigg(-\frac{1}{2}\sigma^{2}_{X}u^{2}+\lambda_{X}\bigg(\bigg(1+\frac{\ii\eta u}{b_{X}}\bigg)^{-1}\bigg(1-\frac{\ii u}{b_{X}\eta}\bigg)^{-1}-1\bigg)\bigg),\quad-b_{X}\eta<\Im u<\frac{b_{X}}{\eta},
\end{equation}
with parameters $\sigma_{X}>0$, $\lambda_{Y}>0$ (shape), $b_{X}>0$ (rate or reciprocal scale), $\eta$ (asymmetry), and
\begin{equation}\label{4.2.2}
  \phi_{Y_{1}}(u)=\exp\bigg(\lambda_{Y}\bigg(\bigg(1-\frac{\ii u}{b_{Y}}\bigg)^{-1}-1\bigg)\bigg),\quad\Im u>-b_{Y},
\end{equation}
with parameters $\lambda_{Y},b_{Y}>0$, similarly.

The L\'{e}vy characteristics of $X$ and $Y$ are readily given by
\begin{align*}
  &\bigg(\mu_{X}=\frac{\lambda_{X}(1-\eta^{2})}{b_{X}\eta},\sigma_{X},\nu_{X}(\dd z)=\lambda_{X}\bigg(\frac{b_{X}\eta}{\eta^{2}+1}\bigg)e^{-b_{X}(z\;\sgn z)\eta^{\sgn z}}\dd z\bigg), \\
  &(0,0,\nu_{Y}(\dd z)=\lambda_{Y}b_{Y}e^{-b_{Y}z}\1_{\{z>0\}}\dd z).
\end{align*}

In the tempered stable case, $X$ is built as a drifted Brownian motion time-changed by an independent L\'{e}vy subordinator -- in particular a tempered stable subordinator regulated by the type-III kernel, while $Y$ is simply a tempered stable subordinator (with separate parametrization). The resulting time-1 characteristic functions are given by
\begin{equation}\label{4.2.3}
  \phi_{X_{1}}(u)=\phi_{Z^{(n)}_{1}}\bigg(\theta u+\frac{\ii u^{2}}{2}\bigg),\quad\sqrt{\theta^{2}+2b_{X}\Gf(n+1)}-\theta<\Im u<\sqrt{\theta^{2}+2b_{X}\Gf(n+1)}+\theta,
\end{equation}
with
\begin{equation*}
  \phi_{Z^{(n)}_{1}}(u)=
  \begin{cases}
    \displaystyle \exp\bigg(a_{X}b^{c_{X}}_{X}\Gf(-c_{X})\bigg({}_{2}\F_{1}\bigg(-c_{X},\frac{1}{n};\frac{1}{n}+1;\frac{\ii u}{b_{X}\Gf(n+1)}\bigg)-1\bigg)\bigg) \\
    \qquad\text{if }c_{X}>0 \\
    \displaystyle \bigg(1-\frac{\ii u}{b_{X}\Gf(n+1)}\bigg)^{-a_{X}}\exp\bigg(-\frac{\ii a_{X}nu}{b_{X}\Gf(n+2)} {}_{2}\F_{1}\bigg(1,\frac{1}{n}+1;\frac{1}{n}+2;\frac{\ii u}{b_{X}\Gf(n+1)}\bigg)\bigg) \\
    \qquad\text{if }c_{X}=0,
  \end{cases}
\end{equation*}
where $_{2}\F_{1}(\cdot,\cdot;\cdot;\cdot)$ denotes Gauss' hypergeometric function and
\begin{equation}\label{4.2.4}
  \phi_{Y_{1}}(u)=
  \begin{cases}
    \displaystyle \exp\big(a_{Y}\Gf(-c_{Y})((b_{Y}-\ii u)^{c_{Y}}-b^{c_{Y}}_{Y})\big)\;&\text{if }c_{Y}>0 \\
    \displaystyle \bigg(1-\frac{\ii u}{b_{Y}}\bigg)^{-a_{Y}}\;&\text{if }c_{Y}=0,
  \end{cases}
  \quad\Im u>-b_{Y}.
\end{equation}
The parameters in this case are $\theta\in\mathbb{R}$ (skewness), $a_{X}>0$ (shape), $b_{X}>0$ (rate), $c_{X}\in[0,1)$ (family), and $n\geq0$ (regulation degree) for $X$, and similarly $a_{Y},b_{Y}>0$, and $c_{Y}\in[0,1)$ for $Y$, but neither the family parameters $c_{X}$ and $c_{Y}$ nor the regulation degree $n$ will take part in optimization due to their dynamical meanings (to be discussed later). Note that if $n=0$, the formula (\ref{4.2.3}) in its limiting case will take the form of (\ref{4.2.4}) (with the subscripts ``$Y$'' changed to ``$X$''). Thus, the characteristic function (\ref{4.2.3}) covers both the variance gamma process (when $c_{X}=n=0$) and the normal inverse Gaussian process (when $c_{X}=1/2$ and $n=0$).

The L\'{e}vy characteristics of $X$ and $Y$ are given by
\begin{align*}
  &\bigg(\mu_{X}=\frac{a_{X}\Gf(1-c_{X})}{b^{1-c_{X}}_{X}\Gf(n+2)},0, \\
  &\qquad\nu_{X}(\dd z)=\int^{\infty}_{0}\dd w\;\frac{a_{X}b^{c_{X}}_{X}(b_{X}\Gf(n+1)w)^{1/n}\Gf(-c_{X}-1/n,b_{X}\Gf(n+1)w)\gamma_{\theta w,w}(\dd z)}{nw}\bigg),\\
  &\bigg(0,0,\nu_{Y}(\dd z)=\frac{a_{Y}e^{-b_{Y}z}}{z^{c_{Y}+1}}\1_{\{z>0\}}\dd z\bigg),
\end{align*}
where $\mu_{X}=\E[X_{1}]$ is computed in the proof of Proposition \ref{pro:3} in \ref{A} and $\gamma_{\theta w,w}$ denotes the Gaussian measure with mean $\theta w$ and variance $w$.

All of the above specified characteristic functions for $X$ and $Y$ are to be put into the conditional characteristic function in (\ref{2.2.3}), after which a pricing engine can be initiated with up to two (parallel) numerical integrals to be computed. With any of the three types of fractional kernels in Section \ref{sec:4.1}, the integral of $H$ has been explicitly given, which leaves us with only one numerical integral. This undoubtedly lays the foundation for efficient pricing--hedging. To go one step further, the next proposition, serving as the main result of this section, gives a perfectly analytical formula for the conditional characteristic function when $h$ happens to be the type-III (fractional) kernel. For a succinct presentation, let us introduce the following notations (for generic $u$):
\begin{align}\label{4.2.5}
  &\varphi(u)=\ii\log\phi_{X_{1}}(u)+u\log\phi_{X_{1}}(-\ii),\quad\varphi_{1}(u)=b_{Y}-\ii\rho u,\quad\varphi_{2}(u)=\frac{\ii\varphi(u)}{\Gf(d+1)}, \nonumber\\
  &\varphi_{3}(u)=\varphi_{1}(u)+\bigg(\frac{1-d}{\kappa}\bigg)^{d}\frac{\ii\varphi(u)}{(1-d)\Gf(d+1)}, \quad\varphi_{4}(u)=\bigg(\frac{1-d}{\kappa}\bigg)^{d}\frac{\ii de^{1-d}\varphi(u)}{(1-d)\Gf(d+1)}.
\end{align}

\begin{proposition}\label{pro:3}
In the context of Proposition \ref{pro:1}, let $h$ be the type-III fractional kernel in (\ref{4.1.2}). Then, with the integral $\int^{t}_{t_{0}}H(t,s)\dd s$ explicitly given in (\ref{4.1.3}), we have the following two assertions. \medskip

(i) If $X$ is an asymmetric Laplace jump-diffusion process and $Y$ is an independent exponential-compound Poisson process, with time-1 characteristic functions in (\ref{4.2.1}) and (\ref{4.2.2}), respectively, then for the conditional characteristic function (\ref{2.2.3}), it holds that
\begin{align}\label{4.2.6}
  &\quad\int^{t}_{t_{0}}\log\phi_{Y_{1}}(\rho u-H(t,s)\varphi(u))\dd s \nonumber\\
  &=
  \begin{cases}
    \displaystyle \lambda_{Y}(t-t_{0})\bigg(\frac{b_{Y}}{\varphi_{1}(u)} {}_{2}\F_{1}\bigg(1,\frac{1}{d};\frac{1}{d}+1;-\frac{(t-t_{0})^{d}\varphi_{2}(u)}{\varphi_{1}(u)}\bigg)-1\bigg) &\displaystyle\text{if }t-t_{0}<\frac{1-d}{\kappa} \\
    \displaystyle \lambda_{Y}\bigg(\frac{1-d}{\kappa}\frac{b_{Y}}{\varphi_{1}(u)} {}_{2}\F_{1}\bigg(1,\frac{1}{d};\frac{1}{d}+1;-\bigg(\frac{1-d}{\kappa}\bigg)^{d}\frac{\varphi_{2}(u)}{\varphi_{1}(u)}\bigg) \\
    \displaystyle \quad+\frac{b_{Y}\log(e^{\kappa s}\varphi_{3}(u)-\varphi_{4}(u))}{\kappa\varphi_{3}(u)}\bigg|^{t-t_{0}}_{s=(1-d)/\kappa}-(t-t_{0})\bigg) &\displaystyle\text{if }t-t_{0}\geq\frac{1-d}{\kappa},
  \end{cases}
\end{align}
and also,
\begin{equation*}
  \E[Y_{1}]=\frac{\lambda_{Y}}{b_{Y}},\quad\Var[Y_{1}]=\frac{2\lambda_{Y}}{b^{2}_{Y}}, \quad\Var[X_{1}]=\frac{2\lambda_{X}(\eta^{4}-\eta^{2}+1)}{b^{2}_{X}\eta}.
\end{equation*}
\medskip

(ii) If $X$ is the Gaussian mixture of a type-III regulating kernel-modulated tempered stable subordinator and $Y$ is an independent tempered stable subordinator, with $\phi_{X_{1}}$ and $\phi_{Y_{1}}$ given in (\ref{4.2.3}) and (\ref{4.2.4}), then for (\ref{2.2.3}) it holds that if $c_{Y}\in(0,1)$,
\begin{align}\label{4.2.7}
  &\quad\int^{t}_{t_{0}}\log\phi_{Y_{1}}(\rho u-H(t,s)\varphi(u))\dd s \nonumber\\
  &=
  \begin{cases}
    \displaystyle a_{Y}\Gf(-c_{Y})(t-t_{0})\bigg(\varphi^{c_{Y}}_{1}(u) {}_{2}\F_{1}\bigg(-c_{Y},\frac{1}{d};\frac{1}{d}+1;-\frac{(t-t_{0})^{d}\varphi_{2}(u)}{\varphi_{1}(u)}\bigg)-b^{c_{Y}}_{Y}\bigg) &\displaystyle\text{if }t-t_{0}<\frac{1-d}{\kappa} \\
    \displaystyle a_{Y}\Gf(-c_{Y})\bigg(\frac{1-d}{\kappa}\varphi^{c_{Y}}_{1}(u) {}_{2}\F_{1}\bigg(-c_{Y},\frac{1}{d};\frac{1}{d}+1;-\bigg(\frac{1-d}{\kappa}\bigg)^{d}\frac{\varphi_{2}(u)}{\varphi_{1}(u)}\bigg) \\
    \displaystyle \quad+\frac{\varphi^{c_{Y}}_{3}(u)}{\kappa c_{Y}}\bigg(\frac{e^{\kappa s}\varphi_{3}(u)}{\varphi_{4}(u)}-1\bigg)\bigg(1-\frac{e^{-\kappa s}\varphi_{4}(u)}{\varphi_{3}(u)}\bigg)^{c_{Y}} \\
    \displaystyle \qquad\times{}_{2}\F_{1}\bigg(1,1;1-c_{Y};\frac{e^{\kappa s}\varphi_{3}(u)}{\varphi_{4}(u)}\bigg)\bigg|^{t-t_{0}}_{s=(1-d)/\kappa}-b^{c_{Y}}_{Y}(t-t_{0})\bigg) &\displaystyle\text{if }t-t_{0}\geq\frac{1-d}{\kappa},
  \end{cases}
\end{align}
while if $c_{Y}=0$,
\begin{align}\label{4.2.8}
  &\quad\int^{t}_{t_{0}}\log\phi_{Y_{1}}(\rho u-H(t,s)\varphi(u))\dd s \nonumber\\
  &=
  \begin{cases}
    \displaystyle a_{Y}(t-t_{0})\bigg(\log b_{Y}-\log(\varphi_{1}(u)+(t-t_{0})^{d}\varphi_{2}(u)) \\
    \displaystyle \quad-d\bigg({}_{2}\F_{1}\bigg(1,\frac{1}{d};\frac{1}{d}+1;-\frac{(t-t_{0})^{d}\varphi_{2}(u)}{\varphi_{1}(u)}\bigg)-1\bigg)\bigg) &\displaystyle\text{if }t-t_{0}<\frac{1-d}{\kappa} \\
    \displaystyle a_{Y}\bigg((t-t_{0})\log b_{Y}-\frac{1-d}{\kappa}\bigg(\log\bigg(\varphi_{1}(u)+\bigg(\frac{1-d}{\kappa}\bigg)^{d}\varphi_{2}(u)\bigg) \\
    \displaystyle \quad+d\bigg({}_{2}\F_{1}\bigg(1,\frac{1}{d};\frac{1}{d}+1;-\bigg(\frac{1-d}{\kappa}\bigg)^{d} \frac{\varphi_{2}(u)}{\varphi_{1}(u)}\bigg)-1\bigg)\bigg) \\
    \displaystyle \quad-\frac{1}{\kappa}\bigg(\kappa s\log\varphi_{3}(u)+\Li_{2}\bigg(\frac{e^{-\kappa s}\varphi_{4}(u)}{\varphi_{3}(u)}\bigg)\bigg)\bigg|^{t-t_{0}}_{s=(1-d)/\kappa}\bigg) &\displaystyle\text{if }t-t_{0}\geq\frac{1-d}{\kappa},
  \end{cases}
\end{align}
where $\Li_{2}(\cdot)$ denotes the dilogarithm. Also, for $c_{X},c_{Y}\in[0,1)$,
\begin{align*}
  &\E[Y_{1}]=\frac{a_{Y}\Gf(1-c_{Y})}{b^{1-c_{Y}}_{Y}},\quad\Var[Y_{1}]=\frac{a_{Y}\Gf(2-c_{Y})}{b^{2-c_{Y}}_{Y}}, \\
  &\Var[X_{1}]=\frac{a_{X}}{b^{2-c_{X}}_{X}}\bigg(\frac{b_{X}\Gf(1-c_{X})}{\Gf(n+2)}+\frac{\theta^{2}\Gf(2-c_{X})}{(2n+1)\Gf^{2}(n+1)}\bigg).
\end{align*}
\end{proposition}

The formulas (\ref{4.2.6}), (\ref{4.2.7}), and (\ref{4.2.8}) as a whole may seem a bit intimidating at first glance. However, let us notice that the computation of $\varphi(u)$ for every $u$ is based solely on the model for $X$, and thus is independent of the choice of fractional kernels; besides, all $\varphi_{i}(u)$'s involve very elementary operations, and regardless of what fractional kernel is in use, this kind of operations will always prevail each time the integrand is computed. For this reason, the benefit from using Proposition \ref{pro:3} with the type-III kernel should be understood in the sense of recasting the numerical integral (of $\log\phi_{Y_{1}}$) into an expression containing only Gauss' hypergeometric functions (and dilogarithms), and all of these functions are readily accessible and efficiently implementable with most standard programming languages (Matlab, Mathematica, Python, etc.).

\medskip

\section{Empirical analysis}\label{sec:5}

In this section, we present a thorough demonstration of the efficacy of our FSV model through meticulous calibration exercises with Bitcoin options. Via comparative analyses against several benchmark models, we not only underscore the superior capacity and flexibility of our FSV model in effectively capturing the key traits of crypto market dynamics but unveil the substantial computational advantage of the type-III fractional kernel relative to other kernel types as well. Notably, to demonstrate the adeptness of the FSV model in accommodating the erratic dynamics of the highly speculative crypto market, it is necessary to evaluate the model performance across different time periods. Towards that end, we draw upon two independent data sets: one representing abnormal market conditions amidst the COVID-19 pandemic, and the other reflecting recent market dynamics. After the calibration, we then extract the calibrated parameters from the two best-performing models to further investigate the impact of the power mechanism on corresponding Quanto inverse options.

\subsection{Data preparation}\label{sec:5.1}

We collect tick-level trade data on Bitcoin options from the Debirit exchange (\underline{deribit.com}) to conduct the calibration exercises. All listed options are of European style. The first data set (labeled 1), directly available from [Xia, 2021, \text{Sect.} 4.2] \cite{X3}, contains 40 call options traded as of July 11, 2020, covering four maturities $T=19,47,166,257$ days. Each maturity covers 10 options with different strike prices ($K$) ranging from \$5,000 to \$32,000. The second data set (labeled 2) is new, consisting of call options traded as of February 19, 2024, also under four maturities, $T=4,39,130,312 $ days, with strike price range $\$[50,000,200,000]\ni K$. The Bitcoin spot prices ($S_{0}$) are also directly sourced from the Deribit exchange and stand at \$9,232.98 and \$52,108.00 for the first and second data sets, respectively. Market prices are determined based on the last prices of the calls.

To ensure calibration stability, we have excluded market prices that present strike arbitrage for the second (new) data set. Specifically, we drop out data points that violate the following monotonicity and convexity conditions:
\begin{enumerate}
  \item $C_0(K, T) \geq C_0(K',T)$ if $K \leq K'$, with $C_{0}(K,T)$ denoting the observed call price with strike price $K$ and maturity $T$;
  \item $C_0(K,T) \leq \delta_2/(\delta_1 + \delta_2)\; C_0(K - \delta_1, T) + \delta_1/(\delta_1 + \delta_2)\; C_0(K + \delta_2, T)$, where $K - \delta_1$ and $K + \delta_2$ are the last and next available strike prices, respectively, for the same maturity $T$.
\end{enumerate}
This procedure only eliminates approximately 10\% of price data, with the majority satisfying the above conditions. As a result, data set 2 also contains 40 calls, with each maturity including exactly 10 prices as in data set 1.

The two data sets were intentionally selected to reflect structurally different market conditions. As can be seen in the forthcoming Figure \ref{fig:3a} and Figure \ref{fig:3b}, data set 1 capturing the COVID-19 period exhibits a pronounced clustering of option prices across comparable maturities, which pattern is indicative of elevated volatility-of-volatility and substantially weaker mean reversion relative to data set 2 corresponding to the post-COVID period. Comprehensibly, the compression of the term structure can be attributed to the significant elevation of short-term uncertainty such that longer maturities offered little additional protection, indicating the extreme levels of market instability at that time in the crypto market.

\subsection{Calibration procedures}\label{sec:5.2}

We shall consider a total of six specifications of the FSV model, using all three types of fractional kernels and both classes of base processes, namely asymmetric Laplace jump-diffusions (ALJD) and Gaussian-mixed regulated tempered stable processes (GMRTS), as detailed in Section \ref{sec:4.2}.  At they stand, the fractional kernels are all governed by two common parameters $(\kappa,d)$, while the ALJD base processes are characterized by six parameters $(\sigma_X,\lambda_X,b_X,\eta,\lambda_Y,b_Y)$ and the GMRTS base processes by eight parameters $(a_X,b_X,c_X,\theta,n,a_Y,b_Y,c_Y)$; there are three additional common parameters $(A_{0},m,\rho)$ introduced by the FSV framework.

To reduce the number of tentative parameters of the GMRTS, we set $c_X=c_Y=1/2$, $n=2$, and $a_Y=b_Y$. This reduction is reasonable considering the dynamic interpretations of the family parameters $c_X$ and $c_Y$, alongside the parameter $n$, which serve as proxies for trading intensity and speculation degree, respectively ([Fei and Xia, 2024] \cite{FX}). Notably, these parameters are typically not directly optimized during calibration but rather estimated statistically ([Todorov and Tauchen, 2011] \cite{TT}) or examined retrospectively, and the specific values chosen for these parameters, namely 1/2 and 2, respectively, are obtained from the findings of the comprehensive time series analysis conducted using daily Bitcoin prices in [Fei and Xia, 2024, \text{Sect.} 6] \cite{FX}. On the other hand, the condition $a_Y=b_Y$ is justified by the fact that the instantaneous activity rate process already has infinitely active jumps in this case, similar to the case of the variance gamma model ([Madan et al., 1998] \cite{MCC}). Additionally, we optionally fix $m=0.1$ to demonstrate that the long-term equilibrium level of the instantaneous activity rate can be completely controlled by the fractional kernel parameters as well as the parameters of $Y$. Therefore, depending on whether $m$ is fixed, there are ten or eleven parameters in total for the FSV-ALJD model and eight parameters for the FSV-GMRTS model that need to be calibrated. The benchmark models that we consider include the Black--Scholes model, the Heston model, and a usual stochastic-volatility (SV) model with jumps obtained by sending $d\nearrow1$ in the FSV models accordingly, as briefly discussed in Section \ref{sec:4.1}. To be clear, the Black--Scholes model is parameterized by a single volatility parameter $\sigma>0$, and the Heston has five parameters: $\kappa>0$ (mean reversion speed), $\rho\in[-1,1]$ (price--volatility correlation), $\varsigma>0$ (volatility-of-volatility scale), $V_{0}>0$ (initial variance), and $m>0$ (mean reversion level -- always calibrated).

The calibration that we run is a joint one taking into account all four maturities and all strike prices simultaneously for each data set. We minimize the average relative pricing error (ARPE) between the market prices of the Bitcoin options and the corresponding model prices. More specifically, the optimization aims to
\begin{equation}\label{5.2.1}
  \min_{\vartheta \in \mathfrak{C}}\sum_{K,T} \frac{| \text{market price} - C_{0}(K,T) |}{\text{market price}},
\end{equation}
where $\vartheta$ is a placeholder for the model parameters, $\mathfrak{C}$ is their corresponding value ranges, as described in Section \ref{sec:4.2}, and $C_{0}$ denotes the model prices computed based on (\ref{3.1.8}). The summation encompasses all 40 available strike prices and maturities, and calibrated parameter values will be denoted as $\hat{\vartheta}$ generically.

\subsection{Calibration results and discussions}\label{sec:5.3}

To efficiently address the optimization problem in (\ref{5.2.1}), we first utilize a genetic algorithm by executing a predetermined number of iterations to generate a set of candidate parameters. Then, we employ a pattern research algorithm, initializing it with these parameters, to iteratively refine our parameter set until convergence is achieved. It is understood that while the genetic algorithm and pattern search algorithm do not guarantee a global minimum or ensure a unique local minimum, our findings suggest that reaching a local minimum alone yields very promising results and are more than sufficient for industrial applications.

\begin{figure}[H]
  \centering
  \begin{minipage}[c]{0.49\linewidth}
  \centering
  \caption*{\footnotesize Data set 1 (July 11, 2020)}
  \includegraphics[scale=0.475]{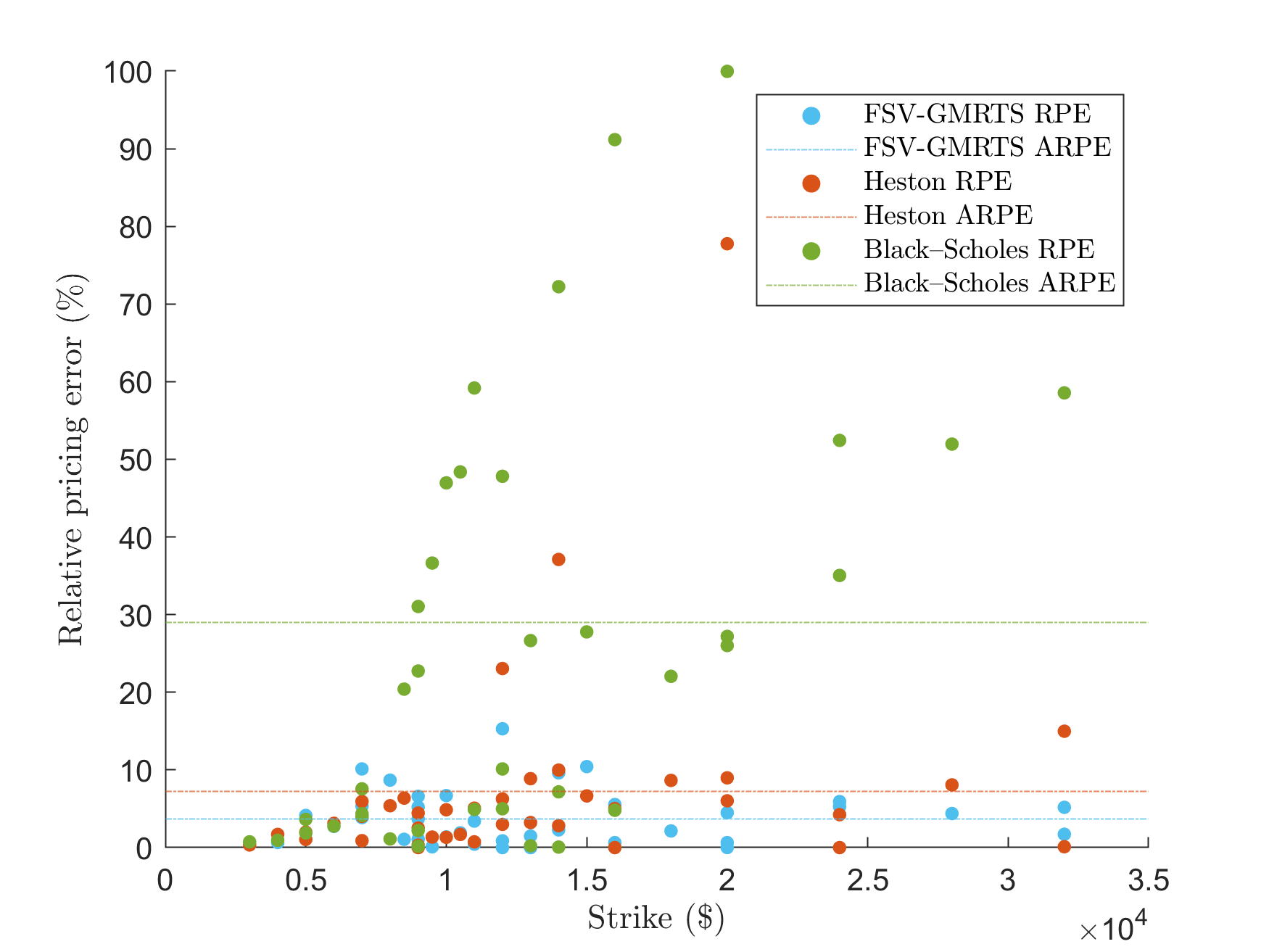}
  \end{minipage}
  \begin{minipage}[c]{0.49\linewidth}
  \centering
  \caption*{\footnotesize Data set 2 (February 19, 2024)}
  \includegraphics[scale=0.475]{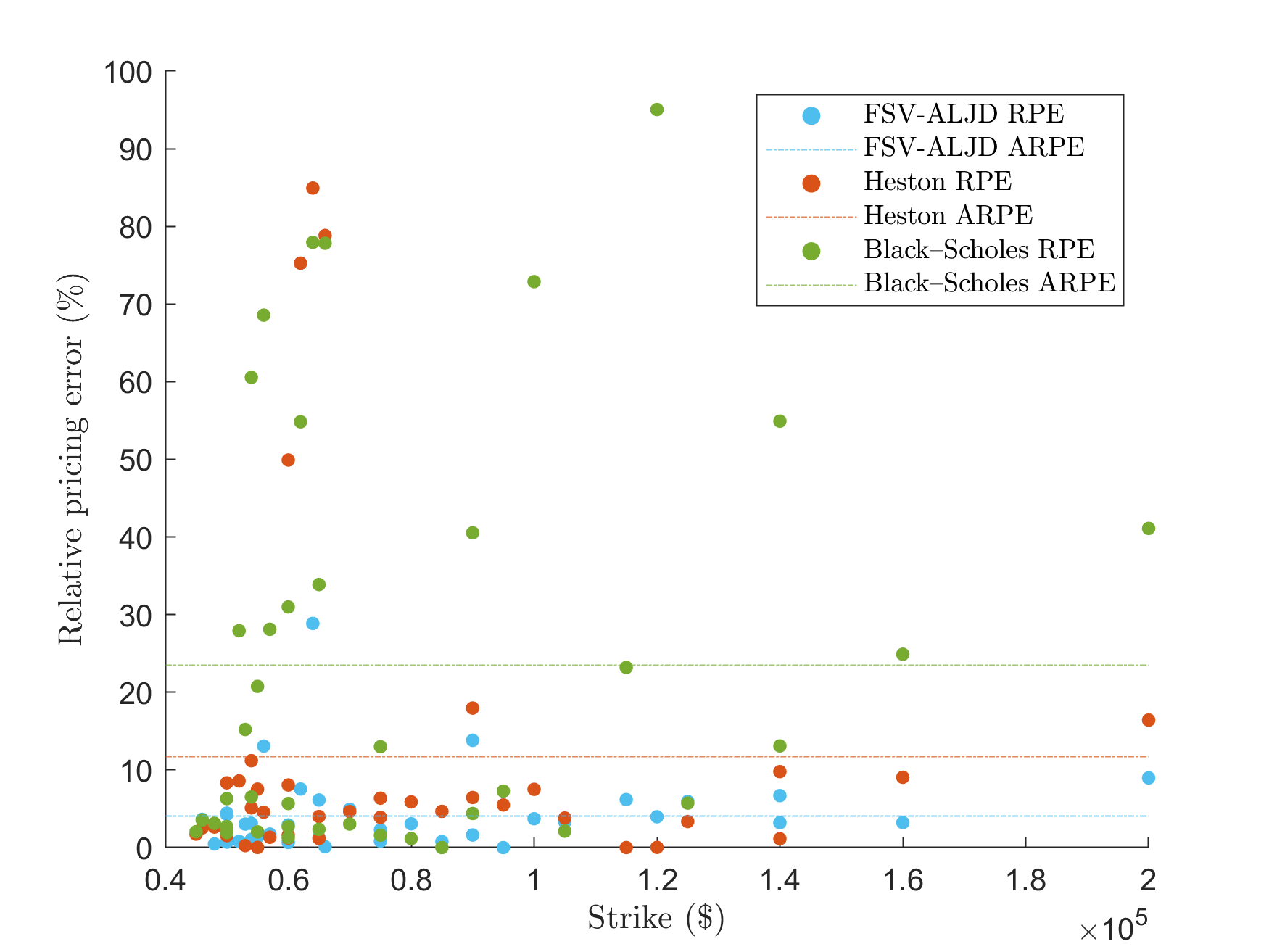}
  \end{minipage}
  \caption{Comparison of relative pricing errors}
  \label{fig:4}
\end{figure}

The calibration results are consolidated in Table \ref{tab:1a}, Table \ref{tab:1b}, Table \ref{tab:1c}, Table \ref{tab:2a}, Table \ref{tab:2b}, and Table \ref{tab:2c}, presenting the calibrated parameter values for the six implemented models individually for each data set, along with the minimized ARPEs. Wall time measurements for the FSV models using different kernel types are also included.\footnote{All calibration programs are coded in MATLAB$^\circledR$ and executed on a personal computer with a single 12th Gen Intel(R) Core(TM) i7-12700H 2.30 GHz processor.} All numerical results are rounded to six significant digits and the best results are marked as `` $\star$ .'' Figure \ref{fig:3a} and Figure \ref{fig:3b} provide visual comparisons between the calibrated model prices and the market prices, for which we only include the best-fitted FSV models and the benchmark models for clarity. Figure \ref{fig:4} further compares the individual relative pricing errors (RPEs) (namely the summands in (\ref{5.2.1})) across the best-fitted FSV models, the Heston model, and the Black--Scholes model.

\begin{table}[H]\footnotesize
  \centering
  \ContinuedFloat*
  \caption{Calibration results for FSV-ALJD model on data set 1}
  \begin{tabular}{c|c|c|c|c|c}
  \hline
    \multirow{4}{*}{\begin{tabular}[c]{@{}c@{}}Type-I kernel \\ ARPE = 5.78514\% \\ Wall time = ($6,001.63+1,031.89$)s\end{tabular}} & $\hat{\sigma}_{X}$ & $\hat{\lambda}_{X}$ & $\hat{b}_{X}$ & $\hat{\eta}$ & $\hat{\lambda}_{Y}$ \\ \cdashline{2-6}
    & 1.63878 & 1.56946 & 8.47621 & 9.03383 & 2.24652 \\  \cline{2-6}
    & $\hat{b}_{Y}$ & $\hat{\kappa}$ & $\hat{d}$ & $\hat{\rho}$ & $\hat{A}_{0}$ \\  \cdashline{2-6}
    & 7.05341 & 4.67423 & 0.65369 & 0.42878 & 0.06162\\ \hline

    \multirow{4}{*}{\begin{tabular}[c]{@{}c@{}}Type-II kernel \\ ARPE = 5.76925\% \\ Wall time = ($2,994.19+136.297$)s\end{tabular}} & $\hat{\sigma}_{X}$ & $\hat{\lambda}_{X}$ & $\hat{b}_{X}$ & $\hat{\eta}$ & $\hat{\lambda}_{Y}$ \\ \cdashline{2-6}
    & 1.61562 &  2.36595 & 4.45261 & 0.87251 & 4.69373 \\  \cline{2-6}
    & $\hat{b}_{Y}$ & $\hat{\kappa}$ & $\hat{d}$ & $\hat{\rho}$ & $\hat{A}_{0}$ \\  \cdashline{2-6}
    & 8.74129 & 3.27911 & 0.66295 & 0.05959 & 0.05187 \\ \hline

    \multirow{4}{*}{\begin{tabular}[c]{@{}c@{}}Type-III kernel \\ ARPE = 5.66093\% $\star$ \\ Wall time = ($1,101.7 + 50.7456$)s $\star$\end{tabular}} & $\hat{\sigma}_{X}$ & $\hat{\lambda}_{X}$ & $\hat{b}_{X}$ & $\hat{\eta}$ & $\hat{\lambda}_{Y}$ \\ \cdashline{2-6}
    & 0.73208 & 0.21292 & 0.98634 & 2.10382 & 8.52514 \\  \cline{2-6}
    & $\hat{b}_{Y}$ & $\hat{\kappa}$ & $\hat{d}$ & $\hat{\rho}$ & $\hat{A}_{0}$ \\  \cdashline{2-6}
    & 4.14291 & 9.70963 &  0.54194 & 0.00641 & 0.24452 \\ \hline
    \end{tabular}
  \label{tab:1a}
\end{table}

\clearpage

\vspace*{0.2in}

\begin{table}[H]\footnotesize
  \centering
  \ContinuedFloat
  \caption{Calibration results for FSV-GMRTS model on data set 1}
  \begin{tabular}{c|c|c|c|c}
  \hline

    \multirow{4}{*}{\begin{tabular}[c]{@{}c@{}}Type-I kernel \\ ARPE = 3.71030\% $\star$ \\ Wall time = ($7,741.48 + 231.343$)s\end{tabular}} & $\hat{a}_{X}$ & $\hat{b}_{X}$ & $\hat{\theta}$ & $\hat{a}_Y$  \\ \cdashline{2-5}
    & 14.2831 & 49.515 & $-0.54990$ & 0.90722 \\  \cline{2-5}
    & $\hat{\kappa}$ & $\hat{d}$ & $\hat{\rho}$ & $ \hat{A}_0$   \\  \cdashline{2-5}
    & 1.59253 & 0.53995 &  0.11765 & 0.24383  \\ \hline

    \multirow{4}{*}{\begin{tabular}[c]{@{}c@{}}Type-II kernel \\ ARPE = 4.130562\% \\ Wall time = ($6,005.08 + 1,318.11$)s\end{tabular}} & $\hat{a}_{X}$ & $\hat{b}_{X}$ & $\hat{\theta}$ & $\hat{a}_Y$  \\ \cdashline{2-5}
    & 11.3645 & 10.9219 & $-0.43454$ & 0.84143  \\  \cline{2-5}
    & $\hat{\kappa}$ & $\hat{d}$ & $\hat{\rho}$ & $ \hat{A}_0$   \\  \cdashline{2-5}
    & 2.47967 & 0.71118 & 0.10131 & 0.17337  \\ \hline

    \multirow{4}{*}{\begin{tabular}[c]{@{}c@{}}Type-III kernel \\ ARPE = 3.83806\% \\ Wall time = ($1,933.46 + 108.595$)s $\star$\end{tabular}} & $\hat{a}_{X}$ & $\hat{b}_{X}$ & $\hat{\theta}$ & $\hat{a}_Y$  \\ \cdashline{2-5}
    & 16.628 & 54.5301 & $-0.48461$ & 0.84964  \\  \cline{2-5}
    & $\hat{\kappa}$ & $\hat{d}$ & $\hat{\rho}$ & $ \hat{A}_0$  \\  \cdashline{2-5}
    & 5.57445 & 0.56133 & 0.0995 &  0.54194  \\ \hline
    \end{tabular}
  \label{tab:1b}
\end{table}

\begin{table}[H]\footnotesize
  \centering
  \ContinuedFloat
  \caption{Calibration results for benchmark models on data set 1}
  \begin{tabular}{c|c|c|c|c|c}
  \hline

    \multirow{4}{*}{\begin{tabular}[c]{@{}c@{}}SV-ALJD model \\ ARPE = 7.11843\% \end{tabular}} & $\hat{\sigma}_{X}$ & $\hat{\lambda}_{X}$ & $\hat{b}_{X}$ & $\hat{\eta}$ & $\hat{\lambda}_{Y}$ \\ \cdashline{2-6}
    &  1.17957 &  1.22495 & $1.70474$ & 8.00241 & 3.58362 \\  \cline{2-6}
    & $\hat{b}_Y$ & $\hat\kappa$ & $\hat{\rho}$ & $ \hat{A}_0$   \\  \cdashline{2-5}
    & 8.31656 &  2.07510  & 0.91946 & 0.11559 \\ \hline

    \multirow{4}{*}{\begin{tabular}[c]{@{}c@{}}SV-GMRTS model \\ ARPE = 5.68418\% \\ \end{tabular}} & $\hat{a}_{X}$ & $\hat{b}_{X}$ & $\hat{\theta}$ & $\hat{a}_Y$ & $\hat{\kappa}$ \\ \cdashline{2-6}
    &  18.0968 & 56.728 & $-0.45752$ & 0.85461 & 1.11153 \\  \cline{2-6}
    &  $\hat{\rho}$ & $ \hat{A}_0$   \\  \cdashline{2-3}
    & 0.2298 & 0.36239   \\ \hline

    \multirow{2}{*}{\begin{tabular}[c]{@{}c@{}}Heston model \\ ARPE = 7.20069\%  \end{tabular}} & $\hat{\kappa}$ & $\hat{\rho}$ & $\hat{\varsigma}$ & $\hat{A_0}$ & $\hat{m}$  \\ \cdashline{2-6}
    & 12.1859 & $-0.15446$ & 10.7303 & 0.20727 & 0.86412 \\  \cline{1-6}

    \multirow{2}{*}{\begin{tabular}[c]{@{}c@{}}Black--Scholes model \\ ARPE = 28.9586\% \end{tabular}} & $\hat{\sigma}$  \\ \cdashline{2-2}
    & 0.75637 \\  \cline{1-6}
    \end{tabular}
  \label{tab:1c}
\end{table}

\begin{table}[H]\footnotesize
  \centering
  \ContinuedFloat*
  \caption{Calibration results for FSV-ALJD model on data set 2}
  \begin{tabular}{c|c|c|c|c|c|c}
  \hline
    \multirow{4}{*}{\begin{tabular}[c]{@{}c@{}}Type-I kernel \\ ARPE = 4.72813\% \\ Wall time = ($ 5,983.49 + 541.187$)s\end{tabular}} & $\hat{\sigma}_{X}$ & $\hat{\lambda}_{X}$ & $\hat{b}_{X}$ & $\hat{\eta}$ & $\hat{\lambda}_{Y}$ & $\hat{m}$ \\ \cdashline{2-7}
    & 0.68929 & 3.6012 & 9.23598 & 2.48601 & 9.60755 & 0.08937 \\  \cline{2-7}
    & $\hat{b}_{Y}$ & $\hat{\kappa}$ & $\hat{d}$ & $\hat{\rho}$ & $\hat{A}_{0}$ \\  \cdashline{2-6}
    & 6.48002 & 4.0997 & 0.70175 & 0.27276 & 0.48370\\ \hline

    \multirow{4}{*}{\begin{tabular}[c]{@{}c@{}}Type-II kernel \\ ARPE = 4.69174\% \\ Wall time = ($7,960.35 + 1,664.99$)s\end{tabular}} & $\hat{\sigma}_{X}$ & $\hat{\lambda}_{X}$ & $\hat{b}_{X}$ & $\hat{\eta}$ & $\hat{\lambda}_{Y}$ & $\hat{m}$\\ \cdashline{2-7}
    & 1.06143 &  8.13196 & 9.67874 & 1.52631 & 6.06906 & 0.20043 \\  \cline{2-7}
    & $\hat{b}_{Y}$ & $\hat{\kappa}$ & $\hat{d}$ & $\hat{\rho}$ & $\hat{A}_{0}$ \\  \cdashline{2-6}
    & 7.1211 & 5.61372 & 0.81249 & 0.38208 & 0.22156 \\ \hline

    \multirow{4}{*}{\begin{tabular}[c]{@{}c@{}}Type-III kernel \\ ARPE = 4.04962\% $\star$ \\ Wall time = ($1,792.78 + 85.6393  $)s $\star$\end{tabular}} & $\hat{\sigma}_{X}$ & $\hat{\lambda}_{X}$ & $\hat{b}_{X}$ & $\hat{\eta}$ & $\hat{\lambda}_{Y}$ & $\hat{m}$ \\ \cdashline{2-7}
    &  1.0805 & 3.29407 &  7.65726 & 1.70052 & 4.1792 & 0.20937 \\  \cline{2-7}
    & $\hat{b}_{Y}$ & $\hat{\kappa}$ & $\hat{d}$ & $\hat{\rho}$ & $\hat{A}_{0}$  \\  \cdashline{2-6}
    & 6.91423 & 8.11425 &  0.80968 & 0.42038 & 0.23813\\ \hline
    \end{tabular}
  \label{tab:2a}
\end{table}

\clearpage

\vspace*{0.9in}

\begin{table}[H]\footnotesize
  \centering
  \ContinuedFloat
  \caption{Calibration results for FSV-GMRTS model on data set 2}
  \begin{tabular}{c|c|c|c|c}
  \hline
    \multirow{4}{*}{\begin{tabular}[c]{@{}c@{}}Type-I kernel \\ ARPE = 5.79969\% \\ Wall time = ($10,332.94 + 638.778$)s\end{tabular}} & $\hat{a}_{X}$ & $\hat{b}_{X}$ & $\hat{\theta}$ & $\hat{a}_Y$  \\ \cdashline{2-5}
    & 13.7251 & 43.3248 & $-0.60290$ & 1.18574  \\  \cline{2-5}
    & $\hat{\kappa}$ & $\hat{d}$ & $\hat{\rho}$ & $ \hat{A_0}$   \\  \cdashline{2-5}
    & 3.24254 & 0.85074 & 0.19028 & 0.51104  \\ \hline

    \multirow{4}{*}{\begin{tabular}[c]{@{}c@{}}Type-II kernel \\ ARPE = 5.02678\% $\star$  \\ Wall time = ($8,912.12 + 776.231$)s\end{tabular}} & $\hat{a}_{X}$ & $\hat{b}_{X}$ & $\hat{\theta}$ & $\hat{a}_Y$  \\ \cdashline{2-5}
    & 40.8464 & 74.3314 & $-0.14679$ & 0.9096  \\  \cline{2-5}
    & $\hat{\kappa}$ & $\hat{d}$ & $\hat{\rho}$ & $ \hat{A_0}$    \\  \cdashline{2-5}
    & 5.97759 & 0.93755 & 0.07887 & 0.249355 \\ \hline

    \multirow{4}{*}{\begin{tabular}[c]{@{}c@{}}Type-III kernel \\ ARPE = 5.16355\%\\ Wall time = ($2,597.51 + 119.115$)s $\star$\end{tabular}} & $\hat{a}_{X}$ & $\hat{b}_{X}$ & $\hat{\theta}$ & $\hat{a}_Y$  \\ \cdashline{2-5}
    & 18.7291 & 81.8525 & $-0.29694$ & 0.96554  \\  \cline{2-5}
    & $\hat{\kappa}$ & $\hat{d}$ & $\hat{\rho}$ & $ \hat{A_0}$    \\  \cdashline{2-5}
    & 6.20742 & 0.66818 & 0.05159&  0.54872  \\ \hline
    \end{tabular}
  \label{tab:2b}
\end{table}

\begin{table}[H]\footnotesize
  \centering
  \ContinuedFloat
  \caption{Calibration results for benchmark models on data set 2}
  \begin{tabular}{c|c|c|c|c|c}
  \hline

    \multirow{4}{*}{\begin{tabular}[c]{@{}c@{}}SV-ALJD model \\ ARPE = 4.92285\% \end{tabular}} & $\hat{\sigma}_{X}$ & $\hat{\lambda}_{X}$ & $\hat{b}_{X}$ & $\hat{\eta}$ & $\hat{\lambda}_{Y}$ \\ \cdashline{2-6}
    & 0.82241 &  2.02401& 6.3415 & 0.91815 & 9.70724 \\  \cline{2-6}
    & $\hat{b}_Y$ & $\hat\kappa$ & $\hat{\rho}$ & $ \hat{A}_0$ & $\hat{m}$ \\  \cdashline{2-6}
    & 8.6606 & 13.5178 & 0.3164 & 0.34524 & 0.57851 \\ \hline

    \multirow{4}{*}{\begin{tabular}[c]{@{}c@{}}SV-GMRTS model \\ ARPE = 9.69829\% \\ \end{tabular}} & $\hat{a}_{X}$ & $\hat{b}_{X}$ & $\hat{\theta}$ & $\hat{a}_Y$ & $\hat{\kappa}$ \\ \cdashline{2-6}
    & 17.7803& 55.1617 & $-0.28331$ & 0.37442 & 1.21200 \\  \cline{2-6}
    &  $\hat{\rho}$ & $ \hat{A}_0$   \\  \cdashline{2-3}
    & 0.0795 & 0.55412  \\ \hline

    \multirow{2}{*}{\begin{tabular}[c]{@{}c@{}}Heston model \\ ARPE = 11.7333\%  \end{tabular}} & $\hat{\kappa}$ & $\hat{\rho}$ & $\hat{\varsigma}$ & $\hat{A_0}$ & $\hat{m}$ \\ \cdashline{2-6}
    & 13.7411 & $-0.28254$ & 7.76294 & 0.39582 & 0.7329  \\  \cline{1-6}

    \multirow{2}{*}{\begin{tabular}[c]{@{}c@{}}Black--Scholes model \\ ARPE = 23.5203\% \end{tabular}} & $\hat{\sigma}$  \\ \cdashline{2-2}
    & 0.72631  \\  \cline{1-6}
    \end{tabular}
  \label{tab:2c}
\end{table}

\clearpage

\vspace*{0.2in}

\begin{figure}[H]
  \centering
  \ContinuedFloat*
  \begin{minipage}[c]{0.49\linewidth}
  \centering
  \caption*{\footnotesize FSV-ALJD type-III kernel}
  \includegraphics[scale=0.475]{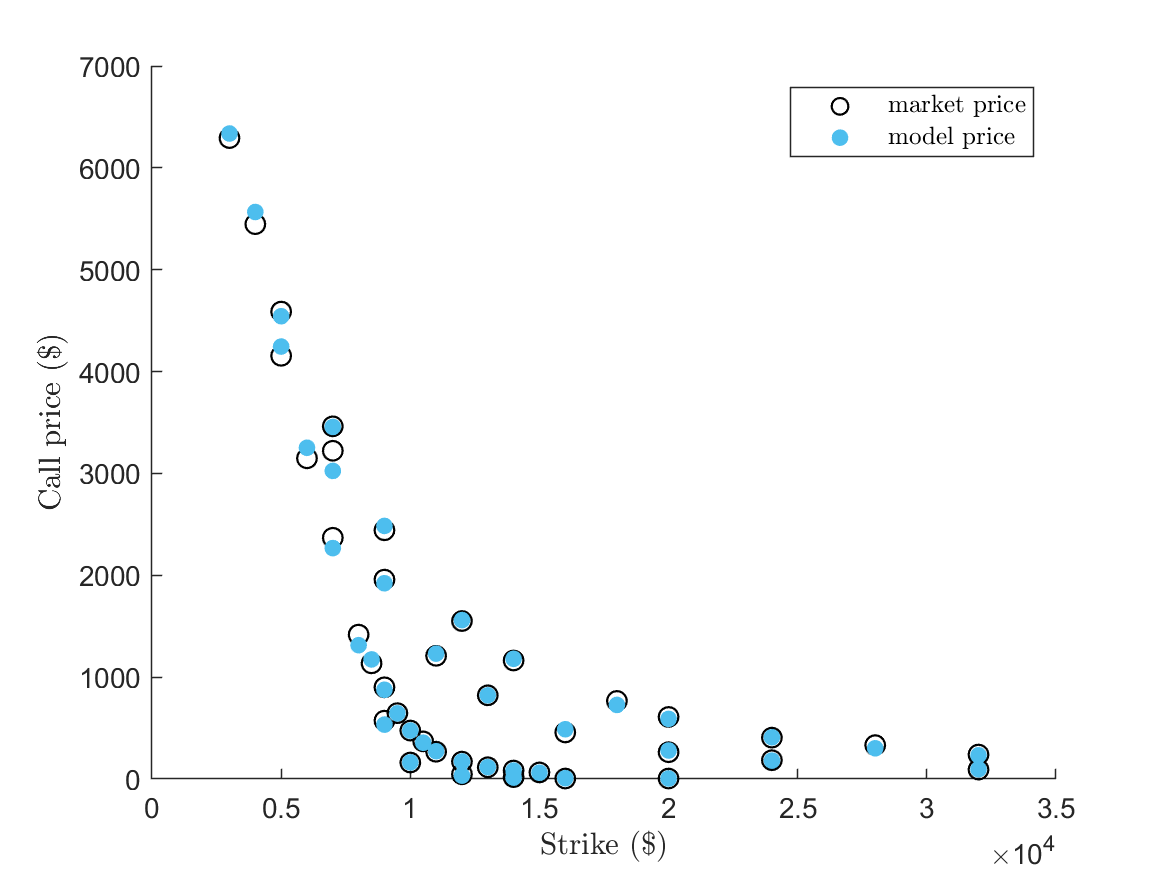}
  \caption*{\footnotesize SV-ALJD}
  \includegraphics[scale=0.475]{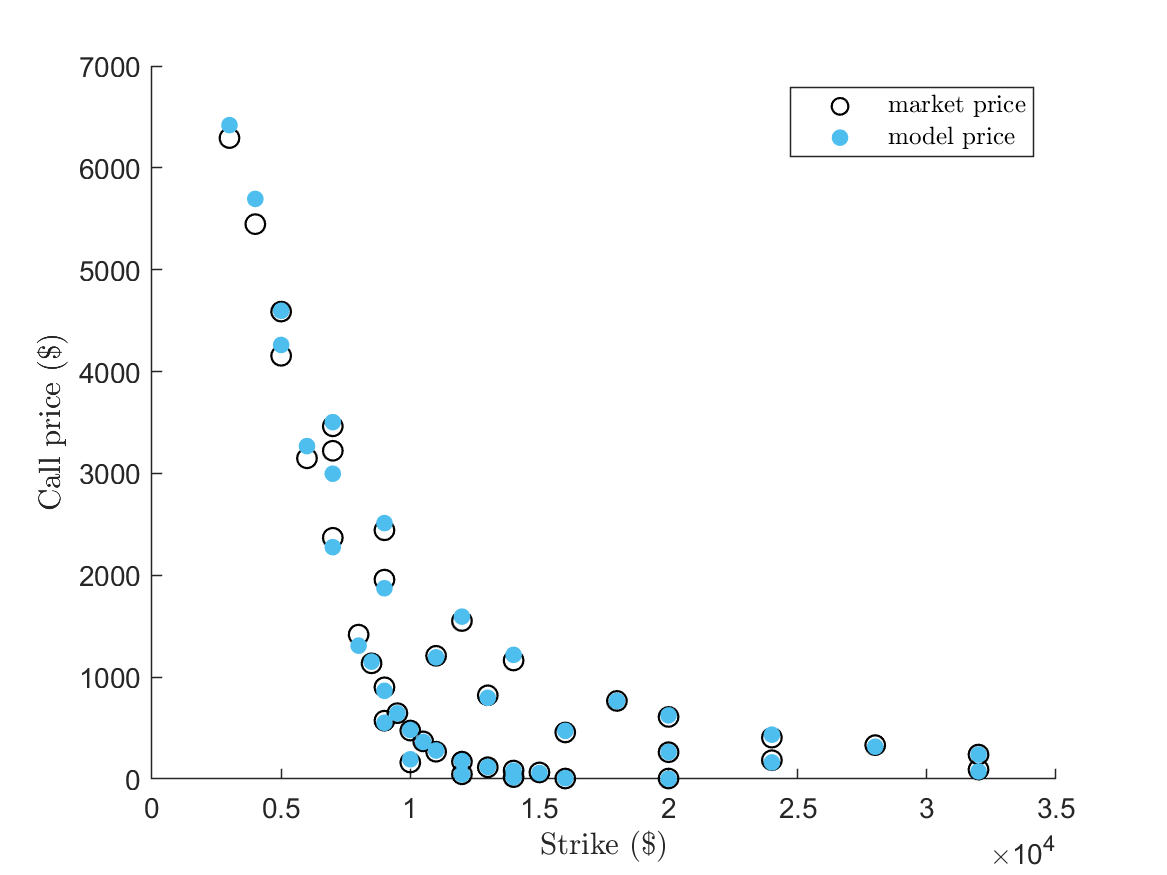}
  \caption*{\footnotesize Heston}
  \includegraphics[scale=0.475]{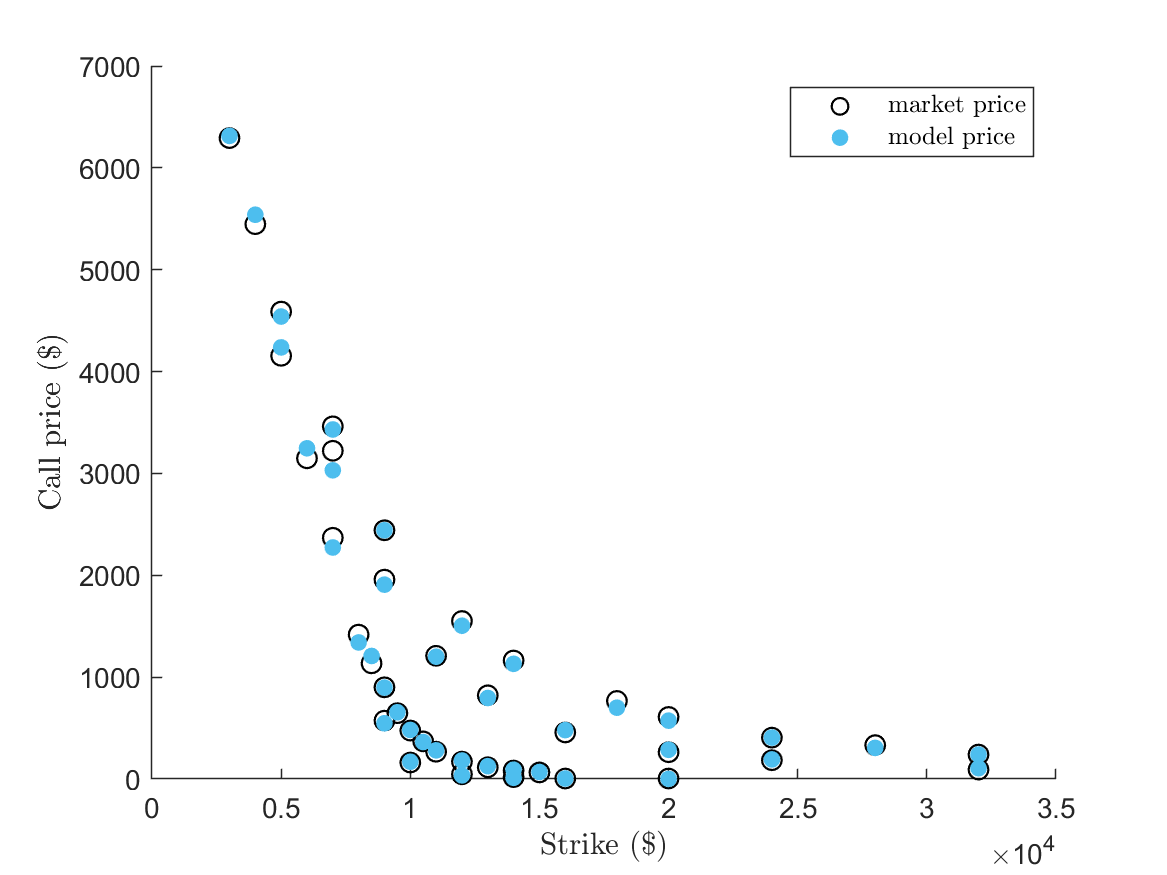}
  \end{minipage}
  \begin{minipage}[c]{0.49\linewidth}
  \centering
  \caption*{\footnotesize FSV-GMRTS type-I kernel}
  \includegraphics[scale=0.475]{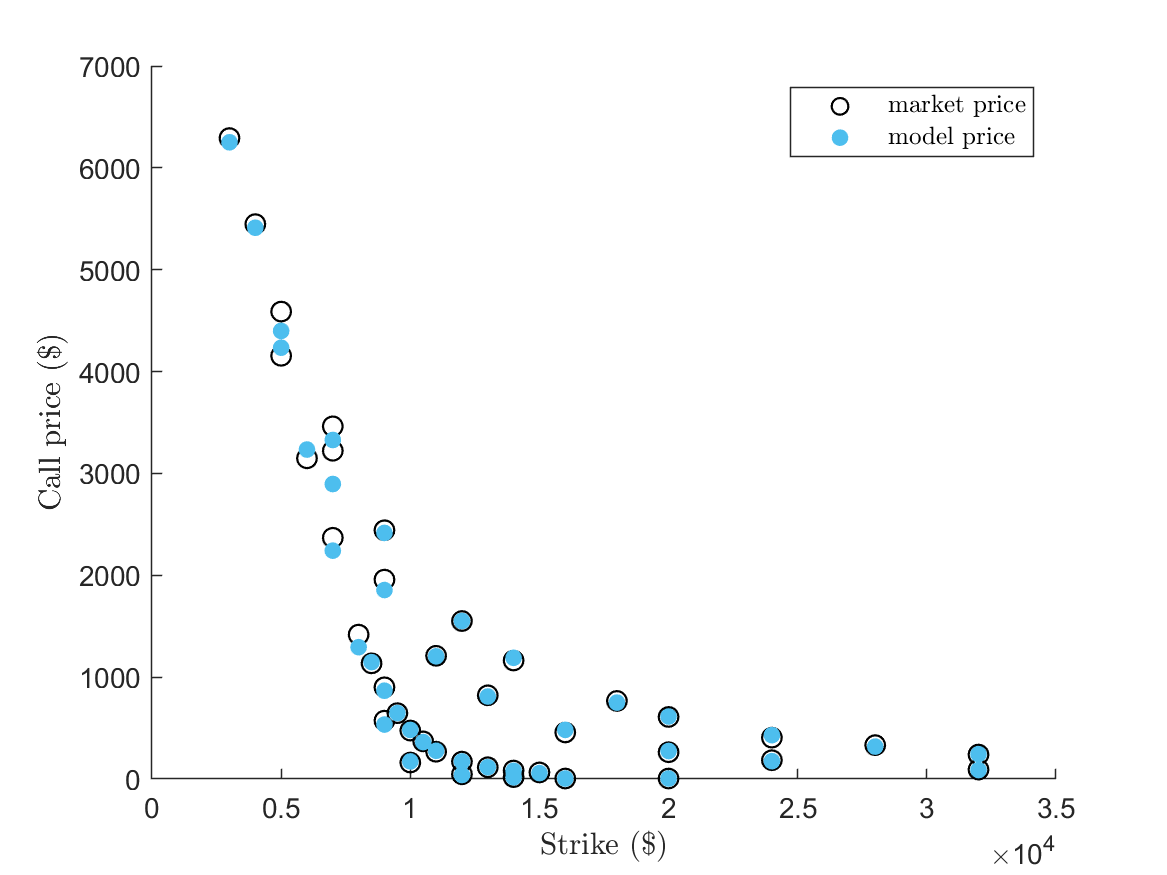}
  \caption*{\footnotesize SV-GMRTS}
  \includegraphics[scale=0.475]{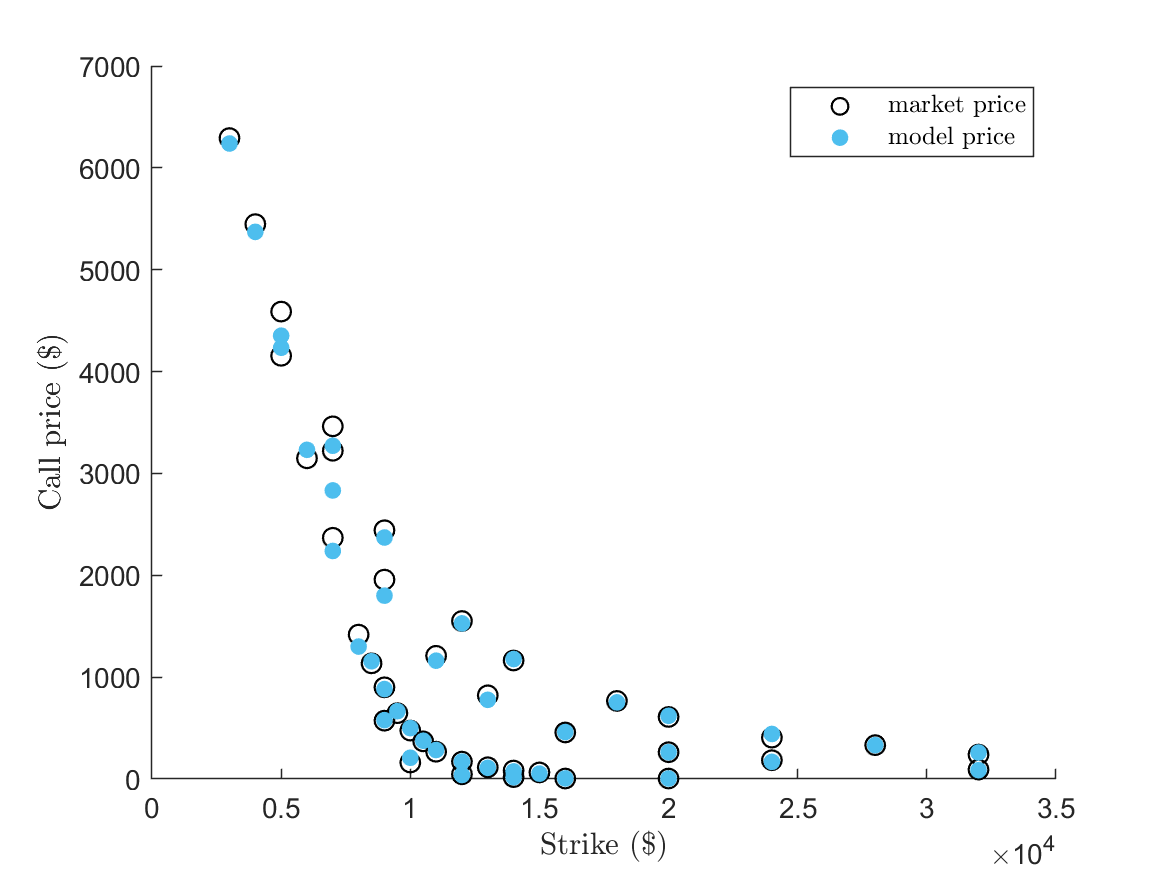}
  \caption*{\footnotesize Black--Scholes}
  \includegraphics[scale=0.475]{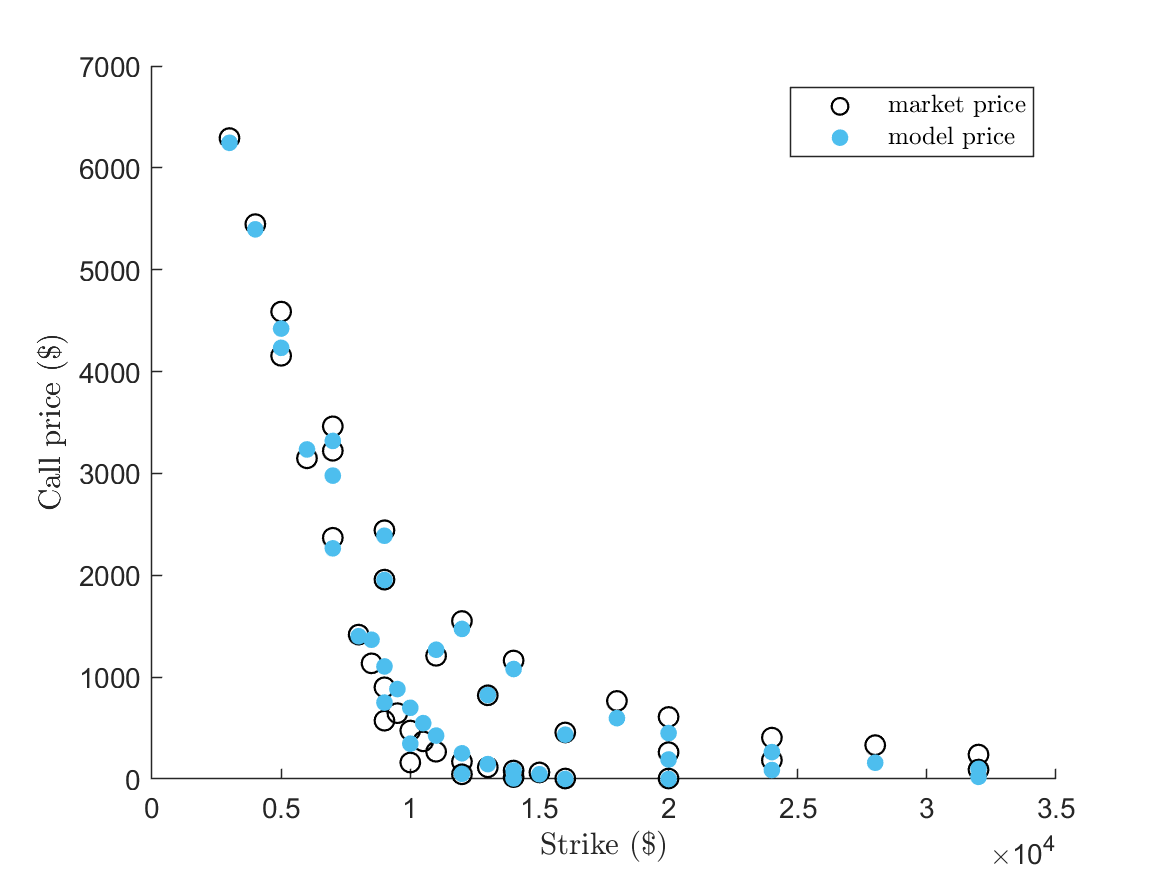}
  \end{minipage}
  \caption{Comparison of market and model prices for data set 1}
  \label{fig:3a}
\end{figure}

\clearpage

\vspace*{0.2in}

\begin{figure}[H]
  \centering
  \ContinuedFloat
  \begin{minipage}[c]{0.49\linewidth}
  \centering
  \caption*{\footnotesize FSV-ALJD type-III kernel}
  \includegraphics[scale=0.475]{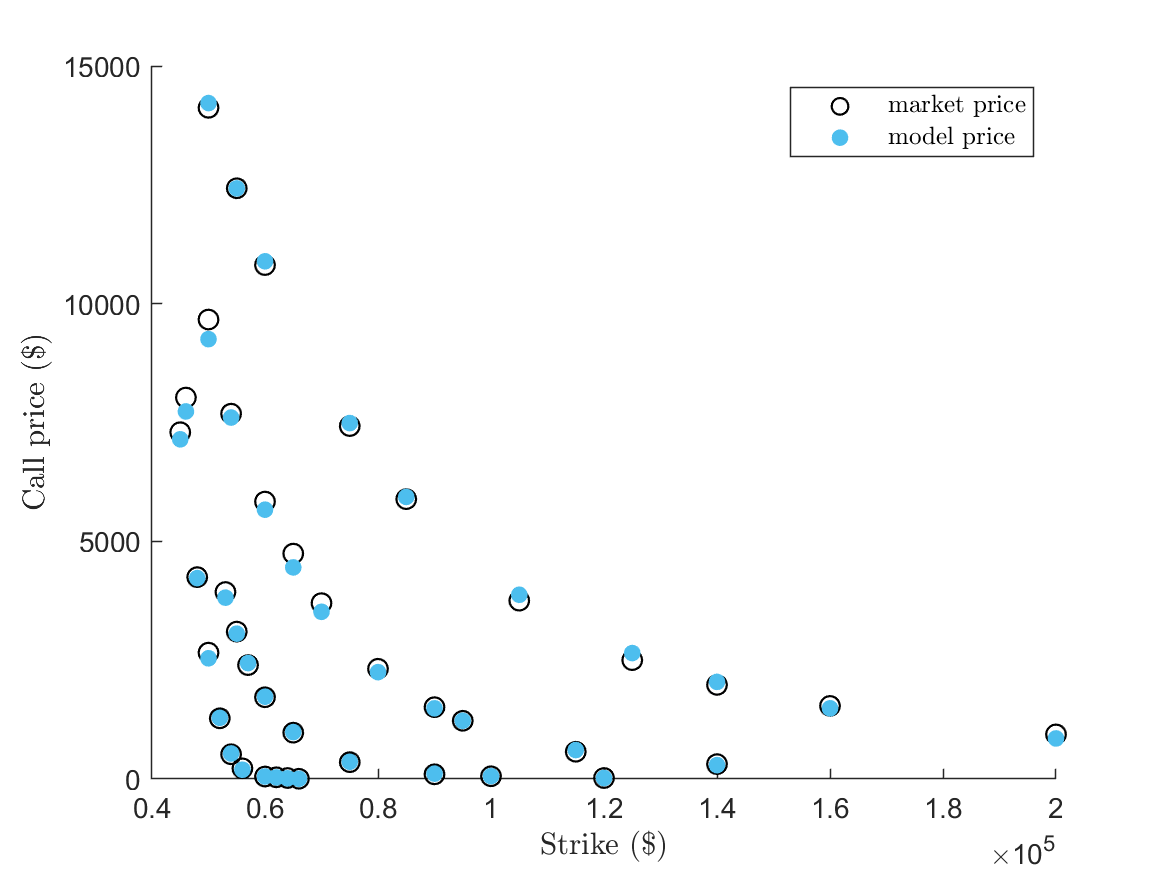}
  \caption*{\footnotesize SV-ALJD}
  \includegraphics[scale=0.475]{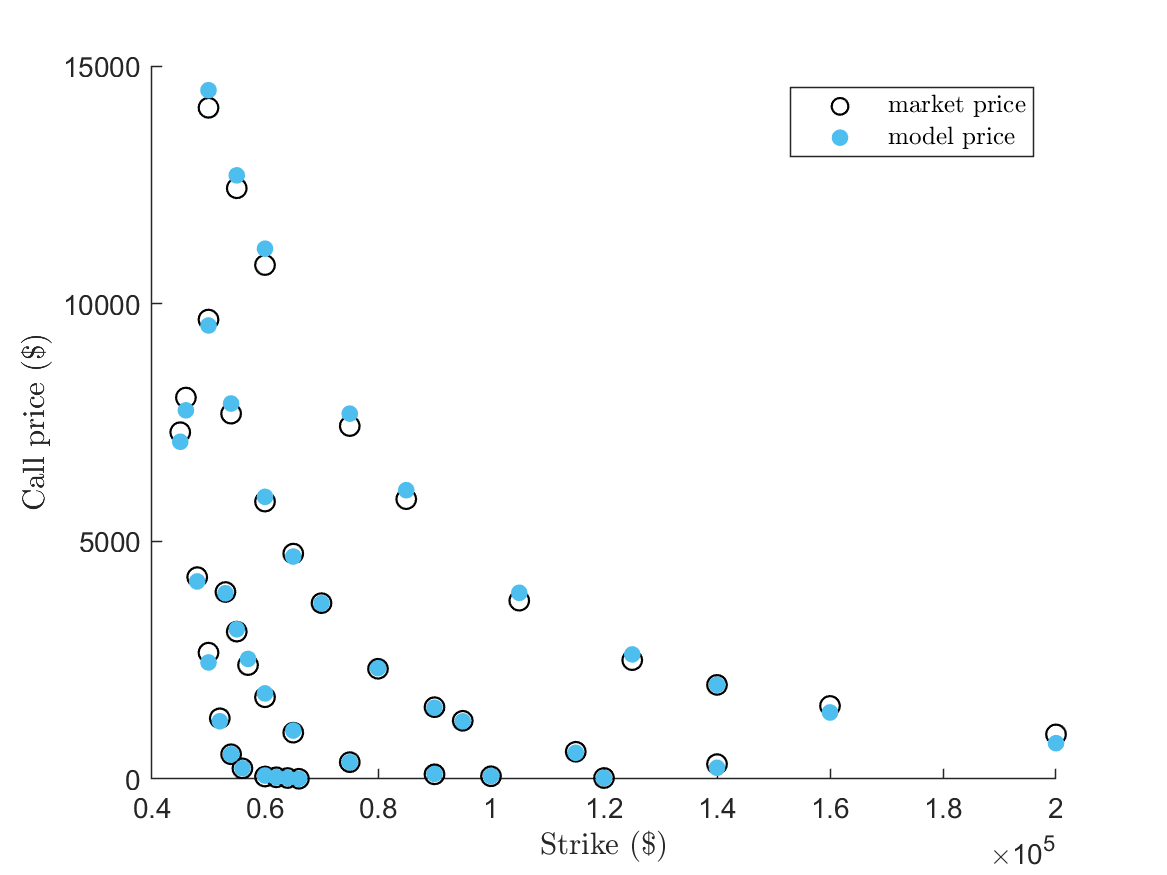}
  \caption*{\footnotesize Heston}
  \includegraphics[scale=0.475]{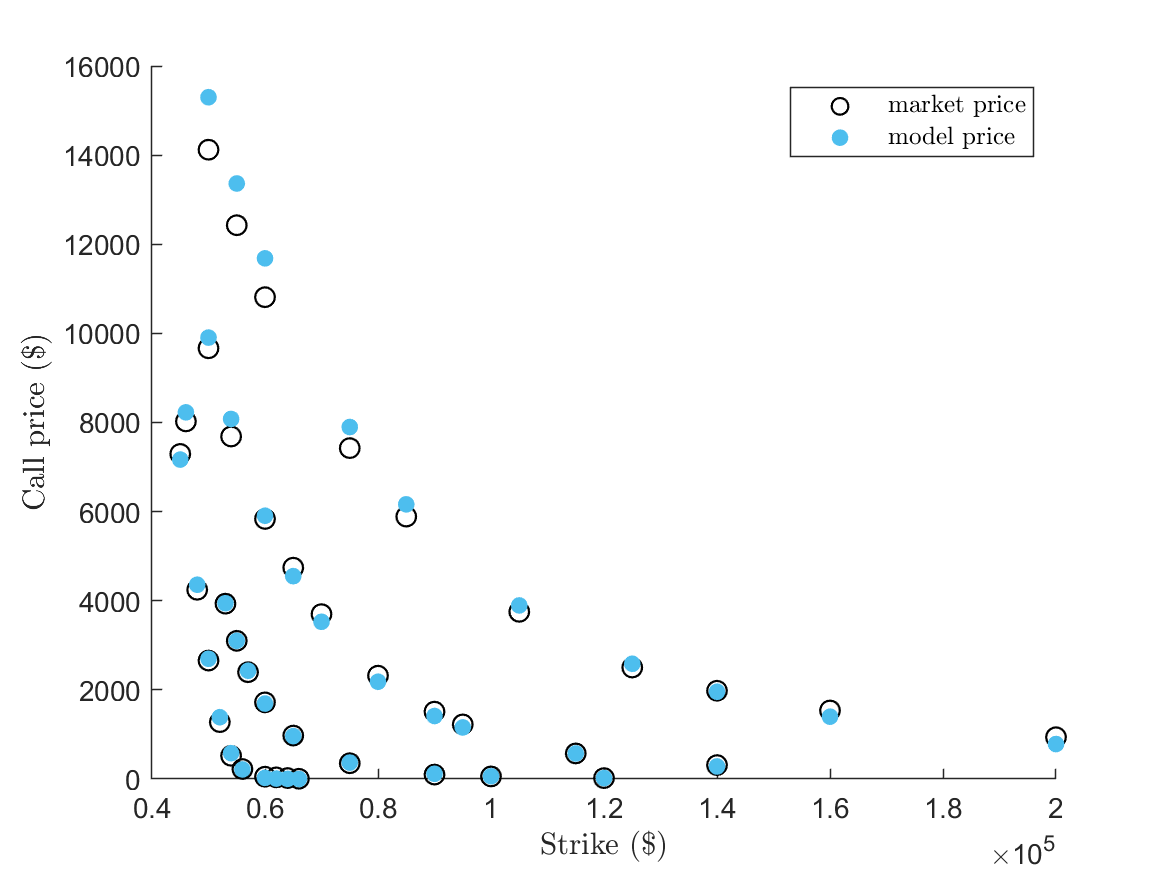}
  \end{minipage}
  \begin{minipage}[c]{0.49\linewidth}
  \centering
  \caption*{\footnotesize FSV-GMRTS type-II kernel}
  \includegraphics[scale=0.475]{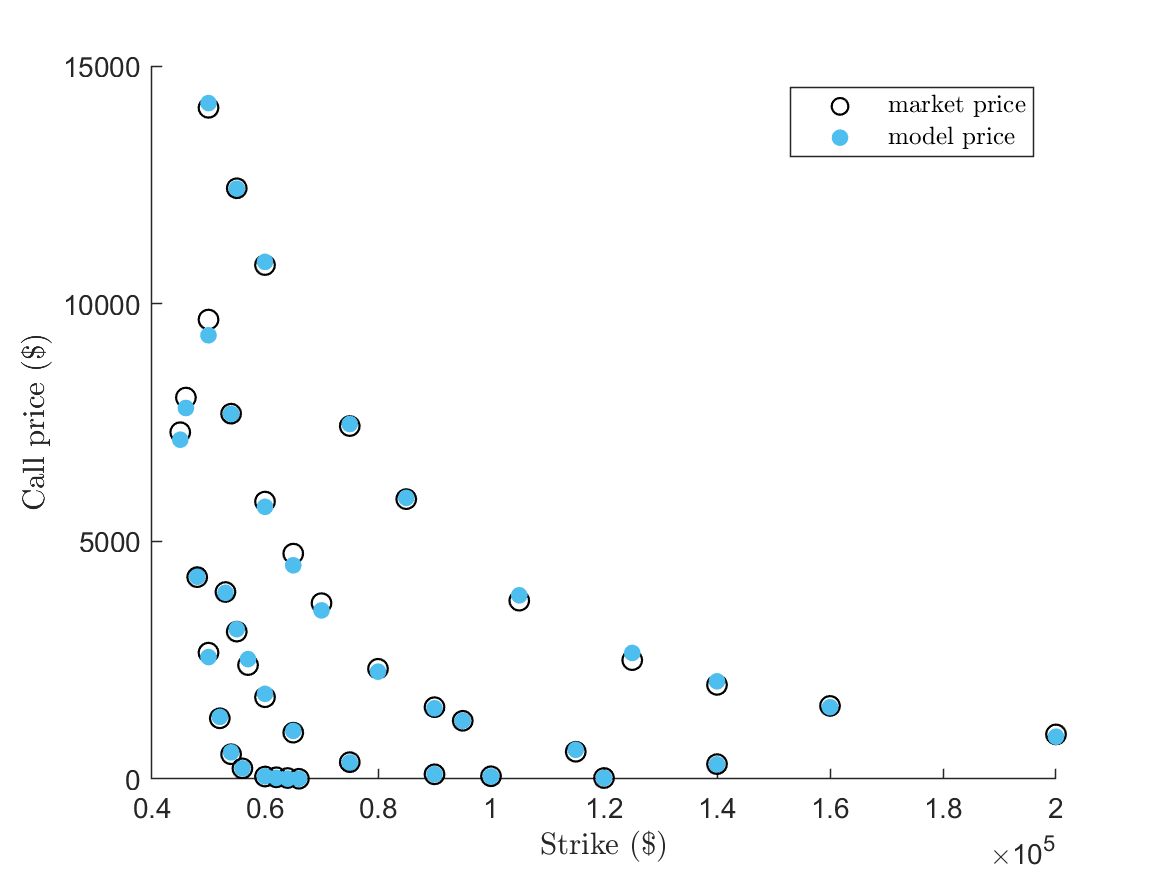}
  \caption*{\footnotesize SV-GMRTS}
  \includegraphics[scale=0.475]{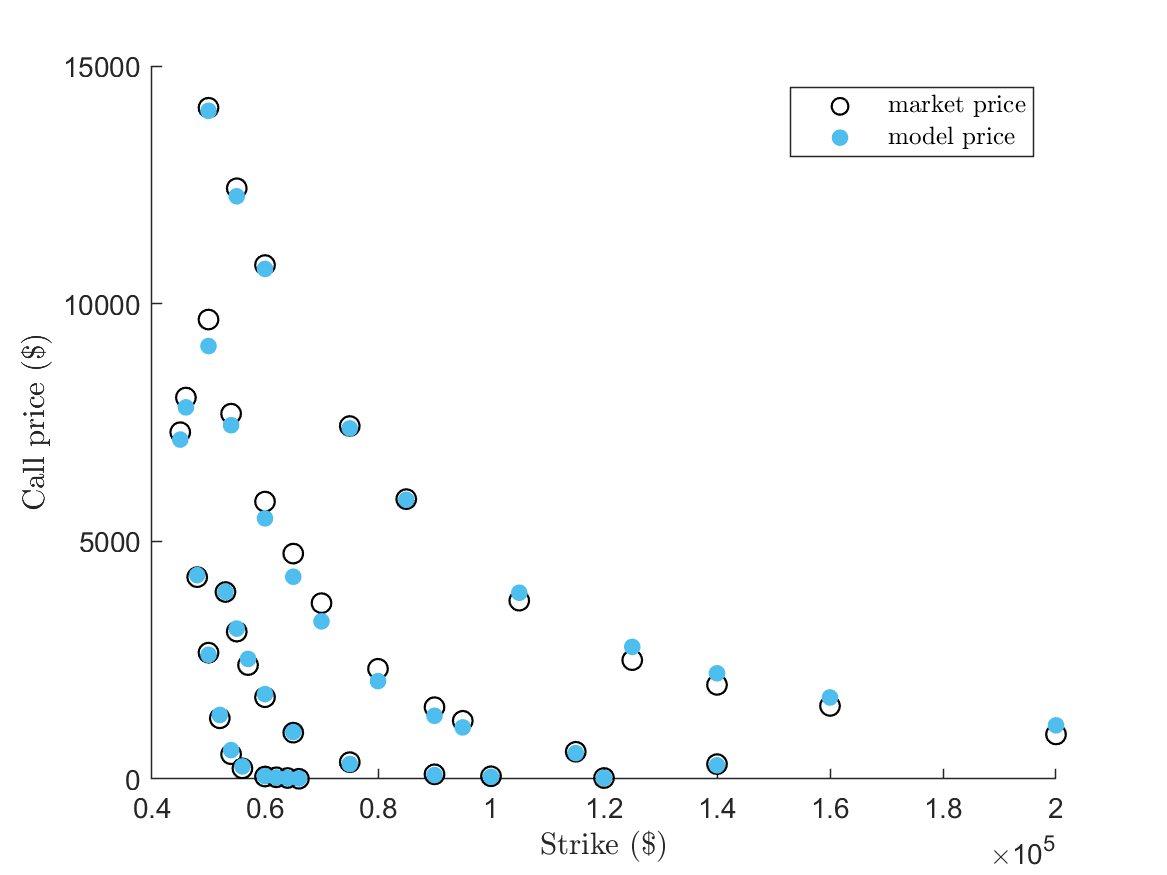}
  \caption*{\footnotesize Black--Scholes}
  \includegraphics[scale=0.475]{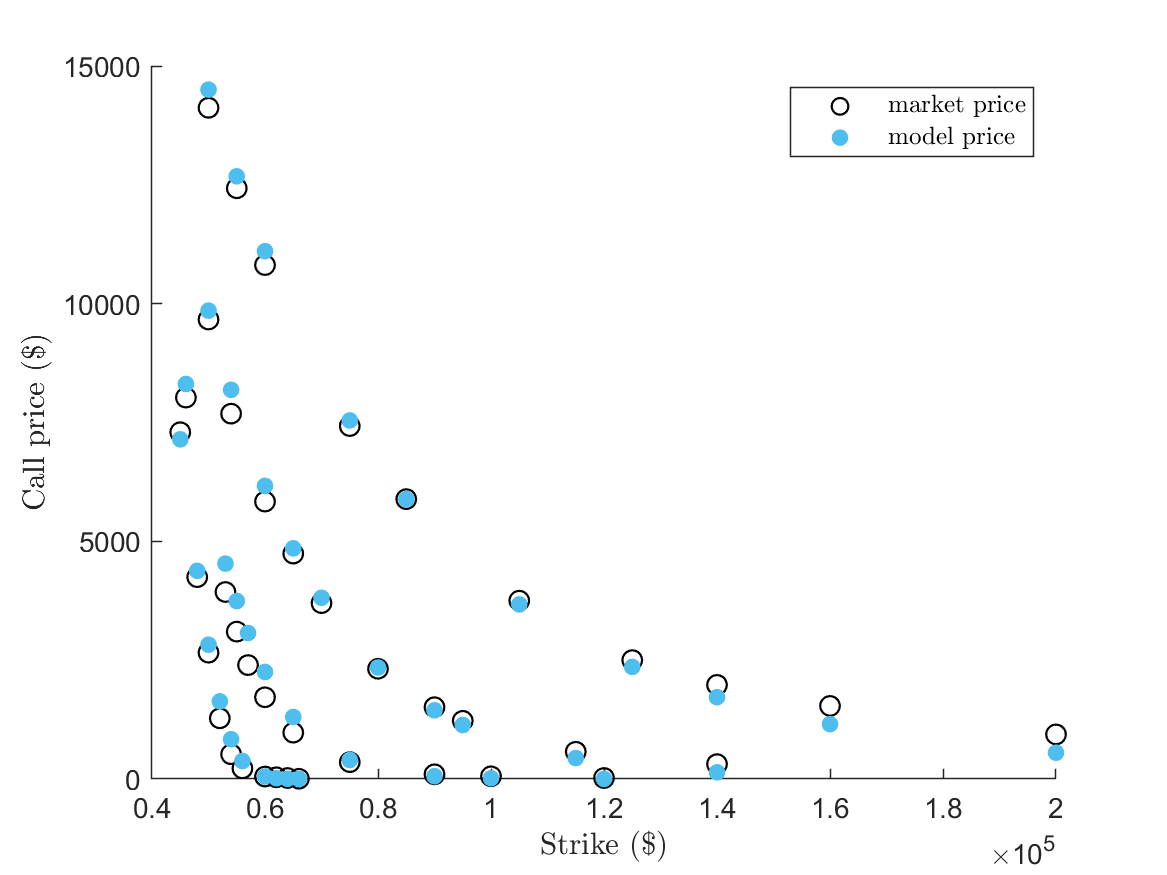}
  \end{minipage}
  \caption{Comparison of market and model prices for data set 2}
  \label{fig:3b}
\end{figure}

\clearpage

On a first look at the calibration results, it is seen that the FSV models easily outperform the popular Black--Sholes model and Heston model as benchmarks -- a direct evidence that correlated jumps in prices and volatility are essential for describing the crypto market dynamics. The observable levels of variability in the calibrated parameters under different types of fractional kernels are mostly due to these kernels exhibiting different local behaviors, despite the same asymptotic behaviors. In terms of model performance, regardless of the base processes in use, all three kernels have demonstrated comparable results, though the type III tends to generate lower ARPEs. Particularly striking is the significantly reduced wall time associated with the type-III kernel, which underscores its suitability for large-scale computational tasks. We stress that this enhanced efficiency comes from the type-III kernel's ability to reexpress the numerical integral for $\log\phi_{Y_1}$ in (\ref{2.2.3}) into a closed-form formula, which can be efficiently implemented thanks to Proposition \ref{pro:3}. Moreover, the comparison of the FSV models against the benchmarks with usual stochastic volatility reveals that the former have consistently outperformed. This observation undoubtedly corroborates the presence of short-range dependence of stochastic volatility in the crypto market, as previously evidenced in [Takaishi, 2020] \cite{T}, underlying its practical necessity. Furthermore, as seen from Figure \ref{fig:4}, for the two data sets, respectively, the maximum relative pricing errors of the FSV models stand at around 15\% and 29\%, significantly lower than those of the Black--Scholes model (100\% and 96\%) and the Heston model (77\% and 86\%). In particular, the FSV models have demonstrated superior performance on data with strike prices above \$10,000 and \$100,000 for the two data sets, respectively, indicating its strong capability in accurately pricing deeply out-of-the-money options.

On a closer look at the calibrated parameters, we see that the fraction parameter $\hat{d}$, directly linked to the degree of short-term dependency, is significantly different from 1 irrespective of the data set under consideration. As discussed in [Wang and Xia, 2022] \cite{WX}, a higher value of $d$ closer to $1$ indicates ``smoother'' sample paths of volatility, while a lower $d$ closer to 0.5 signifies ``rougher'' paths. Hence, this finding suggests the prevalence of short-range volatility dependence for Bitcoin, and by extension, the broader crypto market, which is largely consistent with the empirical evidence presented in [Takaishi, 2020] \cite{T}. Consequently, an important implication is that the proposed FSV framework alone possesses the capability to capture this rough-volatility characteristic in the dynamics of crypto prices. As for the mean reversion-speed parameter $\hat{\kappa}$, we observe from Table \ref{tab:1a} and Table \ref{tab:1b} that the models calibrated on data set 1 demonstrate moderate values ranging from 1 to 5, though the FSV-ALJD model under the type-III kernel comes with a higher value, around 9; in contrast, Table \ref{tab:2a} and Table \ref{tab:2b} generally reveal higher values of $\hat{\kappa}$ calibrated from data set 2 in comparison to the results obtained from the first. This disparity can be attributed to the increasing efficiency of the Bitcoin market in 2024, compared with the conditions prevailing during the COVID-19 pandemic in 2020, for a higher mean-reverting speed typically suggests a rapid dissemination of information and sentiment within the market. This outcome highlights the adeptness of FSV models in capturing the mean-reverting nature of Bitcoin volatility, a stylized fact across equity and commodity markets as well. Regarding the leverage parameter $\hat{\rho}$, which governs the jump correlation between crypto prices and volatility, our findings reveal a consistently positive moderate association across all stochastic-volatility models with jumps, which affirms the existence of the so-called ``inverse leverage effect,'' wherein a sudden increase in price coincides with a surge in volatility. It is crucial to note that this inverse leverage effect is particularly pronounced within jump components, when jump models are being used. On the other hand, the calibrated Heston model, which is entirely diffusive, still suggests negative price--volatility correlations ($\hat{\rho}$), indicating a direct leverage effect.\footnote{Another observation reveals that within the Heston model, the volatility-of-volatility scale ($\hat{\varsigma}$) can soar to as high as 1,500\%, implying that Bitcoin volatility itself exhibits significant volatility, a notion consistent with the findings presented in [Madan et al., 2019, \text{Sect.} 3.4] \cite{MRS}.} This interesting observation aligns precisely with the empirical findings reported by [Huang et al., 2022] \cite{HNX} using crypto time series data; see also [Hou et al, 2020, \text{Tab.} 1] \cite{HWCH}. It is also worth mentioning that this phenomenon is known to prevail in commodity markets ([Trolle and Schwartz, 2009] \cite{TS}) as well as other highly speculative markets such as the Chinese stock market ([Hou, 2013] \cite{H1}). Hence, a pertinent implication is the continued persistence of the inverse leverage effect within the Bitcoin market.

Now we look at the calibrated parameters specific to the base processes. In the ALJD case, particular attention is drawn to the diffusion volatility ($\hat{\sigma}_X$), the jump intensities ($\hat{\lambda}_X$ and $\hat{\lambda}_Y$), and the jump asymmetry parameter ($\hat{\eta}$). As can be seen from Table \ref{tab:1a} and Table \ref{tab:2a}, The calibration results for both data sets indicate a significant volatility of asset returns, exceeding 50\% on the annual basis. Furthermore, it is noted that the jump intensity of volatility (or instantaneous activity) surpasses that of the crypto prices, which suggests the significance of upward jumps in the volatility dynamics of cryptos. Regarding $\hat{\eta}$, it is observed that all optimized values deviate from 1, implying asymmetry in the distribution of Bitcoin price jumps. Specifically, nearly all implemented models yield $\hat{\eta}>1$, which indicates that downward return jumps are more aggressive than upward return jumps on average, except for the model utilizing the type-II kernel calibrated on data set 2. We also observe a substantial difference between the estimates $\hat{\lambda}_{X}$ and $\hat{b}_{X}$, which correspond to jump intensity and (inverse) jump scale, respectively, across the two data sets. In particular, comparing the 2024 period to that of 2020, the estimate $\hat{b}_{X}$ has increased notably, which suggests a potential narrowing in the distribution of jump sizes; at the same time, the higher value of $\hat{\lambda}_{X}$ indicates an increased frequency of return jumps. In other words, the crypto market appears to be experiencing more frequent but smaller jumps, and this shift may signal a more stable trading environment -- one marked by greater market participation and a predominance of milder informational shocks, in contrast to the heightened uncertainty and large jump magnitudes observed during the COVID-19 period.

In the GMRTS case, note that with $c_X=1/2$, $X$ has sample paths of infinite variation, same as the one in the ALJD case where a diffusion component is present, while the parameter $n=2$ has no effect on the path regularity of $X$; we refer to [Fei and Xia, 2024, \text{Sect.} 6] \cite{FX} for a detailed discussion. In this instance, the negative value of the skewness parameter $\hat{\theta}$ validates the left-skewed nature of Bitcoin returns on average, which is precisely in agreement with the implications of the ALJD models. Also, the roles of the shape parameters $a_X$ and $a_Y$ mirror those of the intensity parameters in the ALJD scenario -- they govern the levels of jump activity in the base process $X$ and volatility jump process $Y$ respectively, carrying practical implications akin to trading activity. From the tables, it is evident that $\hat{a}_X$ generally exceeds 1 significantly. When compared to $\hat{a}_Y$, this suggests that Bitcoin tends to experience heightened trading activity despite its elevated volatility and tail risks, which is consistent with the empirical findings presented in [Fei and Xia, 2024] \cite{FX} using time series data. Moreover, the rate parameter $\hat{b}_X$ simply acts as a multiplier dictating the scale of price jumps.

\begin{figure}[H]
  \centering
  \begin{minipage}[c]{0.49\linewidth}
  \centering
  \caption*{\footnotesize Data set 1 (July 11, 2020)}
  \includegraphics[scale=0.31]{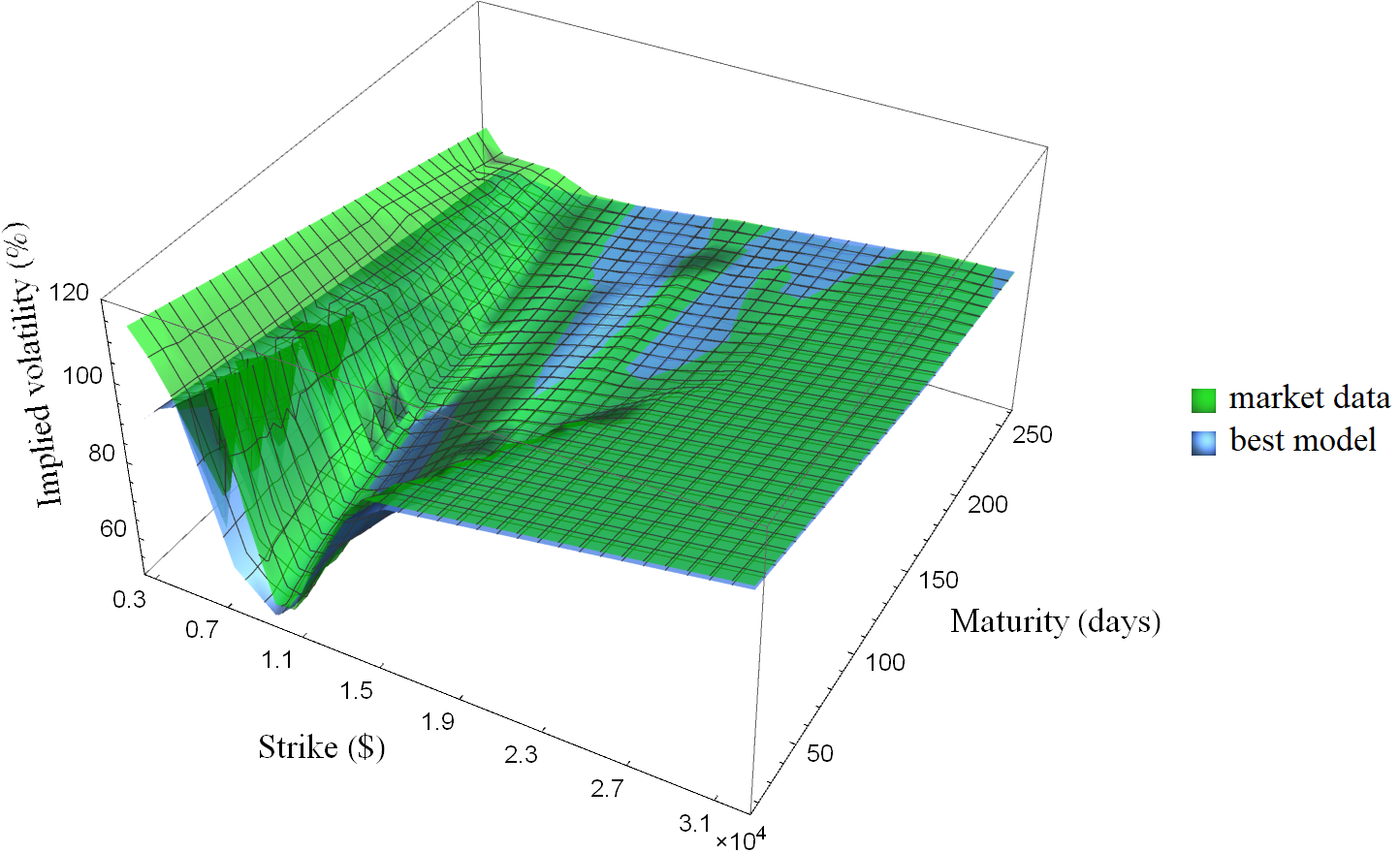}
  \end{minipage}
  \begin{minipage}[c]{0.49\linewidth}
  \centering
  \caption*{\footnotesize Data set 2 (February 19, 2024)}
  \includegraphics[scale=0.31]{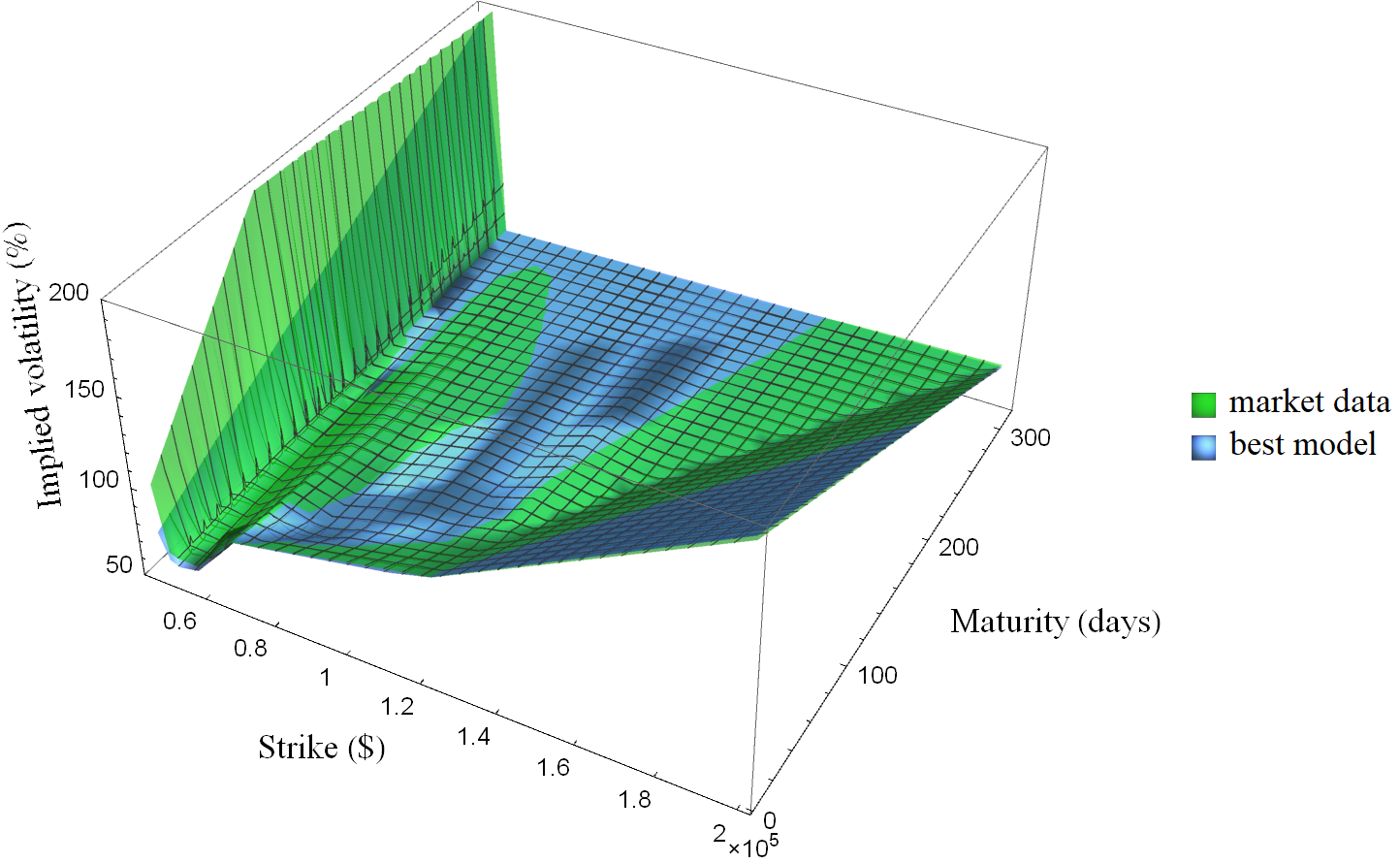}
  \end{minipage}
  \caption{Comparison of market and model implied volatility surfaces}
  \label{fig:5}
\end{figure}

To further illustrate the superior performance of the proposed FSV models in capturing the key dynamics, Figure \ref{fig:5} presents the implied volatility surfaces derived from the Bitcoin option price data and the corresponding best-fitted models (Table \ref{tab:1b} and Table \ref{tab:2a}), identified as the FSV-GMRTS model under the type-I kernel and the FSV-ALJD model under the type-III kernel for the two data sets, respectively. The volatility surfaces are constructed via two-dimensional linear interpolation based on 40 implied volatilities across all strike prices and maturities in each data set. Note that the wide range of implied volatilities (exceeding 200\% in some price regions) reflects the extreme volatility of Bitcoin over the observed periods and, more importantly, despite the markedly different shapes of the implied volatility surfaces (in line with dissimilar market conditions as noted earlier), the best-fitted models successfully replicate the observed surfaces with high accuracy -- particularly in the out-of-the-money regions -- underlying the FSV models robustness across different strike price and maturity levels.

\subsection{Sensitivity analysis for Quanto inverse-power options}\label{sec:5.4}

Using the optimized parameter values from the two models associated with the smallest ARPEs for the two data sets, the FSV-GMRTS model under the type-I kernel and the FSV-ALJD model under the type-III kernel, respectively, we now explore the influence of imposing powers on crypto inverse options, in order to draw insights into the nonlinear leverage effect resulting from the utilization of the power mechanism. Simultaneously, we aim to showcase the general characteristic function-based pricing formulas outlined in Proposition \ref{pro:2}, thereby extending the pricing methodology initially proposed by [Alexander et al., 2023] \cite{ACI} to encompass general stochastic-volatility models. For specificity, in each dataset, we set the conversion rate equal to the spot Bitcoin/USD rate, $R=S_{0}$, and focus on the nearest strike price to $S_{0}$ and the second-longest maturity. The near-the-money Quanto inverse-power option prices are then determined by the model prices obtained from the two best-performing models.

\begin{figure}[H]
  \centering
  \begin{minipage}[c]{0.49\linewidth}
  \centering
  \caption*{\footnotesize Data set 1 (July 11, 2020)}
  \includegraphics[scale=0.475]{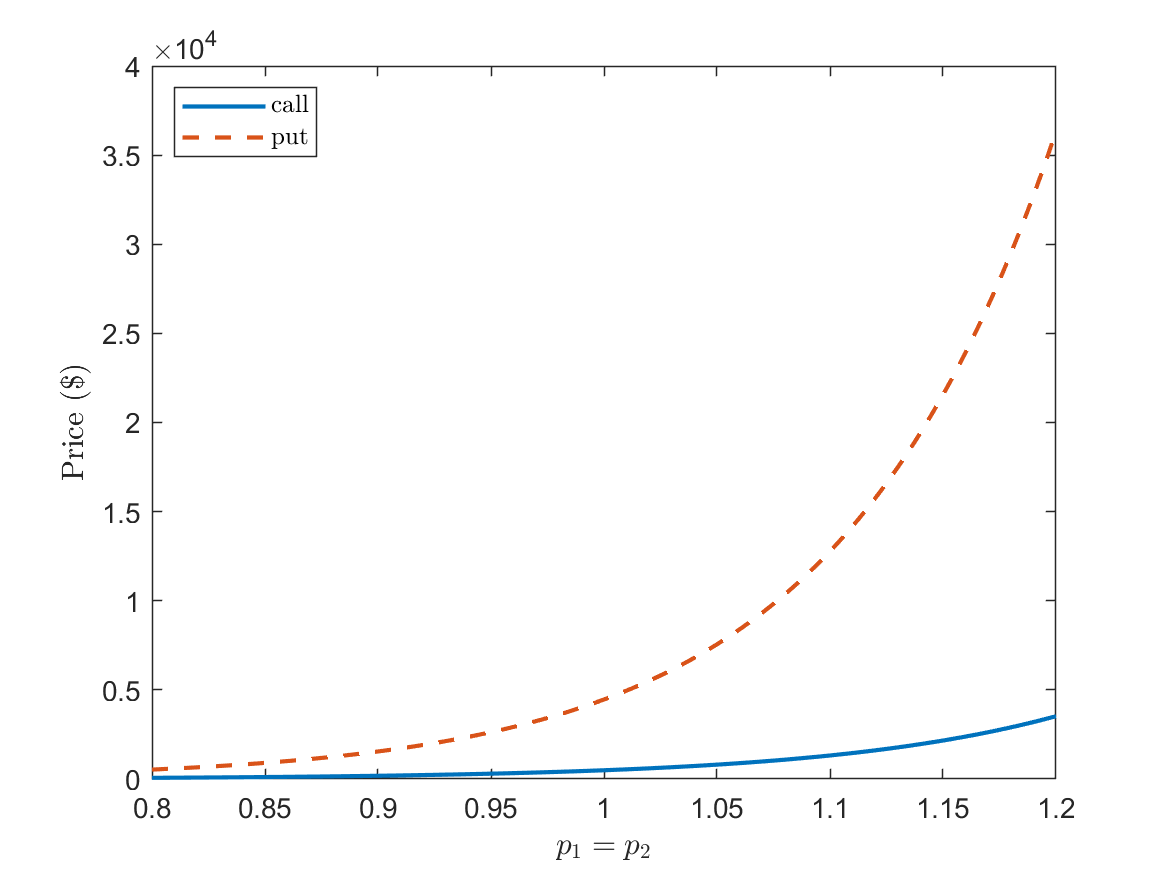}
  \end{minipage}
  \begin{minipage}[c]{0.49\linewidth}
  \centering
  \caption*{\footnotesize Data set 2 (February 19, 2024)}
  \includegraphics[scale=0.475]{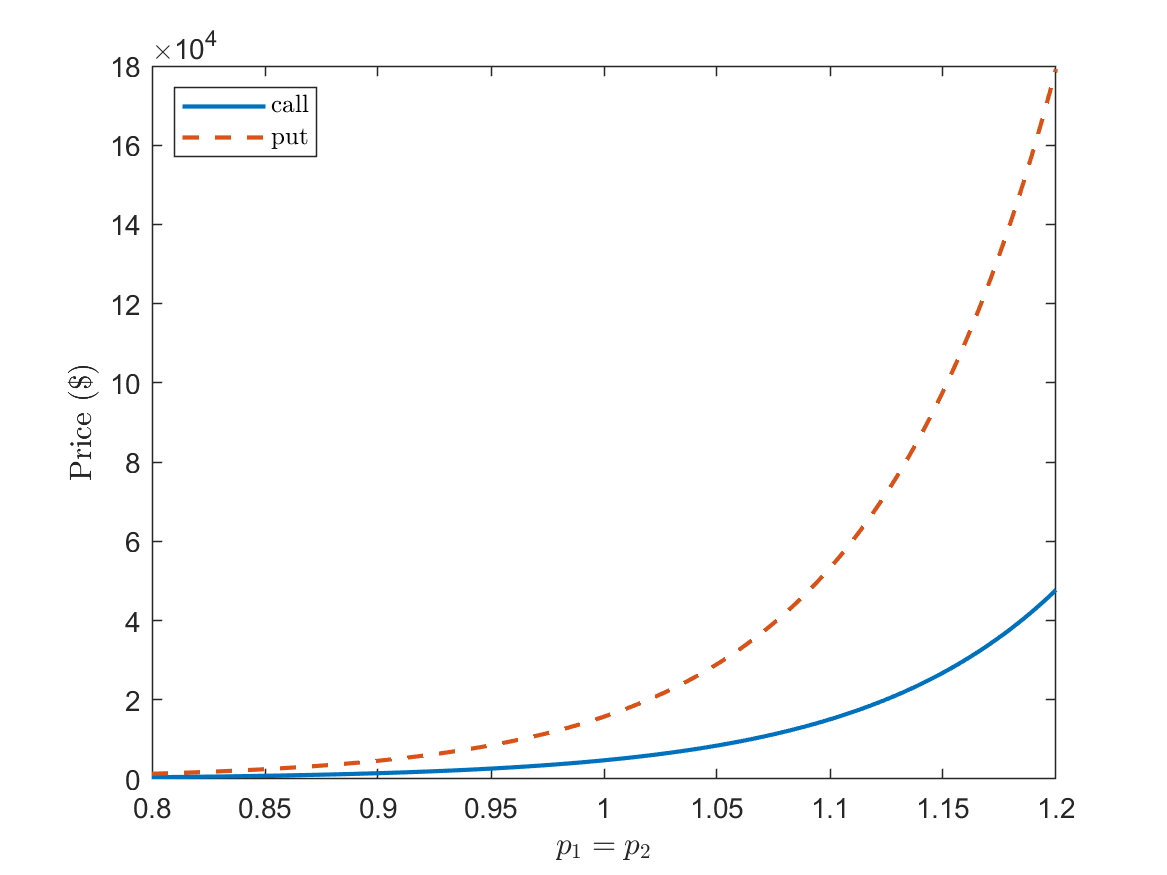}
  \end{minipage}
  \caption{Power curves for Quanto inverse Bitcoin options}
  \label{fig:6}
\end{figure}

\begin{figure}[H]
  \centering
  \caption*{\footnotesize Data set 1 (July 11, 2020)}
  \includegraphics[scale=0.475]{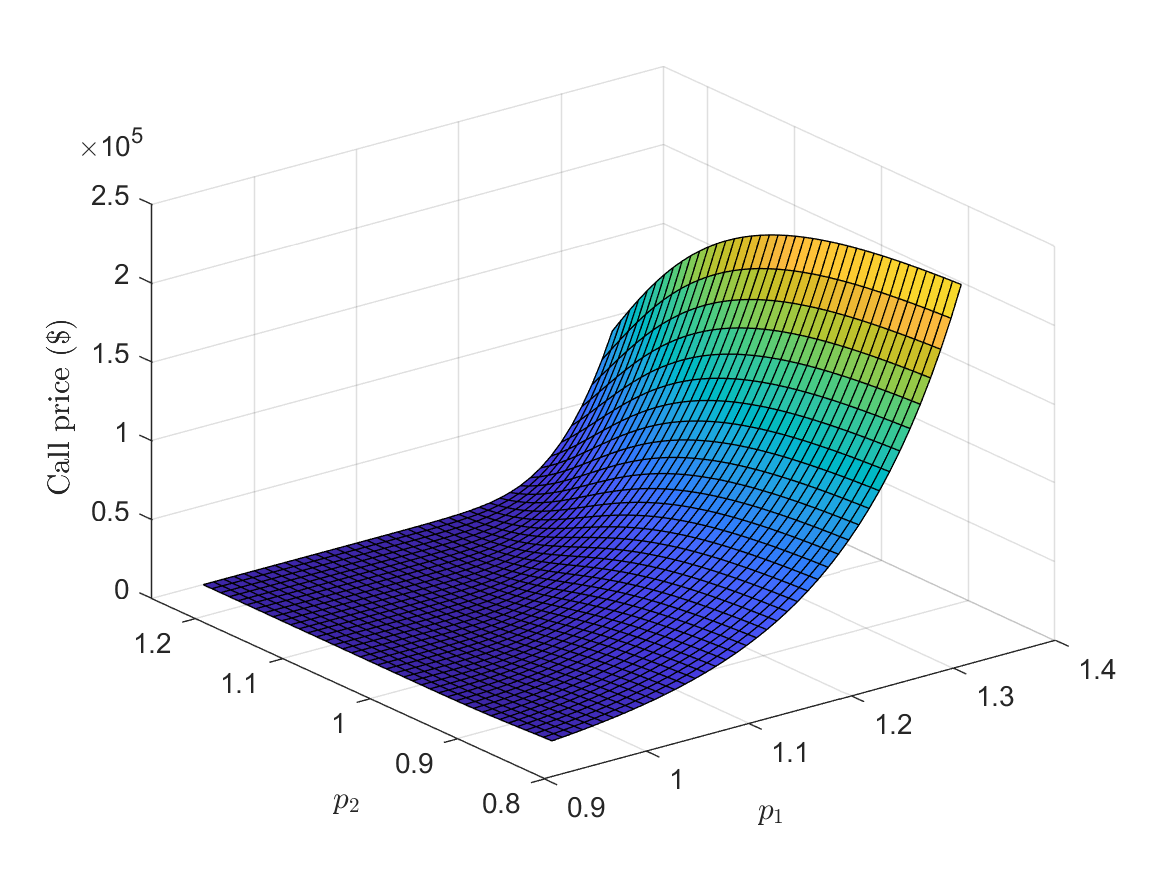}
  \includegraphics[scale=0.475]{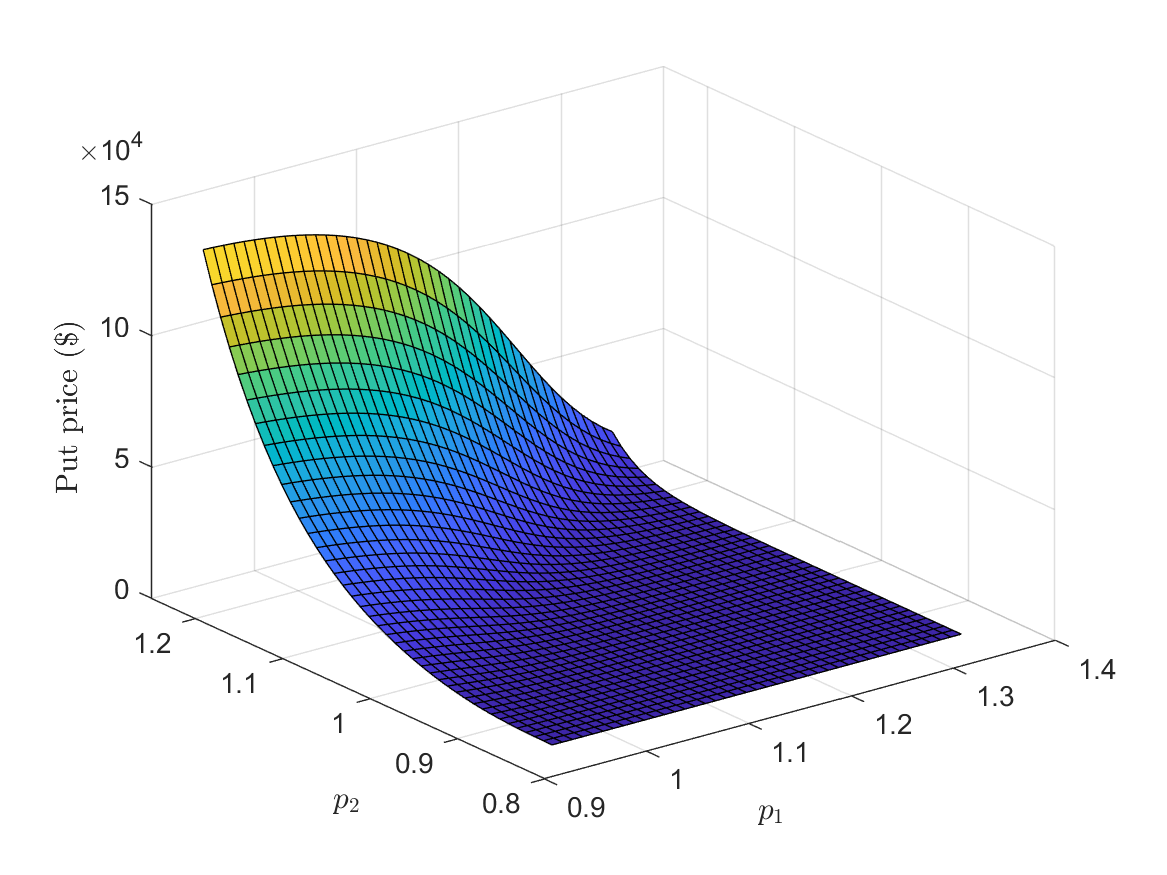}\\
  \caption*{\footnotesize Data set 2 (February 19, 2024)}
  \includegraphics[scale=0.475]{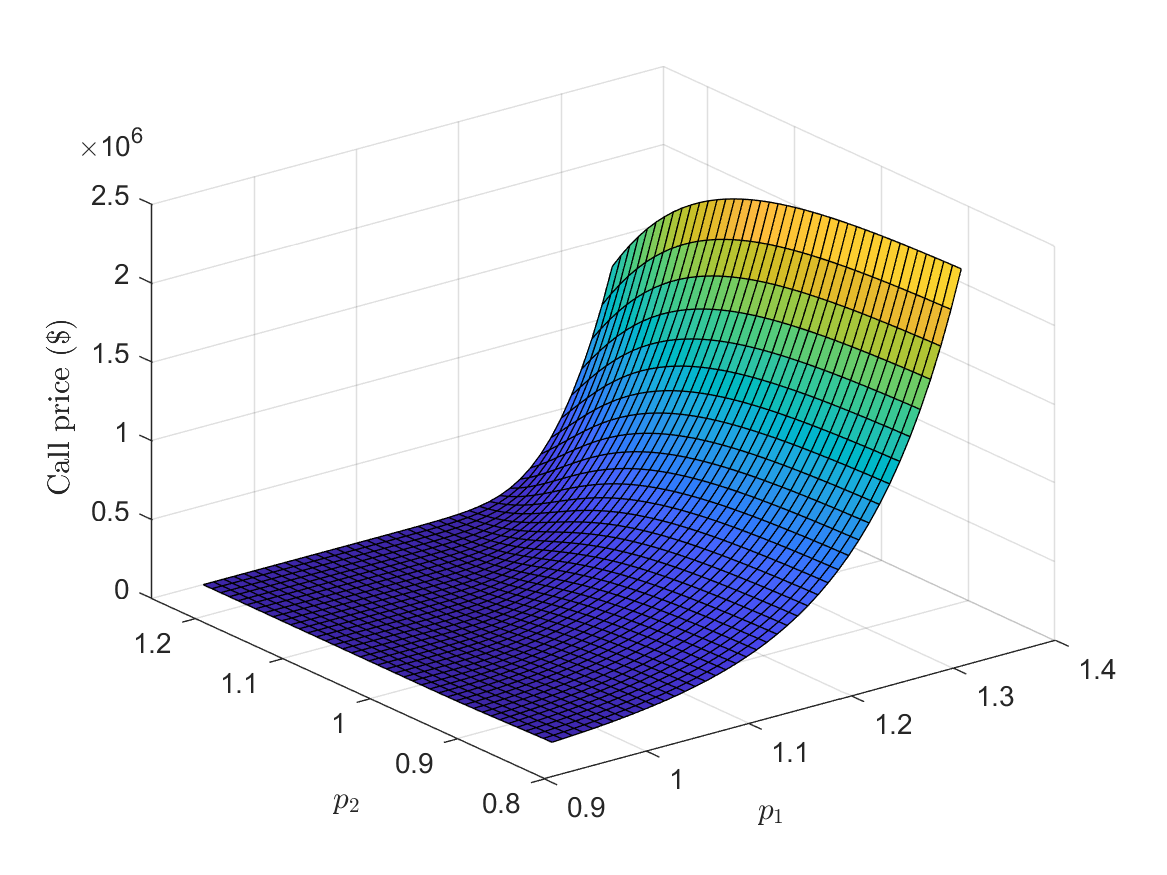}
  \includegraphics[scale=0.475]{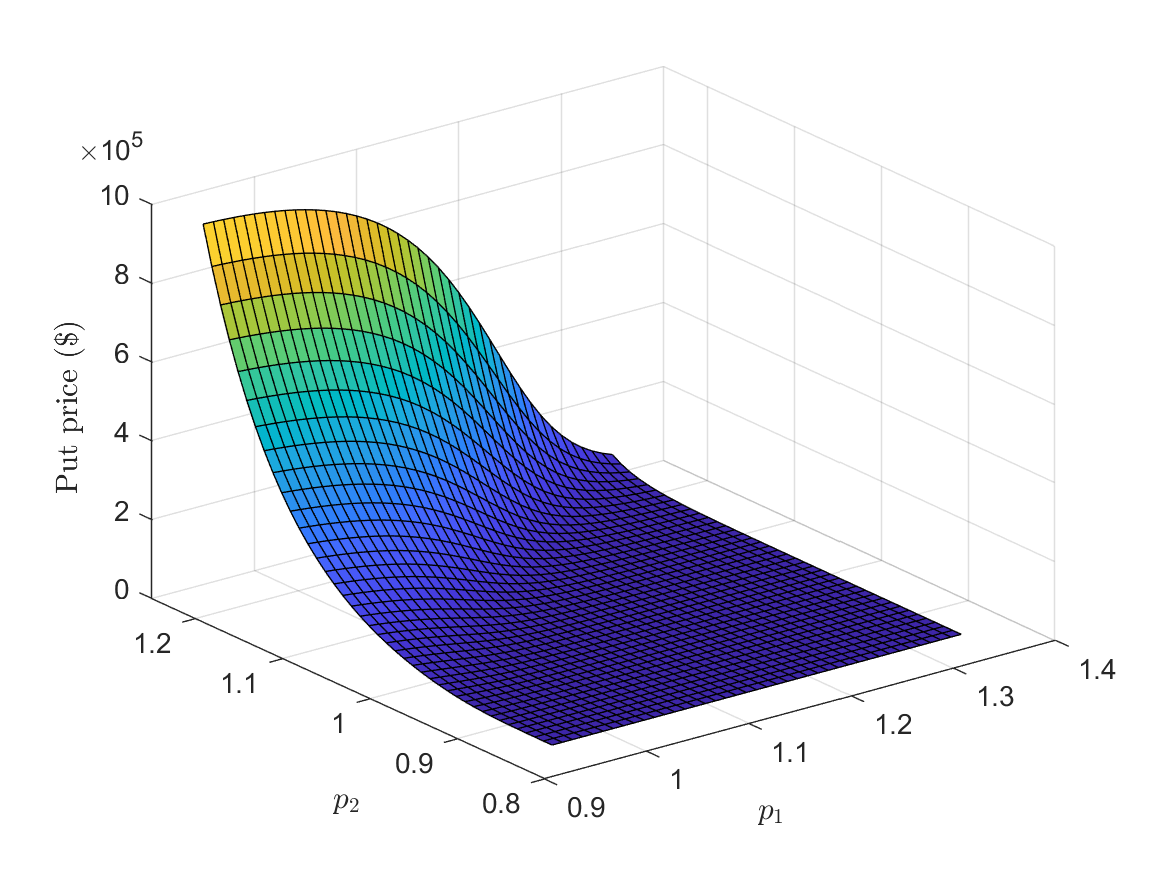}\\
  \caption{Power surfaces for Quanto inverse Bitcoin options}
  \label{fig:7}
\end{figure}

The first illustration is given by imposing equal power coefficients $p_{1}=p_{2}\in[0.8,1.2]$, aligning with the depiction of the power impact on inverse option payoffs as illustrated in Figure \ref{fig:1}. In the second illustration, we allow the two power coefficients to vary independently from 0.8 to 1.4, which enables us to generate inverse-power surfaces and thus elucidate the distinct effects of engineering the power coefficients. Results are shown in Figure \ref{fig:6} and Figure \ref{fig:7}.

Based on Figure \ref{fig:6}, it is seen that across both periods, the put option values are more susceptible to variations in the power coefficients compared to the call options under the best-fitted models. One plausible explanation is that in theory, the (Quanto) inverse put options have the potential to pay an infinite amount of dollars to the holder as the underlying value approaches zero, which circumstance is further amplified by the power impact; see Figure \ref{fig:1} (right panel) and also [Alexander et al., 2023, \text{Fig.} 5] \cite{ACI}. On the other hand, from Figure \ref{fig:7} wherein the two power coefficients are independently adjusted, we observe that when the difference between the power coefficients is substantial, their influence on the call option surpasses that on the put option. This observation resonates with direct options, for which there is no upper limit to the call option payoff as the underlying value continues to rise. However, the power surfaces cease to exhibit convexity due to the inverse feature; compare, for instance, [Xia, 2019, \text{Ex.} 4] \cite{X2}. This in turn clarifies the nuanced influence of the power mechanism on Quanto inverse options, particularly in terms of leveraging the investor's risk exposure, which has intricate contingency on the way in which the power coefficients are designed and adjusted.

\clearpage

\medskip

\section{Conclusions}\label{sec:6}

Motivated by the pronounced presence of price and volatility jumps within the cryptocurrency (crypto) market, in this paper we have proposed a versatile fractional stochastic-volatility (FSV) model designed to comprehensively capture various fundamental characteristics of crypto market dynamics. The core concept lies in employing a random time-change argument ([Carr et al., 2003] \cite{CGMY}) to induce volatility clustering, for which the instantaneous activity rate process is nonetheless modeled using a generalized Ornstein--Uhlenbeck process with square-integrable kernels (comparable to [Wolpert and Taqqu, 2004] \cite{WT}), enabling the depiction of short-range dependence via kernel singularity. While the notion of employing such a generalized Ornstein--Uhlenbeck process was previously explored in [Wang and Xia, 2022] \cite{WX} to model the evolution of instantaneous variance, the use of time change provides a substantial advantage in terms of analytical tractability, while still effectively capturing the essence of ``rough volatility.'' The resulting model framework can be understood as a generalized Barndorff-Nielsen--Shephard model ([Barndorff-Nielsen et al., 2014] \cite{B-NBPV}) with heavy-tailed noises in both returns and volatility. The fractional nature of the activity process accounts for short-range dependence in volatility, rendering the model capable of describing a wide spectrum of jump leverage effects -- either direct or inverse -- as well as price--volatility co-jumps.

Generally speaking, the benefits from using the FSV model framework are threefold. First, the time-change argument by itself provides significant flexibility in modeling the joint dynamics of crypto price and volatility by combining multifarious L\'{e}vy processes. One particular strength is that it does not require price innovations (represented by the L\'{e}vy process $X$) to have a Brownian component. This feature enables the model to be formulated as a wide array of Gaussian mixture models, with enhanced interpretability regarding the price--volume relationship and allowing for further improvement through techniques to address large speculations within the crypto market ([Fei and Xia, 2024] \cite{FX}). Second, with the leveraged L\'{e}vy process $\rho Y$, the FSV model proficiently captures substantial upward volatility jumps and their immediate impact on crypto prices -- a common occurrence over extended periods. When specified, this capability particularly enhances the resulting models' ability to accurately price crypto derivatives and manage risk, including in turbulent market conditions. Third, the Ornstein--Uhlenbeck structure is compatible with a variety of fractional kernels of Volterra type ($h$). These kernels only need to exhibit certain tail behaviors (see (\ref{2.1.2})) to generate the desired mean reversion and volatility dependence, and as long as their further tail integrals (up to order 2), namely $H$ and $J$, have explicit expressions,\footnote{This is a very loose condition which is true for plenty of combinations of powers and exponential functions as well as gamma functions.} the model characteristic function (for the log-price) can be written in semi-closed form involving up to one numerical integral, which underlies its amenability to computation at scale. To our knowledge, the only case where a closed-form characteristic function is available (Proposition \ref{pro:3}) is with the type-III (piecewise) fractional kernel (\ref{4.1.3}), which ensures complete explicitness for a wide range of L\'{e}vy models with explicit characteristic functions.

Since the majority of actively traded crypto options are of the inverse type through the Deribit exchange, USD-denominated investors naturally face significant currency risk when holding these options. This risk can be effectively mitigated through the utilization of Quanto inverse options, as introduced by [Alexander et al., 2023] \cite{ACI}, by employing predetermined conversion rates. While emphasizing the essential functionalities of Quanto inverse options, the present paper also introduces generalizations featuring nonlinear power payoffs. The resulting Quanto inverse-power options offer investors the flexibility to further adjust their exposure to the inherent risk of cryptos. It is important to note that all Quanto inverse-power options are valued in fiat currencies like USD. We have developed (in Section \ref{sec:3}) efficient Fourier transform-type pricing--hedging formulas for this broader class of exotic crypto options, all formulas being model-independent and requiring none but a straightforward characteristic function. Noteworthily, these formulas are presented in a conditional format, allowing for rapid implementation in dynamic pricing and hedging scenarios where frequent model updating may be undesirable.

Our empirical analysis utilizing recent Bitcoin option price data has consistently demonstrated the efficiency of FSV-type models across various specifications, permitting parameter updates on a frequent basis. On the whole, these models have significantly outperformed many existing models with jumps and stochastic volatility, even amidst the abnormal market conditions during the COVID-19 pandemic. The calibrated parameter values align closely with stylized facts in the crypto market ([Madan et al., 2019] \cite{MRS}, [Hou et al., 2020] \cite{HWCH}, [Takaishi, 2020] \cite{T}, [Huang et al., 2022] \cite{HNX}, etc.), highlighting the importance of jumps in both crypto prices and volatility, volatility anti-persistence, and a notable inverse leverage effect. These findings have critical implications for selecting appropriate models in the valuation and risk management of crypto-based financial instruments and the design of related optimal investment strategies. In the meantime, the empirical results have revealed a significant reduction in computation time when utilizing the type-III kernel, while maintaining pricing accuracy at the same level. This compelling outcome beyond doubt underlines the potential for widespread adoption of the type-III kernel in constructing fractional models for volatility analysis.

This paper also offers opportunities for further research in various directions. One notable limitation of the current framework is its inability to generate fractional stochastic volatility-of-volatility, or more precisely, ``rough volatility-of-volatility,'' a phenomenon observed in the equity market (see [Da Fonseca and Zhang, 2019] \cite{DZ}). A potential solution to overcome this limitation was suggested in [Wang and Xia, 2022, \text{Sect.} 6.2] \cite{WX}, via an additional time-change argument at the volatility level. However, doing so would entail evaluating two additional numerical integrals; with the type-III kernel ensuring closed-form characteristic functions within the current FSV model framework, incorporating stochastic volatility-of-volatility would hence significantly increase computational complexity due to the need to evaluate two numerical integrals in the characteristic function, unless a separate time series-based estimation is performed. An alternative approach could involve exploring the rough Heston model ([El Euch and Rosenbaum, 2019] \cite{ER2}) with the piecewise kernel. Nevertheless, this approach would inherently compromise the model's ability to handle volatility jumps and consequently the jump leverage effect, presenting a challenging trade-off. Additionally, the FSV model framework embodies a general concept that can be tailored to different models and subsequently applied to various markets or different asset classes.

\vspace{0.2in}

\appendix
\gdef\thesection{\Alph{section}}

\renewcommand{\theequation}{A.\arabic{equation}}

\section{Mathematical proofs}\label{A}

\textbf{Proposition \ref{pro:1}}

\begin{proof}
Let $t_{0}<t$ be fixed in $[0,T]$. First, it follows directly from (\ref{2.1.1}), (\ref{2.1.5}), and integration-by-parts that
\begin{align}\label{A.1}
  \T_{t}-\T_{t_{0}}=\int^{t}_{t_{0}}A_{s}\dd s&=B(t_{0},t)+\bigg(\int^{t}_{0}-\int^{t_{0}}_{0}\bigg)\int^{v}_{0}h(v,s)\dd Y_{s}\dd v \nonumber\\
  &=B(t_{0},t)+\int^{t}_{0}H(t,s)\dd Y_{s}-\int^{t_{0}}_{0}H(t_{0},s)\dd Y_{s},
\end{align}
where $B(t_{0},t)=(A_{0}-m)(e^{-\kappa t_{0}}-e^{-\kappa t})/\kappa+m(t-t_{0})$ is deterministic and $H(t,s)$ is as given in (\ref{2.2.4}).

The stochastic independence between $X$ and $Y$ allows to apply the independence lemma and the law of iterated expectations given the sigma field $\upsigma(Y_{s}:s\in[0,t])$. Upon plugging in the dynamics of $\log S$ in (\ref{2.2.1}), we may write
\begin{align}\label{A.2}
  \phi_{\log S_{t}|t_{0}}(u)&=\E_{t_{0}}\big[S^{\ii u}_{t_{0}}\exp\big(\ii u\big(X_{\T_{t}}-X_{\T_{t_{0}}}+\rho(Y_{t}-Y_{t_{0}})-\log\phi_{X_{1}}(-\ii)(\T_{t}-\T_{t_{0}}) \nonumber\\
  &\quad-\log\phi_{Y_{1}}(-\ii\rho)(t-t_{0})\big)\big)\big] \nonumber\\
  &=\frac{S^{\ii u}_{t_{0}}}{\phi^{\ii u(t-t_{0})}_{Y_{1}}(-\ii\rho)}\E_{t_{0}}\big[\E\big[e^{\ii u(X_{\T_{t}}-X_{\T_{t_{0}}}+\rho(Y_{t}-Y_{t_{0}})-\log\phi_{X_{1}}(-\ii)(\T_{t}-\T_{t_{0}}))}\big|\upsigma(Y_{s}:s\in[0,t]) \vee\mathcal{F}_{t_{0}}\big]\big] \nonumber\\
  &=\frac{S^{\ii u}_{t_{0}}}{\phi^{\ii u(t-t_{0})}_{Y_{1}}(-\ii\rho)}\E_{t_{0}}\big[e^{\log\phi_{X_{1}}(u)(\T_{t}-\T_{t_{0}})-\ii u\log\phi_{X_{1}}(-\ii)(\T_{t}-\T_{t_{0}})+\ii\rho u(Y_{t}-Y_{t_{0}})}\big],\quad u\in\mathbb{R},
\end{align}
where the last equality follows from the infinite divisibility of $\mathcal{L}(X_{1})$. We remark that $\mathbb{F}$ contains the $\T$-stopped filtration of $X$, which is well-defined because of the path continuity of $\T$. By substituting the expression for $\T$ in terms of $Y$ and writing $B(t_{0},t)=(A_{0}-m)(e^{-\kappa t_{0}}-e^{-\kappa t})/\kappa+m(t-t_{0})$, the exponent in the last equality in (\ref{A.2}) is equal to
\begin{align*}
  &\quad\log\phi_{X_{1}}(u)(\T_{t}-\T_{t_{0}})-\ii u\log\phi_{X_{1}}(-\ii)(\T_{t}-\T_{t_{0}})+\ii\rho u(Y_{t}-Y_{t_{0}}) \\
  &=(\log\phi_{X_{1}}(u)-\ii u\log\phi_{X_{1}}(-\ii))\bigg(B(t_{0},t)+\int^{t}_{t_{0}}H(t,s)\dd Y_{s}+\int^{t_{0}}_{0}(H(t,s)-H(t_{0},s))\dd Y_{s}\bigg) \\
  &\quad\quad+\ii\rho u(Y_{t}-Y_{t_{0}}) \\
  &=(\log\phi_{X_{1}}(u)-\ii u\log\phi_{X_{1}}(-\ii))\int^{t}_{t_{0}}H(t,s)\dd Y_{s}+\ii\rho u(Y_{t}-Y_{t_{0}}) \\
  &\quad\quad+(\log\phi_{X_{1}}(u)-\ii u\log\phi_{X_{1}}(-\ii))\bigg(B(t_{0},t)+\int^{t_{0}}_{0}(H(t,s)-H(t_{0},s))\dd Y_{s}\bigg),
\end{align*}
where the first equality uses (\ref{A.1}). The last result is the sum of two terms, one independent of $\mathcal{F}_{t_{0}}$ and the other measurable with respect to $\mathcal{F}_{t_{0}}$. Then, using further the infinite divisibility of $\mathcal{L}(Y_{1})$ we have for $u\in\mathbb{R}$,
\begin{align}\label{A.3}
  \phi_{\log S_{t}|t_{0}}(u)&=\frac{S^{\ii u}_{t_{0}}}{\phi^{\ii u(t-t_{0})}_{Y_{1}}(-\ii\rho)}\prod^{t}_{t_{0}}\E\bigg[\exp\bigg(((\log\phi_{X_{1}}(u)-\ii u\log\phi_{X_{1}}(-\ii))H(t,s)+\ii\rho u)Y_{1}\bigg)\bigg]^{\dd s} \nonumber\\
  &\quad\times\exp\bigg((\log\phi_{X_{1}}(u)-\ii u\log\phi_{X_{1}}(-\ii))\bigg(B(t_{0},t)+\int^{t_{0}}_{0}(H(t,s)-H(t_{0},s))\dd Y_{s}\bigg)\bigg) \nonumber\\
  &=\exp\bigg(\ii u(\log S_{t_{0}}-(t-t_{0})\log\phi_{Y_{1}}(-\ii\rho)) \nonumber\\
  &\quad+\int^{t}_{t_{0}}\log\phi_{Y_{1}}\big(\rho u-H(t,s)(\ii\log\phi_{X_{1}}(u)+u\log\phi_{X_{1}}(-\ii))\big)\dd s \nonumber\\
  &\quad+(\log\phi_{X_{1}}(u)-\ii u\log\phi_{X_{1}}(-\ii))\bigg(B(t_{0},t)+\int^{t_{0}}_{0}(H(t,s)-H(t_{0},s))\dd Y_{s}\bigg)\bigg),
\end{align}
where $\prod^{t}_{t_{0}}$ denotes a geometric integral ([Slav\'{i}k, 2007] \cite{S2}). This gives an alternative expression for the conditional characteristic function of the log-price.

Since the quadratic variation $[Y,Y]$ is a L\'{e}vy process,
\begin{equation*}
  \E_{t_{0}}[[Y,Y]_{t}-[Y,Y]_{t_{0}}]=(t-t_{0})\E[[Y,Y]_{1}].
\end{equation*}
Also, according to [Jacod, 1979, \text{Thm.} 10.17] \cite{J}, the path continuity of $\T$ implies that $[X_{\T},X_{\T}]_{t}=[X,X]_{\T_{t}}$, $\Q$-a.s., and hence
\begin{align*}
  \E_{t_{0}}[[X_{\T},X_{\T}]_{t}-[X_{\T},X_{\T}]_{t_{0}}]&=\E_{t_{0}}[\E_{t_{0}}[[X,X]_{\T_{t}}-[X,X]_{\T_{t_{0}}} |\upsigma(Y_{s}:s\in[0,t])\vee\mathcal{F}_{t_{0}}]] \\
  &=\E_{t_{0}}[\T_{t}-\T_{t_{0}}]\E[[X,X]_{1}] \\
  &=\bigg(B(t_{0},t)+\int^{t}_{t_{0}}H(t,s)\dd s\E[Y_{1}]+\int^{t_{0}}_{0}(H(t,s)-H(t_{0},s))\dd Y_{s}\bigg) \\
  &\quad\times\E[[X,X]_{1}].
\end{align*}
For the cross-variation, the claim is that $[X_{\T},Y]_{t}=0$, $\Q$-a.s., despite the dependence between $\T$ and $Y$. To see this, recall that $\T$ is a continuous strictly increasing process by (\ref{2.1.1}) and the strict positivity of $A$, so that $\T^{-1}$ exists $\Q$-\text{a.s.} and is also strictly increasing, thus being a valid time change. Reusing the previous argument based on path continuity, we have that $[X_{\T},Y]_{t}=[X,Y_{\T^{-1}}]_{\T_{t}}$ $\Q$-a.s., but since $t\in(0,T]$ is arbitrary and $[X,Y_{\T^{-1}}]$ has c\`{a}dl\`{a}g sample paths by construction, this means that $[X_{\T},Y]_{t}=0$ if and only if $[X,Y_{\T^{-1}}]_{t}=0$. As $Y$ is purely discontinuous, it suffices to consider the case where $X$ is also purely discontinuous, or has no Brownian part, with
\begin{equation*}
  [X,Y_{\T^{-1}}]_{t}=\sum_{s\leq t}\Delta X_{s}\Delta Y_{\T^{-1}_{s}},
\end{equation*}
where $\Delta$ stands for the jump component. The stochastic independence between $X$ and $Y_{\T^{-1}}$ implies that their sets of jump times within $[0,t]$ have no common elements $\Q$-a.s. Hence, $[X,Y_{\T^{-1}}]_{t}=0$ and
\begin{equation*}
  \E_{t_{0}}[[X_{\T},Y]_{t}-[X_{\T},Y]_{t_{0}}]=0.
\end{equation*}

Therefore, with $\imath$ denoting the identity map on time, we conclude that
\begin{align}\label{A.4}
  &\quad\E_{t_{0}}\bigg[\bigg[\log\frac{S}{S_{0}},\log\frac{S}{S_{0}}\bigg]_{t}-\bigg[\log\frac{S}{S_{0}},\log\frac{S}{S_{0}}\bigg]_{t_{0}}\bigg] \nonumber\\
  &=\E_{t_{0}}\big[[X_{\T} -\log\phi_{X_{1}}(-\ii)\T+\rho Y-\phi_{Y_{1}}(-\ii\rho)\imath,X_{\T} -\log\phi_{X_{1}}(-\ii)\T+\rho Y-\phi_{Y_{1}}(-\ii\rho)\imath]_{t} \nonumber\\
  &\quad-[X_{\T}-\log\phi_{X_{1}}(-\ii)\T+\rho Y-\phi_{Y_{1}}(-\ii\rho)\imath,X_{\T}-\log\phi_{X_{1}}(-\ii)\T+\rho Y-\phi_{Y_{1}}(-\ii\rho)\imath]_{t_{0}}\big] \nonumber\\
  &=\E_{t_{0}}\big[[X,X]_{\T_{t}}-[X,X]_{\T_{t_{0}}}+\rho^{2}([Y,Y]_{t}-[Y,Y]_{t_{0}})\big] \nonumber\\
  &=\bigg(B(t_{0},t)+\int^{t}_{t_{0}}H(t,s)\dd s\E[Y_{1}]+\int^{t_{0}}_{0}(H(t,s)-H(t_{0},s))\dd Y_{s}\bigg)\E[[X,X]_{1}] \nonumber\\
  &\quad+\rho^{2}(t-t_{0})\E[[Y,Y]_{1}],
\end{align}
which implies that
\begin{align}\label{A.5}
  &\quad\int^{t_{0}}_{0}(H(t,s)-H(t_{0},s))\dd Y_{s} \nonumber\\
  &=\frac{\E_{t_{0}}[[\log(S/S_{0}),\log(S/S_{0})]_{t}] -[\log(S/S_{0}),\log(S/S_{0})]_{t_{0}}-\rho^{2}(t-t_{0})\E[[Y,Y]_{1}]}{\E[[X,X]_{1}]} \nonumber\\
  &\quad-B(t_{0},t)-\int^{t}_{t_{0}}H(t,s)\dd s\E[Y_{1}].
\end{align}
Lastly, note that $\E[[X,X]_{1}]=\Var[X_{1}]$ and $\E[[Y,Y]_{1}]=\Var[Y_{1}]$ because $X$ and $Y$ are both L\'{e}vy processes. Putting (\ref{A.5}) back in (\ref{A.3}) gives the desired result upon rearrangement.
\end{proof}

\medskip

\noindent \textbf{Theorem \ref{thm:1}}

\begin{proof}
Let $t\in(0,T]$ be fixed and let $g_{X_{t}}(x)$, $x\in\mathbb{R}$, denote the cumulative distribution function of $X_{t}$. Since either $\sigma_{X}>0$ or $\nu_{X}(\mathbb{R})=\infty$, it is known (see, e.g., [Kohatsu-Higa and Takeuchi, 2019, \text{Thm.} 6.3.4] \cite{K-HT}) that $\mathcal{L}(X_{t})$ is absolutely continuous and the density function $g'_{X_{t}}(x)=f_{X_{t}}(x)$ exists. Then, by the total probability law,
\begin{equation}\label{A.6}
  \Q[X_{\T_{t}}+\rho Y_{t}\leq x]=\E[g_{X_{\T_{t}}+\rho Y_{t}}(x)].
\end{equation}
Here, $g_{X_{\T_{t}}+\rho Y_{t}}(x)$ should be understood as being conditional on the random vector $(\T_{t},Y_{t})$, and so $g'_{X_{\T_{t}}+\rho Y_{t}}(x)=f_{X_{\T_{t}}+\rho Y_{t}}(x)$ is a valid density function on $\mathbb{R}$, which can be viewed as a random variable for every fixed $x\in\mathbb{R}$. Since $\T_{t}$ is by construction nonzero $\Q$-\text{a.s.} and $g_{X_{\T_{t}}+\rho Y_{t}}(x)\geq0$, $\forall x\in\mathbb{R}$, an application of the Fubini--Tonelli theorem yields that the distribution of the compound variable $X_{\T_{t}}+\rho Y_{t}$ admits a density function equal to
\begin{align*}
  \frac{\dd}{\dd x}\Q[X_{\T_{t}}+\rho Y_{t}\leq x]&=\frac{\dd}{\dd x}\int_{\mathbb{R}_{+}\times\mathbb{R}_{++}}\int^{x}_{-\infty}f_{X_{\tau}+\rho y}(w)\dd w\dd\Q[\T_{t}\leq\tau,Y_{t}\leq y] \\
  &=\int_{\mathbb{R}_{+}\times\mathbb{R}_{++}}f_{X_{\tau}+\rho y}(x)\dd\Q[\T_{t}\leq\tau,Y_{t}\leq y] \\
  &=\E[f_{X_{\T_{t}}}(x-\rho Y_{t})].
\end{align*}

Based on (\ref{2.2.1}), the distribution $\mathcal{L}(\log S_{t})$ is also absolutely continuous by the continuous mapping theorem. Using the conditional representation in (\ref{A.2}) directly yields the continuity property for the conditional distribution on $\mathcal{F}_{t_{0}}$, given $t_{0}\in[0,t)$.
\end{proof}

\medskip

\noindent \textbf{Proposition \ref{pro:2}}

\begin{proof}
As before, fix $t_{0}\in[0,T)$ and let $g_{\log S_{T}|t_{0}}(x)$, $x\in\mathbb{R}$, denote the cumulative distribution function of $\log S_{T}$ conditional on $\mathcal{F}_{t_{0}}$, which also has a density function $f_{\log S_{T}|t_{0}}(x)$, $x\in\mathbb{R}$, due to Theorem \ref{thm:1}.

For convenience, we fix the conversion rate at $R=1$ as it is just a scalar multiple.\footnote{Notice that with $R=1$ the resulting pricing formulas also work for the inverse-power options with crypto-valued payoffs in (\ref{3.1.1}) and (\ref{3.1.2}).} For the inverse-power call option, the time-$t_{0}$ value is then
\begin{align*}
  C^{\rm (qip)}_{t_{0}}&=\E_{t_{0}}\bigg[\frac{(S^{p_{1}}_{T}-K^{p_{2}})^{+}}{S^{p_{1}}_{T}}\bigg] \\
  &=\Q_{t_{0}}[S^{p_{1}}_{T}\geq K^{p_{2}}]-\E_{t_{0}}\bigg[\frac{K^{p_{2}}}{S^{p_{1}}_{T}}\1_{\{S^{p_{1}}_{T}\geq K^{p_{2}}\}}\bigg] \\
  &=1-g_{\log S_{T}|t_{0}}\bigg(\frac{p_{2}}{p_{1}}\log K\bigg)-K^{p_{2}}\int^{\infty}_{p_{2}/p_{1}\log K}e^{-p_{1}x}f_{\log S_{T}|t_{0}}(x)\dd x.
\end{align*}
Applying the inverse Fourier transform with the Hermitian property of the characteristic function it follows that
\begin{equation*}
  \tilde{\Pi}_{2}=1-g_{\log S_{T}|t_{0}}\bigg(\frac{p_{2}}{p_{1}}\log K\bigg)=\frac{1}{2}+\frac{1}{\pi}\int^{\infty}_{0}\Re\bigg[\frac{K^{-\ii up_{2}/p_{1}}}{\ii u}\phi_{\log S_{T}|t_{0}}(u)\bigg]\dd u.
\end{equation*}
Also, substituting $f_{\log S_{T}|t_{0}}(x)=1/\pi\int^{\infty}_{0}\Re[e^{-\ii ux}\phi_{\log S_{T}|t_{0}}(u)]\dd u$ we have
\begin{align*}
  \tilde{\Pi}_{3}&=K^{p_{2}}\int^{\infty}_{p_{2}/p_{1}\log K}e^{-p_{1}x}\frac{1}{\pi}\int^{\infty}_{0}\Re[e^{-\ii ux}\phi_{\log S_{T}|t_{0}}(u)]\dd u\dd x \\
  &=K^{p_{2}}\int^{\infty}_{0}e^{-p_{1}(x+p_{2}/p_{1}\log K)}\frac{1}{\pi}\int^{\infty}_{0}\Re\big[e^{-\ii u(x+p_{2}/p_{1}\log K)}\phi_{\log S_{T}|t_{0}}(u)\big]\dd u\dd x \\
  &=\frac{1}{\pi}\int^{\infty}_{0}\Re\bigg[K^{-\ii up_{2}/p_{1}}\phi_{\log S_{T}|t_{0}}(u)\int^{\infty}_{0}e^{-p_{1}x-\ii ux}\dd x\bigg]\dd u \\
  &=\frac{1}{\pi}\int^{\infty}_{0}\Re\bigg[\frac{K^{-\ii up_{2}/p_{1}}\phi_{\log S_{T}|t_{0}}(u)}{\ii u+p_{1}}\bigg]\dd u,
\end{align*}
where the third equality uses the Fubini theorem, which is applicable provided $p_{1}>0$.

The valuation formula for the similar inverse-power put option stems from the parity argument that
\begin{equation*}
  \frac{(S^{p_{1}}_{T}-K^{p_{2}})^{+}}{S^{p_{1}}_{T}}-\frac{(K^{p_{2}}-S^{p_{1}}_{T})^{+}}{S^{p_{1}}_{T}}=1-\frac{K^{p_{2}}}{S^{p_{1}}_{T}},
\end{equation*}
along with the value of the inverse-power forward, $\E_{t_{0}}[S^{-p_{1}}_{T}]=\phi_{\log S_{T}|t_{0}}(\ii p_{1})$.
\end{proof}

\medskip

\noindent \textbf{Corollary \ref{cor:1}}

\begin{proof}
Note that we only consider the results for the Quanto inverse-power call option, from which those for the put option follow immediately. Since the process $V_{S}(\imath,T)$ is of finite variation, we have $\langle V^{\rm c}_{S},V^{\rm c}_{S}\rangle=0$ and $\langle S^{\rm c},V^{\rm c}\rangle=0$, $\Q$-a.s. A straightforward application of It\^{o}'s formula for general semimartingales (see, e.g., [Jacod and Shiryaev, 2010, \text{Thm.} 4.57] \cite{JS}) yields
\begin{align*}
  S_{t}&=S_{0}+\int^{t}_{0}S_{s-}\dd X_{\T_{s}}+\rho\int^{t}_{0}S_{s-}\dd Y_{s}+\int^{t}_{0}S_{s}\bigg(\bigg(\frac{1}{2}\sigma^{2}_{X}-\log\phi_{X_{1}}(-\ii)\bigg)A_{s}-\log\phi_{Y_{1}}(-\ii\rho)\bigg)\dd s \\
  &\quad+\sum_{s\leq t}S_{s-}\big(e^{\Delta X_{\T_{s}}+\rho\Delta Y_{s}}-1-\Delta X_{\T_{s}}-\rho\Delta Y_{s}\big),\quad t\geq0,
\end{align*}
where $\langle\cdot,\cdot\rangle$ acts as the dual predictable projection of $[\cdot,\cdot]$ for $(\mathbb{F},\Q)$-local martingales. Because $X$ has characteristics $(\mu_{X},\sigma_{X},\nu_{X})$, it follows that ($\Q$-\text{a.s.} for every $s\in[0,t]$)
\begin{equation*}
  \langle S^{\rm c},S^{\rm c}\rangle_{s}=\sigma^{2}_{X}\int^{t}_{0}S^{2}_{s}A_{s}\dd s,\quad\Delta S_{s}=S_{s-}(e^{\Delta X_{\T_{s}}+\rho\Delta Y_{s}}-1).
\end{equation*}

By treating the value of inverse power call option as a function of time $t$, the contemporaneous cryptocurrency price $S$, and the variance swap (inception) price $V_{S}\equiv V_{S}(\imath,T)$, namely $C^{\rm (qip)}_{t}\equiv C^{\rm (qip)}(t,S,V_{S})$, we have by using It\^{o}'s formula again that
\begin{align*}
  C^{\rm (qip)}_{t}&=C^{\rm (qip)}_{0}+\int^{t}_{0}\dot{C}^{\rm (qip)}_{s}\dd s+\int^{t}_{0}(\pd_{S}C^{\rm (qip)})_{s-}\dd S_{s}+\int^{t}_{0}(\pd_{V_{S}}C^{\rm (qip)})_{s-}\dd V_{S}(s,T) \\
  &\quad+\frac{1}{2}\int^{t}_{0}(\pd_{SS}C^{\rm (qip)})_{s}\dd\langle S^{\rm c},S^{\rm c}\rangle_{s}+\sum_{s\leq t}\big(\Delta C^{\rm (qip)}_{s}-\pd_{S}C^{\rm (qip)}_{s-}\Delta S_{s}-\pd_{V_{S}}C^{\rm (qip)}_{s-}\Delta V_{S}(s,T)\big)
\end{align*}
for $t\in[0,T]$. According to the pricing formulas in Proposition \ref{pro:1}, assuming the validity of doing so, taking the corresponding partial derivatives of (\ref{3.1.9}) comes down to differentiating the conditional characteristic function in (\ref{2.2.3}) and in (\ref{A.3}). After some simplifications we obtain that for every fixed $u\in\mathbb{R}$,
\begin{align}\label{A.7}
  (\dot{\phi}_{\log S_{T}|\cdot})_{s}(u)&=\phi_{\log S_{T}|s}(u)\bigg(\ii u\log\phi_{Y_{1}}(-\ii\rho)-\log\phi_{Y_{1}}(\rho u-H(T,s)(\ii\log\phi_{X_{1}}(u)+u\log\phi_{X_{1}}(-\ii))) \nonumber\\
  &\quad+(\log\phi_{X_{1}}(u)-\ii u\log\phi_{X_{1}}(-\ii))\bigg(\frac{\rho^{2}\Var[Y_{1}]}{\Var[X_{1}]}+H(T,s)\E[Y_{1}]\bigg)\bigg), \nonumber\\
  (\pd_{S}\phi_{\log S_{T}|\cdot})_{s}(u)&=\frac{\ii u\phi_{\log S_{T}|s}(u)}{S_{s}}, \nonumber\\
  (\pd_{SS}\phi_{\log S_{T}|\cdot})_{s}(u)&=\frac{(\ii u-1)\phi_{\log S_{T}|s}(u)}{S^{2}_{s}}, \nonumber\\
  (\pd_{V_{S}}\phi_{\log S_{T}|\cdot})_{s}(u)&=\frac{(\log\phi_{X_{1}}(u)-\ii u\log\phi_{X_{1}}(-\ii))\phi_{\log S_{T}|s}(u)}{\Var[X_{1}]},
\end{align}
and, with $\Delta V_{S}(s,T)=H(T,s)\Delta Y_{s}\Var[X_{1}]$ from (\ref{2.2.2}) and (\ref{A.4}),
\begin{equation}\label{A.8}
  (\Delta\phi_{\log S_{T}|\cdot})_{s}(u)=\phi_{\log S_{T}|s-}(u)\big(\exp\big(\ii u(\Delta X_{\T_{s}}+\rho\Delta Y_{s})+(\log\phi_{X_{1}}(u)-\ii u\phi_{X_{1}}(-\ii))H(T,s)\Delta Y_{s}\big)-1\big).
\end{equation}
By substituting (\ref{A.7}) and (\ref{A.8}) into (\ref{3.1.9}), all the items in (\ref{3.2.2}) follow immediately.

It remains to justify for (\ref{3.1.9}) the validity of differentiation under the integral sign, while passing it inside the real part is allowed by the Hermitian property of characteristic functions. Indeed, since $|K^{-\ii up_{2}/p_{1}}|=1$ and $\int^{\infty}_{0}\Re[\phi_{\log S_{T}|s}(u)]\dd u$ converges by the absolute continuity of $\mathcal{L}(\log S_{T}|\mathcal{F}_{s})$ (Theorem \ref{thm:1}), we only need to consider the tail integrability of the multiplier of $K^{-\ii up_{2}/p_{1}}\phi_{\log S_{T}|s}(u)$ in each item in (\ref{3.2.2}).

For example, for $(\pd_{S}C^{\rm (qip)})_{s}$, integrability is immediate because $1/(\ii u+p_{1})=O(1)$ and $1/(\ii u+p_{1})=O(u^{-1})$ as $u\searrow0$ and $u\rightarrow\infty$, respectively. For $(\pd_{V_{S}}C^{\rm (qip)})_{s}$, by the L\'{e}vy--Khintchine formula,
\begin{equation*}
  \log\phi_{X_{1}}(u)=\ii\mu_{X}u-\frac{1}{2}\sigma^{2}_{X}u^{2}+\int_{\mathbb{R}\setminus\{0\}}(e^{\ii uz}-1-\ii uz)\nu_{X}(\dd z)=O(u^{2}),\quad\text{as }u\rightarrow\infty,
\end{equation*}
so that $\log\phi_{X_{1}}(u)/(\ii p_{1}u-u^{2})=O(1)$ as $u\rightarrow\infty$, implying the desired tail integrability as well. The tail integrability for all the other items (namely $\dot{C}^{\rm (qip)}_{s}$, $(\pd^{2}_{S}C^{\rm (qip)})_{s}$, and $\Delta C^{\rm (qip)}_{s}$) can be justified similarly.
\end{proof}

\medskip

\noindent \textbf{Proposition \ref{pro:3}}

\begin{proof}
It is merely a matter of computation to establish the results. For conciseness, we only demonstrate the second (tempered stable) case. The first (asymmetric Laplace jump-diffusion) case follows the exact same idea and is also much easier to handle.

First, assume $c_{Y}>0$, so that
\begin{equation*}
  \log\phi_{Y_{1}}(u)=a_{Y}\Gf(-c_{Y})((b_{Y}-\ii u)^{c_{Y}}-b^{c_{Y}}_{Y}),\quad\Im u>-b_{Y}
\end{equation*}
from (\ref{4.2.4}). To compute the integral in (\ref{2.2.3}) involving $\log\phi_{Y_{1}}$, it suffices to focus on the integral of $(b_{Y}-\ii(\rho u+H_{3}(t-s)\varphi(u)))^{c_{Y}}$, where $\varphi$ is as defined in (\ref{4.2.5}). By letting $H_{3,-}$ and $H_{3,+}$ be the first piece and second piece, respectively, of $H$ in (\ref{4.1.3}), direct integration yields
\begin{align}\label{A.9}
  &\quad\int^{t}_{t_{0}}(b_{Y}-\ii(\rho u-H_{3}(t-s)\varphi(u)))^{c_{Y}}\dd s \nonumber\\
  &=
  \begin{cases}
    \displaystyle \int^{t}_{t_{0}}(\varphi_{1}(u)+H_{3,-}(t-s)\varphi(u)))^{c_{Y}}\dd s &\displaystyle\text{if }t-t_{0}<\frac{1-d}{\kappa} \\
    \displaystyle \int^{t}_{t-(1-d)/\kappa}(\varphi_{1}(u)+H_{3,-}(t-s)\varphi(u)))^{c_{Y}}\dd s \\
    \displaystyle \quad+\int^{t-(1-d)/\kappa}_{t_{0}}(\varphi_{1}(u)+H_{3,+}(t-s)\varphi(u)))^{c_{Y}}\dd s &\displaystyle\text{if }t-t_{0}\geq\frac{1-d}{\kappa}
  \end{cases} \nonumber\\
  &=
  \begin{cases}
    \displaystyle \int^{t-t_{0}}_{0}(\varphi_{1}(u)+s^{d}\varphi_{2}(u)))^{c_{Y}}\dd s &\displaystyle\text{if }t-t_{0}<\frac{1-d}{\kappa} \\
    \displaystyle \int^{(1-d)/\kappa}_{0}(\varphi_{1}(u)+s^{d}\varphi_{2}(u)))^{c_{Y}}\dd s \\
    \displaystyle \quad+\int^{t-t_{0}}_{(1-d)/\kappa}(\varphi_{3}(u)-e^{-\kappa s}\varphi_{4}(u))^{c_{Y}}\dd s &\displaystyle\text{if }t-t_{0}\geq\frac{1-d}{\kappa},
  \end{cases}
\end{align}
where $\varphi_{i}$, for $i\in\{1,2,3,4\}$, are defined in (\ref{4.2.5}).

For the first two integrals in the last equality of (\ref{A.9}), take a placeholder $\spadesuit\in\{t-t_{0},(1-d)/\kappa\}\subsetneq\mathbb{R}_{++}$; the substitution $s^{d}\mapsto s$ then gives
\begin{align*}
  \int^{\spadesuit}_{0}(\varphi_{1}(u)+s^{d}\varphi_{2}(u))^{c_{Y}}\dd s&=\frac{\varphi^{c_{Y}}_{1}(u)}{d}\int^{\spadesuit^{d}}_{0}s^{1/d-1}\bigg(1+\frac{s\varphi_{2}(u)}{\varphi_{1}(u)}\bigg)^{c_{Y}}\dd s \\
  &=\spadesuit\varphi^{c_{Y}}_{1}(u){}_{2}\F_{1}\bigg(-c_{Y},\frac{1}{d};\frac{1}{d}+1;-\frac{\spadesuit^{d}\varphi_{2}(u)}{\varphi_{1}(u)}\bigg),
\end{align*}
where the second equality uses [Gradshteyn and Ryzhik, 2007, \text{Eq.} 3.194.1] \cite{GR}. For the third integral, if $t-t_{0}\geq(1-d)/\kappa$, the substitution $e^{-\kappa s}\mapsto s$ gives
\begin{align*}
  \int^{t-t_{0}}_{(1-d)/\kappa}(\varphi_{3}(u)-e^{-\kappa s}\varphi_{4}(u))^{c_{Y}}\dd s&=\frac{\varphi^{c_{Y}}_{3}(u)}{\kappa}\int^{e^{-(1-d)}}_{e^{-\kappa(t-t_{0})}}s^{-1} \bigg(1-\frac{s\varphi_{4}(u)}{\varphi_{3}(u)}\bigg)^{c_{Y}}\dd s \\
  &=-\frac{\varphi^{c_{Y}}_{3}(u)}{\kappa}\bigg(\frac{\varphi_{3}(u)/(s\varphi_{4}(u))-1}{1-\varphi_{3}(u)/(s\varphi_{4}(u))}\bigg)^{c_{Y}} \\
  &\qquad\times\mathrm{B}_{\varphi_{3}(u)/(s\varphi_{4}(u))}(-c_{Y},c_{Y}+1)\bigg|^{e^{-(1-d)}}_{s=e^{-\kappa(t-t_{0})}} \\
  &=\frac{\varphi^{c_{Y}}_{3}(u)}{\kappa c_{Y}}\bigg(1-\frac{\varphi_{3}(u)}{s\varphi_{4}(u)}\bigg)\bigg(1-\frac{s\varphi_{4}(u)}{\varphi_{3}(u)}\bigg)^{c_{Y}} \\
  &\qquad\times{}_{2}\F_{1}\bigg(1,1;1-c_{Y};\frac{\varphi_{3}(u)}{s\varphi_{4}(u)}\bigg)\bigg|^{e^{-(1-d)}}_{s=e^{-\kappa(t-t_{0})}},
\end{align*}
where the second and third equalities follow from the definition of the incomplete beta function $\mathrm{B}\equiv\mathrm{B}_{\cdot}(\cdot,\cdot)$ and its fundamental relation to Gauss' hypergeometric function (see [Gradshteyn and Ryzhik, 2007, \text{Eq.} 8.391] \cite{GR}) along with Euler's transformation.

For the case $c_{Y}=0$, it is quite cumbersome to derive the results as direct limits of the above; instead, we perform a separate computation. With $c_{Y}=0$ and the procedures in (\ref{A.9}) in mind, we are led to evaluate the integrals of $\log(\varphi_{1}(u)+s^{d}\varphi_{2}(u))$ and $\log(\varphi_{3}(u)-e^{-\kappa s}\varphi_{4}(u))$, respectively. First, by expanding the logarithm around $s=0$, we have
\begin{align*}
  \int^{\spadesuit}_{0}\log(\varphi_{1}(u)+s^{d}\varphi_{2}(u))\dd s&=\frac{1}{d}\int^{\spadesuit^{d}}_{0}s^{1/d-1}\log(\varphi_{1}(u)+s\varphi_{2}(u))\dd s \\
  &=\frac{1}{d}\int^{\spadesuit^{d}}_{0}s^{1/d-1}\Bigg(\log\varphi_{1}(u)+\sum^{\infty}_{k=1}\frac{(-1)^{k}s^{k}}{k} \bigg(\frac{\varphi_{2}(u)}{\varphi_{1}(u)}\bigg)^{k}\Bigg)\dd s \\
  &=\spadesuit\log\varphi_{1}(u)-\sum^{\infty}_{k=1}\frac{(-1)^{k}\spadesuit^{kd+1}}{k(kd+1)}\bigg(\frac{\varphi_{2}(u)}{\varphi_{1}(u)}\bigg)^{k} \\
  &=\spadesuit\bigg(\log(\varphi_{1}(u)+\spadesuit^{d}\varphi_{2}(u)) +d\bigg({}_{2}\F_{1}\bigg(1,\frac{1}{d};\frac{1}{d}+1;-\frac{\spadesuit^{d}\varphi_{2}(u)}{\varphi_{1}(u)}\bigg)-1\bigg)\bigg),
\end{align*}
where the third equality results from interchanging integration and summation, the validity of which is justified by analytic continuation. In a similar fashion, with $t-t_{0}\geq(1-d)/\kappa$,
\begin{align*}
  \int^{t-t_{0}}_{(1-d)/\kappa}\log(\varphi_{3}(u)-e^{-\kappa s}\varphi_{4}(u))\dd s&=\frac{1}{\kappa}\int^{e^{-(1-d)}}_{e^{-\kappa(t-t_{0})}}s^{-1}\log(\varphi_{3}(u)-s\varphi_{4}(u))\dd s \\
  &=\frac{1}{\kappa}\Bigg(\log s\log\varphi_{3}(u)-\sum^{\infty}_{k=0}\frac{s^{k}}{k^{2}}\bigg(\frac{\varphi_{4}(u)}{\varphi_{3}(u)}\bigg)^{k}\Bigg) \bigg|^{e^{-(1-d)}}_{s=e^{-\kappa(t-t_{0})}} \\
  &=\frac{1}{\kappa}\bigg(\log s\log\varphi_{3}(u)-\Li_{2}\bigg(\frac{s\varphi_{4}(u)}{\varphi_{3}(u)}\bigg)\bigg)\bigg|^{e^{-(1-d)}}_{s=e^{-\kappa(t-t_{0})}}.
\end{align*}
This concludes the computation of the required integral upon simplification.

It remains to specify the involved moments of $X_{1}$ and $Y_{1}$. According to [K\"{u}chler and Tappe, 2013, \text{Eq.} 2.20 and \text{Eq.} 2.21] \cite{KT}, we have
\begin{equation*}
  \E[Y_{1}]=\frac{a_{Y}\Gf(1-c_{Y})}{b^{1-c_{Y}}_{Y}},\quad\Var[Y_{1}]=\frac{a_{Y}\Gf(2-c_{Y})}{b^{2-c_{Y}}_{Y}},
\end{equation*}
and similarly, by the law of total variance,
\begin{equation*}
  \Var[X_{1}]=\E[Z^{(n)}_{1}]+\theta^{2}\Var[Z^{(n)}_{1}]=\frac{a_{X}}{b^{2-c_{X}}_{X}}\bigg(\frac{b_{X}\Gf(1-c_{X})}{\Gf(n+2)} +\frac{\theta^{2}\Gf(2-c_{X})}{(2n+1)\Gf^{2}(n+1)}\bigg),
\end{equation*}
where the expectation and variance are directly gleaned from [Fei and Xia, 2024, \text{Eq.} 16] \cite{FX}. Putting everything together and arranging terms we arrive at the desired formula.
\end{proof}

\end{document}